\documentclass[prl,amssymb,reprint,superscriptaddress,showpacs,longbibliography]{revtex4-1}
\usepackage{amsmath}
\usepackage{amsfonts}
\usepackage{amssymb}
\usepackage{epsfig}
\usepackage{graphicx}
\usepackage[normalem]{ulem}
\renewcommand{\phi}{\varphi}
\newcommand{\be}{\begin{equation}}
\newcommand{\ee}{\end{equation}}
\newcommand{\bea}{\begin{equnaray}}
\newcommand{\eea}{\end{equnaray}}
\newcommand{\ba}{\begin{align}}
\newcommand{\ea}{\end{align}}

\usepackage{color}
\usepackage{comment}

\definecolor{green}{rgb}{0.0, 0.44, 0.0}
\definecolor{red}{rgb}{1.0, 0.13, 0.32}
\definecolor{blue}{rgb}{0.06, 0.2, 0.65}
\definecolor{magenta}{rgb}{1.0, 0.0, 1.00}
\definecolor{purple}{rgb}{0.7, 0.0, 0.7}
\definecolor{cyan}{rgb}{0.0, 1.0, 1.0}
\newcommand{\MO}[1]{{\bf \color{red} $^{MO}$ #1}}

\begin{document}

\title{Elasticity, Facilitation and Dynamic Heterogeneity in Glass Forming liquids}

\author{Misaki Ozawa}

\affiliation{Laboratoire de Physique de l'Ecole Normale Sup{\'e}rieure, ENS,
Universit\'e PSL, CNRS, Sorbonne Universit\'e, Universit\'e de Paris, F-75005 Paris, France}

\author{Giulio Biroli}
\affiliation{Laboratoire de Physique de l'Ecole Normale Sup{\'e}rieure, ENS,
Universit\'e PSL, CNRS, Sorbonne Universit\'e, Universit\'e de Paris, F-75005 Paris, France}

\begin{abstract}

We study the role of elasticity-induced facilitation on the dynamics of glass-forming liquids by a coarse-grained two-dimensional model in which local relaxation events, taking place by thermal activation, can trigger new relaxations by long-range elastically-mediated interactions. By simulations and an analytical theory, we show that the model reproduces the main salient facts associated with dynamic heterogeneity and offers a mechanism to explain the emergence of dynamical correlations at the glass transition. We also discuss how it can be generalized and combined with current theories.

\begin{comment}
We investigate the dynamics of glass-forming
liquids by using phenomenological modeling at a coarse-grained scale. Elastoplastic models (EPM) that have
been extensively studied in the rheology of amorphous materials are now applied to supercooled liquids
in thermal equilibrium as the phenomenological modeling. This EPM modeling naturally
explains the facilitated heterogeneous dynamics and elastic responses observed in recent molecular dynamics simulation studies and non-trivial growth of higher order fluctuations measured in real experiments. Moreover, a mean-field theory for the EPM explains qualitatively activation process observed in numerical simulations.
\end{comment}

\end{abstract}

\maketitle

Glass forming liquids display a huge slowing down of the dynamics, characterized by a relaxation time that grows by more than fourteen orders of magnitude when the temperature is reduced by just $2/3$ from its value at melting \cite{berthier2011theoretical}. Whereas it is very difficult to find signatures of this dramatic change of behavior in static correlation functions, a clear signal emerges in dynamical spatial correlations and length-scales \cite{berthier2011dynamical}. The associated concomitant growth of time and length-scales is reminiscent of critical slowing down at second-order phase transitions, and it is a hint of the collective nature of the relaxation processes underpinning glassy dynamics. This phenomenon, called dynamic heterogeneity (DH), has been a central one in the research on the glass transition both theoretically and experimentally \cite{berthier2011dynamical}. However, a full understanding of DH, especially close to the glass transition, is still lacking. 

In this respect, an important aspect is certainly dynamic facilitation \cite{berthier2011dynamical}. This is the property by which a local region that undergoes relaxational motion in a super-cooled liquid, or in a similar slow relaxing material, gives rise to or facilitates a neighboring local region to move and relax subsequently. Clearly, facilitation must play an important role in the emergence of spatial dynamical correlations. Some theories advocate that facilitation provides a complete explanation of DH \cite{chandler2010dynamics}, whereas others suggest that it is part of a more complex dynamical process associated with the growth of a static length-scale \cite{bhattacharyya2008facilitation} (see also the recent discussion in Ref.~\cite{biroli2022rfot}). Despite its important role and the fact that facilitation has been observed in numerical simulations and experiments \cite{berthier2011dynamical}, 
the cause of dynamic facilitation in real systems has not been fully elucidated yet.
A theory of glassy dynamics \cite{chandler2010dynamics,garrahan2011kinetically,keys2011excitations} posits the emergence of kinetic constraints and describes dynamical slowing down using Kinetically Constrained Models (KCM) \cite{ritort2003glassy}. In this case, facilitation is the main mechanism at play, however, its effect on dynamics is very dependent on the kind of kinetic constraint chosen, and a first-principle study of the mechanisms that would lead to specific kinetic constraints is currently lacking.

In this work, we address these issues by envisioning super-cooled liquids as solids that flow~\cite{dyre2006colloquium,lemaitre2014structural}. 
We consider that close to the glass transition dynamics proceeds by local events that take place in a surrounding matrix which is solid on the time-scale over which the local event takes place. In consequence, the local relaxation event causes an elastic deformation in the surrounding. Such deformation changes 
the arrangements of the particles and can then make some nearby regions more prone to relaxation. The recent work \cite{chacko2021elastoplasticity} has shown that this elastic mechanism indeed leads to dynamic facilitation. Reference~\cite{hasyim2021theory} has used elasticity theory to obtain an estimate of the energy scale associated with the dynamic facilitation theoretical scenario of Ref.~\cite{chandler2010dynamics}. Another series of works \cite{schoenholz2016structural,tah2022fragility} have concentrated on "softness" and elasticity as the fields that mediate facilitation. Here we study a coarse-grained model that encodes this "elasticity-induced facilitation". We show by numerical simulations and theoretical analysis that this mechanism allows us to capture the main salient facts associated with DH. Our model offers a starting point for a quantitative theory of dynamical correlations in glass-forming liquids, and it sheds new light on the theories of the glass transition. 

Two different lines of research merge in our approach. The first one encompasses the numerical studies which have demonstrated the existence and the importance of dynamic facilitation in the slow dynamics of atomistic models of glass-forming liquids. These results were first obtained at temperatures close to the Mode-Coupling-Theory cross-over \cite{candelier2010spatiotemporal,keys2011excitations} and then recently  
at lower temperatures (closer to the glass transition)  \cite{chacko2021elastoplasticity,guiselin2021microscopic,scalliet2022thirty}. 
The authors of Refs.~\cite{guiselin2021microscopic,scalliet2022thirty} have shown that dynamic facilitation is enhanced at lower temperatures and very likely plays an important role in the asymmetric relaxation spectra found in molecular liquid experiments~\cite{menon1992wide}. Note that the mean-field facilitated trap models introduced in ~\cite{bouchaud1995dynamical,scalliet2021excess} shares important conceptual similarities with the one studied in this work. 

The other line of research to which our work is related highlights the role of elasticity and plasticity, two mechanisms characteristic of the solid phase, in the equilibrium dynamics of super-cooled liquids. 
Besides the works \cite{chacko2021elastoplasticity,hasyim2021theory} that we already mentioned, there is growing numerical evidence that long-range, anisotropic stress correlations emerge in isotropic quenched liquid~\cite{lemaitre2014structural,lemaitre2015tensorial,chowdhury2016long,tong2020emergent,nishikawa2022relaxation}. Such Eshelby-like patterns are expected to be screened by thermal fluctuations at finite temperatures with some length-scale which increases with decreasing temperature~\cite{wu2015anisotropic,maier2017emergence,steffen2022molecular}. When this scale is large enough, one can envision the equilibrium dynamics of a super-cooled liquid as a solid that flows thanks to thermal fluctuations. In this scenario, which has been investigated numerically in Ref.~\cite{lerbinger2021relevance} (see also Ref.~\cite{li2022local}), plastic events activated by thermal fluctuations are responsible for equilibration and flow.

Glassy phenomenologies, in particular dynamic heterogeneity, have been universally observed in a wide variety of glassy materials such as metallic glasses, molecular liquids, polymers, and colloids, granular materials regardless of the details of the microscopic interactions~\cite{berthier2011dynamical} (and also for non time-reversible dynamics such as the one of granular media). This strongly suggests the essential physical ingredients that cause glassy dynamics emerge at a coarse-grained scale and can be captured within a coarse-grained simplified description.  
In consequence, we aim to study a coarse-grained model which encodes in the simplest way all these previous findings about elasticity, plasticity, thermal activation, and dynamic facilitation. 
The set of models that naturally fit this requirement are the elastoplastic models (EPMs). They have been extensively studied in the context of the rheology of amorphous solids under external loadings~\cite{nicolas2018deformation}.
EPMs are very successful, even quantitatively, in describing rheology and yield transitions of amorphous materials, in particular, under steady-state shear~\cite{lin2014scaling,nicolas2018deformation}.

In this work, we focus on a scalar EPM in which plastic relaxation is not induced by external shear but by thermal activation. We study the observables used to probe equilibrium dynamics and DH of super-cooled liquids \cite{berthier2011dynamical}. To the best of our knowledge, this research direction has not been explored, with the exception of a work by Bulatove and Argon~\cite{bulatov1994stochastic}, which considered only the energetic features of glass-forming liquids. 
\begin{figure}[t]
\includegraphics[width=0.95\columnwidth]{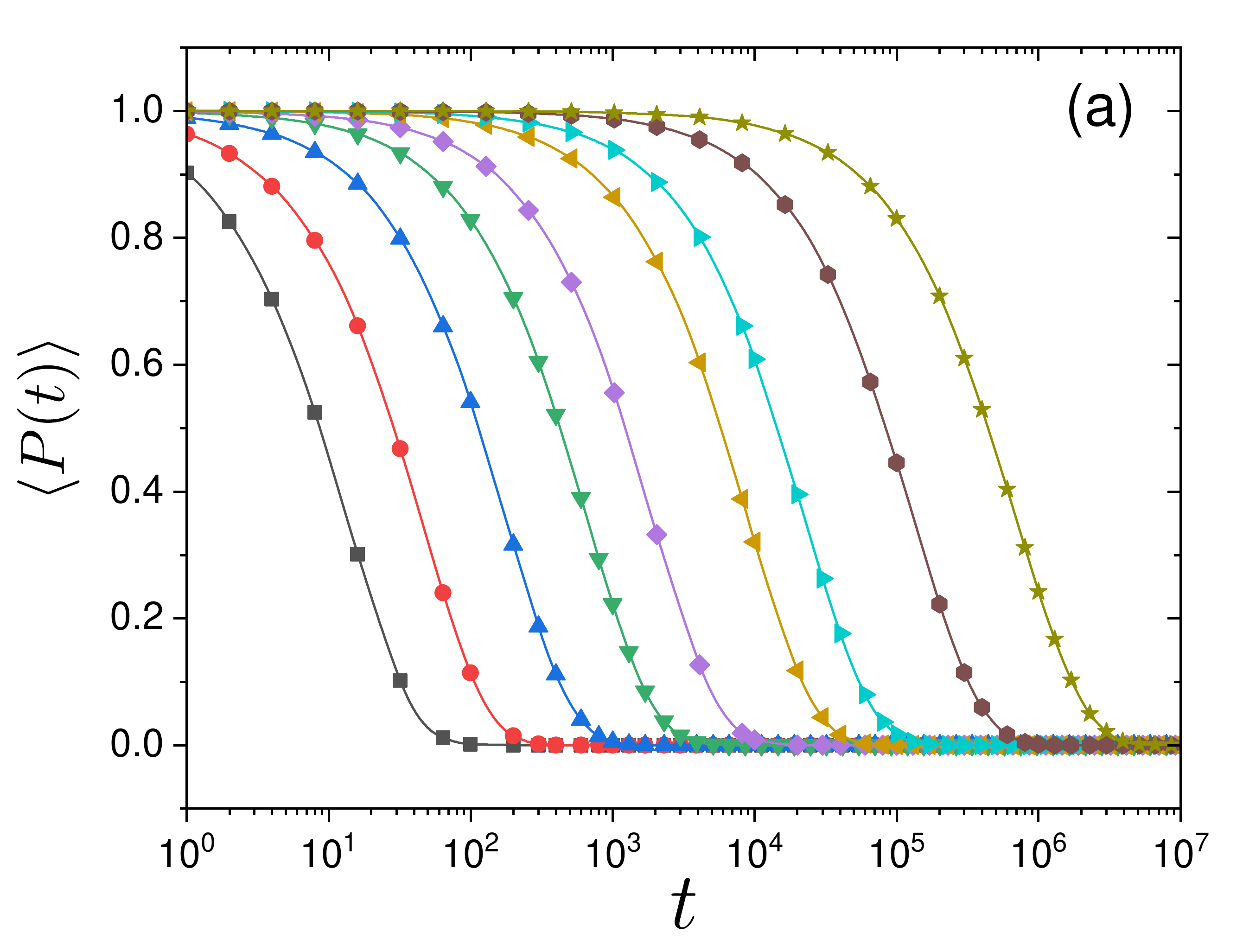}
\includegraphics[width=0.95\columnwidth]{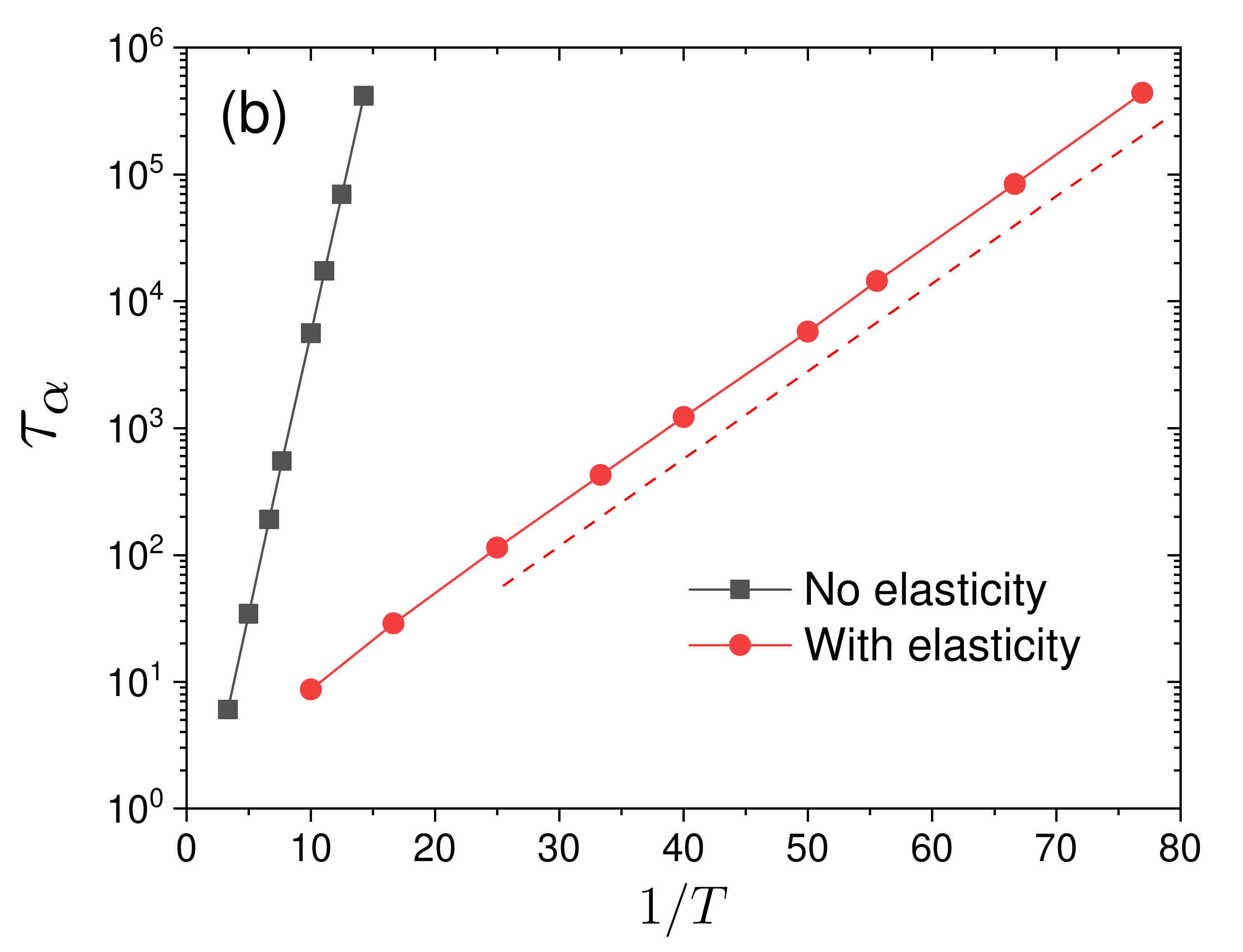}
\caption{(a): The average persistence function, $\langle P(t)\rangle$, for $T=0.100$, $0.060$, $0.040$, $0.030$, $0.025$, $0.020$, $0.018$, $0.015$, and $0.013$ (from left to right). (b): The relaxation time $\tau_{\alpha}$ for the models with (circle) and without (square) elastic interaction.
$\tau_{\alpha}$ is defined by $\langle P(\tau_{\alpha}) \rangle=1/2$. The dashed straight line defines an (average) activation energy barrier, $\Delta E(\overline \sigma^{\rm act})$.
}
\label{fig:dynamics}
\end{figure} 
The model we focus on, that we call EPM-Q, is defined on a $L \times L$ square lattice where each site represents a mesoscopic coarse-grained region of the equilibrium super-cooled liquid. 
To each site $i$ we associate a local stress $\sigma_i$, a { local} energy barrier $\Delta E(\sigma_i)$ activating the plastic relaxation event~\cite{ferrero2014relaxation,popovic2020thermally,ferrero2021yielding}, and an orientation $\psi_i \in (0, \pi/2]$ for the Eshelby elastic interaction.
The following dynamical rules of the model encode the effects of local thermal activation, local plastic rearrangement, and long-range and anisotropic elastic interaction.
At each time-step, we pick a site $i$ at random uniformly among the $L^2$ sites. If $|\sigma_i|$ is greater than or equal to a threshold value $\sigma_c>0$ then the site $i$ rearranges with probability one, whereas if $|\sigma_i|<\sigma_c$ then the site $i$ rearranges with probability $e^{-\Delta E(\sigma_i)/T}$, where $T$ is the temperature.
We set $\sigma_c=1$ in this study.
Because of this local plastic event, $\sigma_i$ is updated by a local stress drop: $\sigma_i \to \sigma_i - \delta \sigma_i$, where $\delta \sigma_i=(z+|\sigma_i|-\sigma_c) {\rm sgn}(\sigma_i)$; $z >0$ is a random number drawn by a distribution $\rho(z)$. The sign function ${\rm sgn}(x)$ takes into account that if $\sigma_i > 0$ (or $\sigma_i < 0$) local yielding is activated by a barrier at $\sigma_c$ (or $-\sigma_c$). 
The stress drop $\delta\sigma_i$ at site $i$ is then redistributed on the surrounding sites using the Eshelby kernel~\cite{picard2004elastic} with the (random) orientation $\psi_i$.  A new orientation is drawn uniformly at random after each plastic event.
We repeat the above attempt $L^2$ times, which corresponds to the unit of time.
Our choice of $\rho(z)$ and $\Delta E(\sigma)$ is based on previous literature that suggest the following forms~\cite{barbot2018local,popovic2020thermally,lerbinger2021relevance,ferrero2021yielding}:
\begin{equation}\label{eq:local-t}
    \rho(z)=\frac{1}{z_0} e^{-z/z_0}\,\,,\,\,\, \Delta E(\sigma)=K (\sigma_c-|\sigma|)^a,
\end{equation}
where $a=1.5$~\cite{maloney2006energy,fan2014thermally}, and the mean value $z_0$ and the generalized stiffness $K$ are set to one for simplicity~\footnote{ 
Note that the dynamics of EPM-Q does not verify time-reversal symmetry (or detailed balance). This is not necessarily a major concern as no spontaneous activity is created at small temperature. Moreover, the phenomenology of DH is observed also for non time-reversible systems, e.g. granular media \cite{berthier2011dynamical}. However, it is certainly a property that would be worth  
restoring in a more complete version of the model}. 
We study systems with $L=64$ and $128$, and we mainly show the results from $L=64$ unless otherwise stated.
All results presented in this paper are obtained in the stationary state.
More detailed model descriptions are presented in the supplementary information (SI).
We first present results on the bulk stationary dynamics. As in usual investigations of glassy dynamics, we have studied the intermediate scattering function \cite{kob1995testing}. To make the connection with studies on KCMs \cite{garrahan2011kinetically}, and since it is particularly well-suited for lattice systems, we have also focused on the average of the persistence function $P(t)=\frac{1}{L^2}\sum_i p_i(t)$, where $p_i(t)$ is equal to one if the site $i$ did not relax from time zero to time $t$ and zero otherwise~\cite{whitelam2005renormalization}. Both correlation functions behave in a qualitatively and quantitatively analogous way. We show the latter in Fig.~\ref{fig:dynamics}(a) and the former in the SI. 

\begin{figure}
\includegraphics[width=0.48\columnwidth]{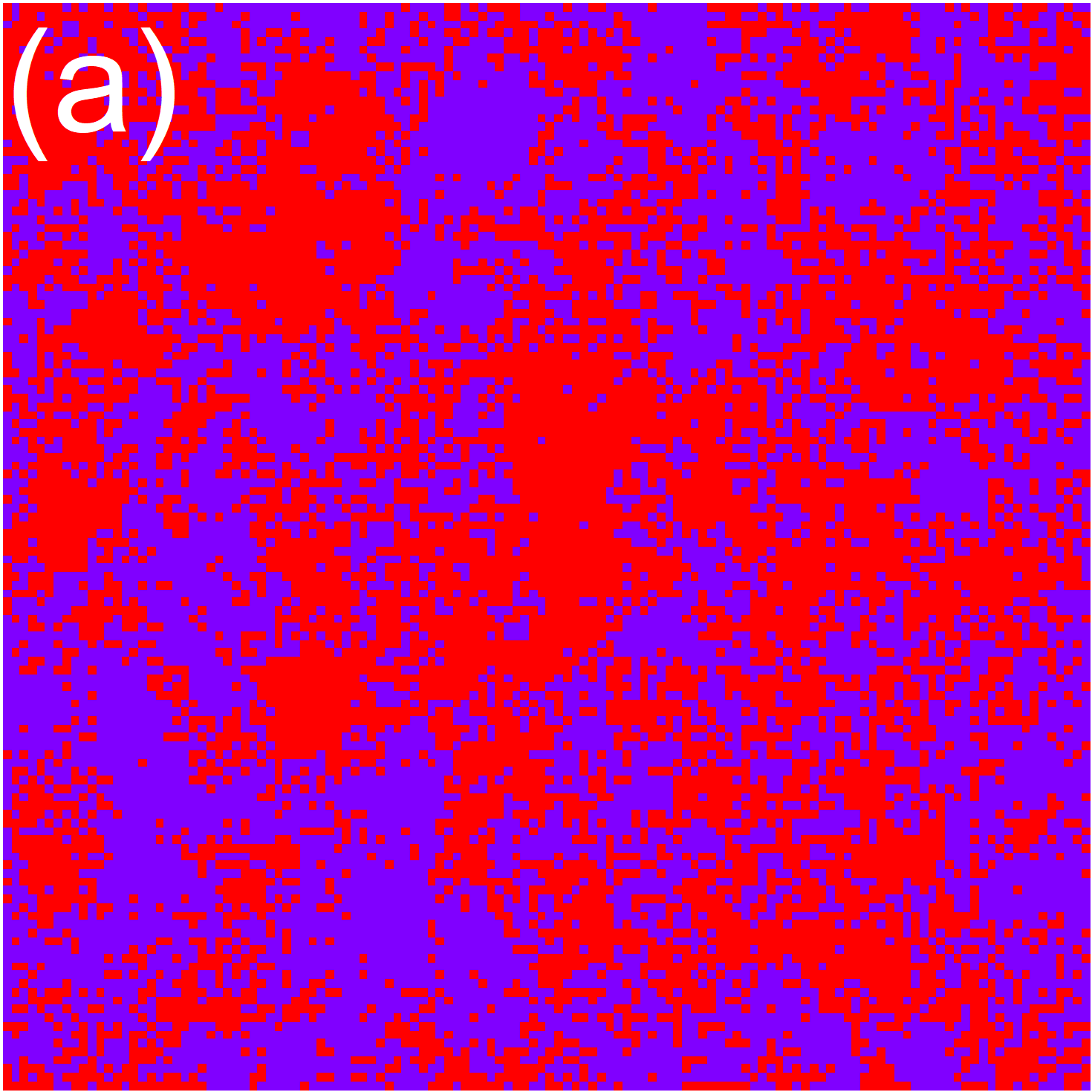}
\includegraphics[width=0.48\columnwidth]{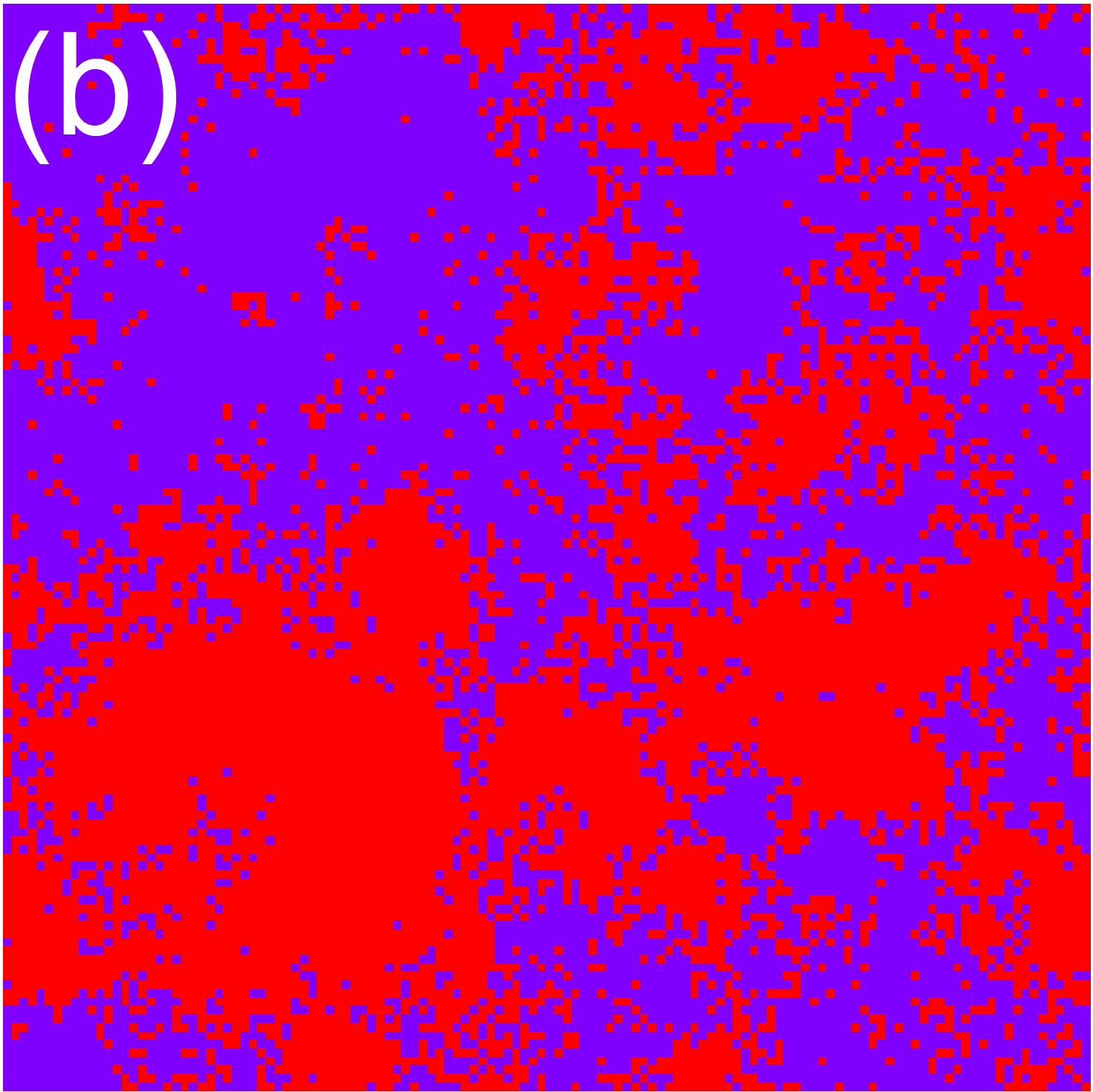}
\caption{Snapshots for local persistence, $p_i(\tau_{\alpha})$, when $P(\tau_{\alpha}) \approx 1/2$ for $T=0.040$ (a) and $T=0.013$ (b), respectively.
The system size is $L=128$.
Red and blue sites correspond to mobile ($p_i(\tau_{\alpha})=0$) and immobile ($p_i(\tau_{\alpha})=1$) sites, respectively.  
}
\label{fig:snap}
\end{figure}

Similarly to what is found for dynamical correlation functions in super-cooled liquids, $\langle P(t) \rangle$, where $\langle \cdots \rangle$ is the time average at the stationary state, decays in an increasingly sluggish way the more one decreases the temperature $T$, thus capturing the slowing down of the dynamics. 
The shape of the relaxation function is simpler than the one of realistic liquid models. This is due to the simplicity of the model and can be cured by generalizing it, as we shall discuss later. By plotting the relaxation time $\tau_{\alpha}$ (defined as $\langle P(\tau_{\alpha}) \rangle=1/2$) as a function of $1/T$ in Fig.~\ref{fig:dynamics}(b) we find that $\tau_{\alpha}$ diverges in an Arrhenius way when lowering the temperature. For comparison, we also plot $\tau_{\alpha}$ obtained from the model without elastic interactions, i.e., in the absence of stress redistribution. Remarkably, this relaxation time is larger, showing that elastic interactions substantially {\it diminish the 
energy barrier}. This is a direct evidence that elastic interactions facilitate and accelerate dynamics in the model. 
This conclusion is achieved thanks to the coarse-grained model approach we employ, where we can turn elasticity on and off. This is virtually impossible for molecular simulations since elasticity is an emergent property of a material.
The second direct evidence is provided by studying the morphology of dynamical correlations. Figure~\ref{fig:snap} shows the patterns formed by the local persistence $p_i(t)$ at two different temperatures: clearly, the dynamics is spatially heterogeneous over lengths that increase when lowering the temperature. The patterns in Fig.~\ref{fig:snap} strongly resemble the ones found in realistic (atomistic, colloidal, granular) systems \cite{berthier2011dynamical}.
The counterparts of Fig.~\ref{fig:snap} in the absence of elastic interactions (not shown) display no spacial dynamical correlations at all.

\begin{comment}
which is well-suited for lattice models and was studied thoroughly for KCM
Figure~\ref{fig:dynamics}(a) shows the persistent time autocorrelation function, $\langle P(t) \rangle$ often used in KCMs~\cite{garrahan2011kinetically}. 
$P(t)=\frac{1}{N}\sum_i p_i(t)$.
As with the usual two-point correlation functions, $\langle P(t) \rangle$ decays slowly with decreasing temperature $T$.
To evaluate the role of the elastic interaction on the dynamics, we plot the relaxation time $\tau_{\alpha}$ for two models with and without elastic interactions in Fig.~\ref{fig:dynamics}(b). Remarkably, the presence of the elastic interaction significantly accelerates or facilitates the dynamics. We also visually confirm the emergence of facilitated heterogeneous dynamics by watching simulation movies. 

\end{comment}

\begin{figure}
\includegraphics[width=0.95\columnwidth]{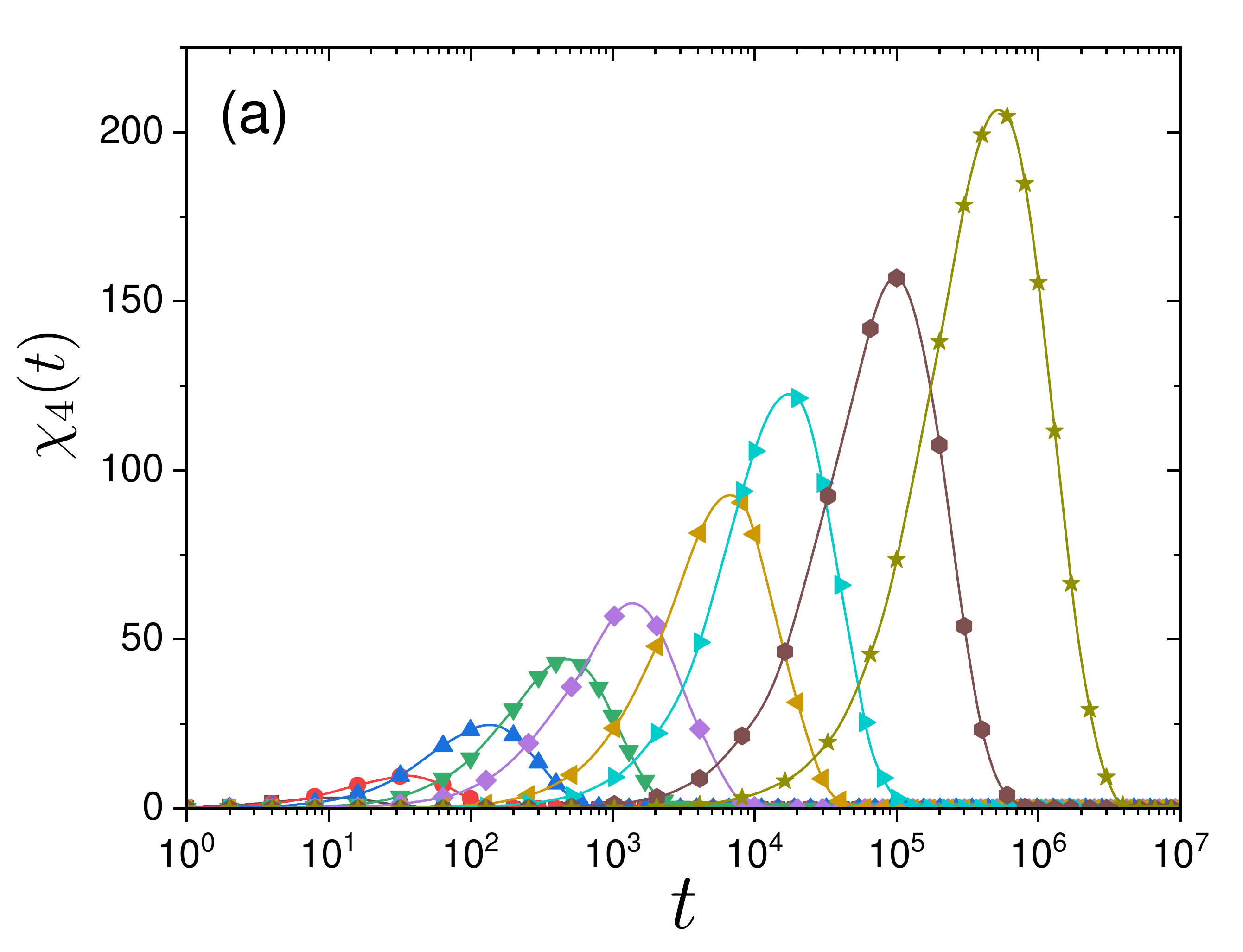}
\includegraphics[width=0.95\columnwidth]{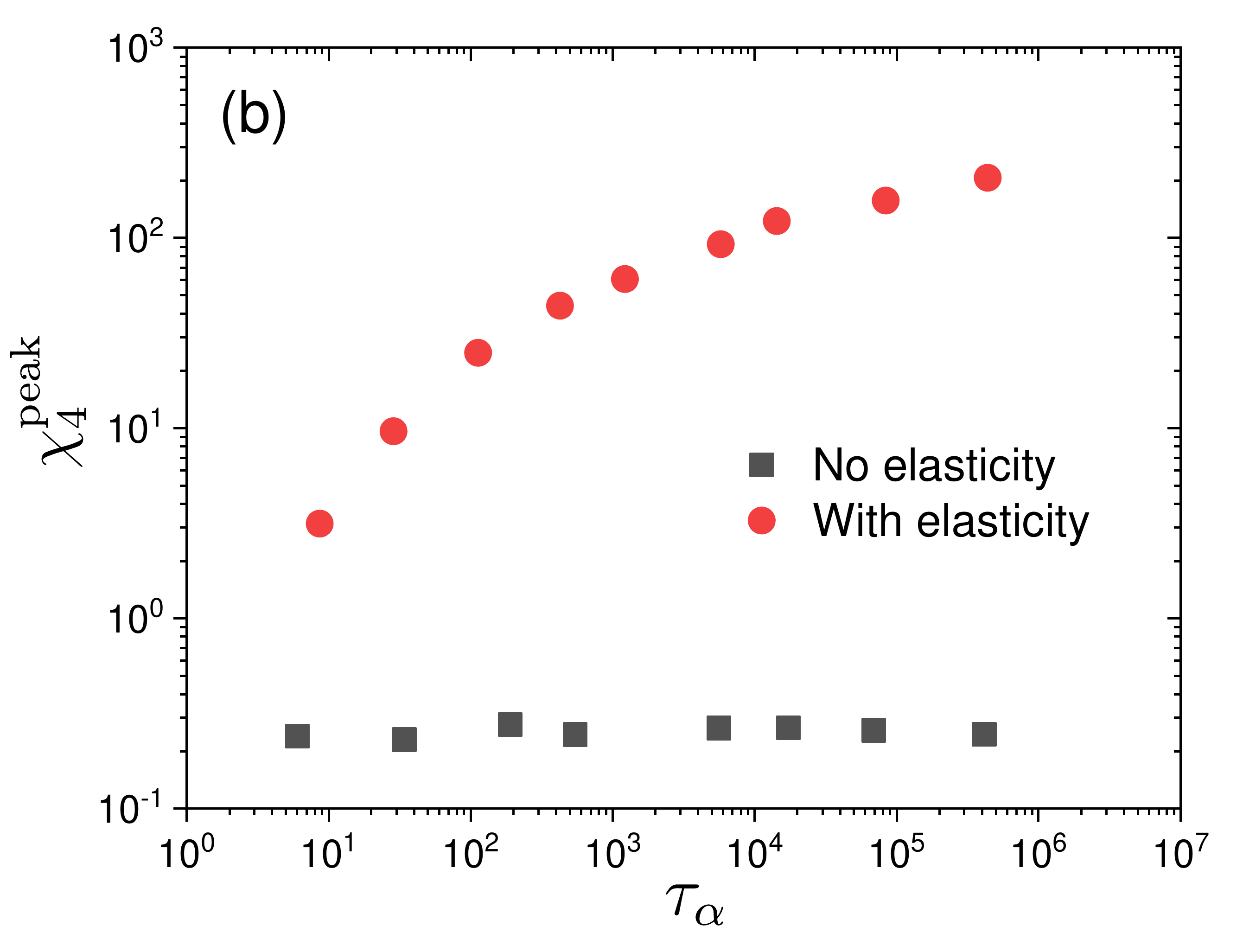}
\caption{(a): Time and temperature evolution of $\chi_4(t)$ for $T=0.100$, $0.060$, $0.040$, $0.030$, $0.025$, $0.020$, $0.018$, $0.015$, and $0.013$ (from left to right). (b): The peak value of $\chi_4$ as a function of $\tau_{\alpha}$ for the models with (circle) and without (square) elastic interactions.}
\label{fig:chi4}
\end{figure}

To quantify dynamic heterogeneity we measure the dynamical susceptibility~\cite{donati2002theory}, $\chi_4(t)=L^2\left( \langle P^2(t) \rangle -\langle P(t) \rangle^2 \right)$, in Fig.~\ref{fig:chi4}(a). We observe essentially the same time and temperature evolution found in molecular dynamics simulation of super-cooled liquids~\cite{berthier2011dynamical}. To study the relationship between time and length scales, we plot the peak of $\chi_4$ as a function of the logarithm of  $\tau_{\alpha}$ in Fig.~\ref{fig:chi4}(b). This curve displays a striking similarity with the ones obtained from experimental data~\cite{dalle2007spatial}: after a fast increase during the first decades of slowing down, the increase of the dynamical correlation length becomes slower, possibly logarithmic, with respect to $\tau_{\alpha}$.  
This is a highly non-trivial result that can be found only in some tailored KCMs \cite{garrahan2011kinetically} and it has been argued to hold for the Random First Order Transition theory \cite{lubchenko2007theory,biroli2012random}. In both cases, the bending shown in Fig.~\ref{fig:chi4}(b) is associated with cooperative dynamics. As we shall explain later, in our model, the reason is different (although it shares some similarities with KCMs), and it is purely due to facilitation.

\begin{figure}
\includegraphics[width=0.95\columnwidth]{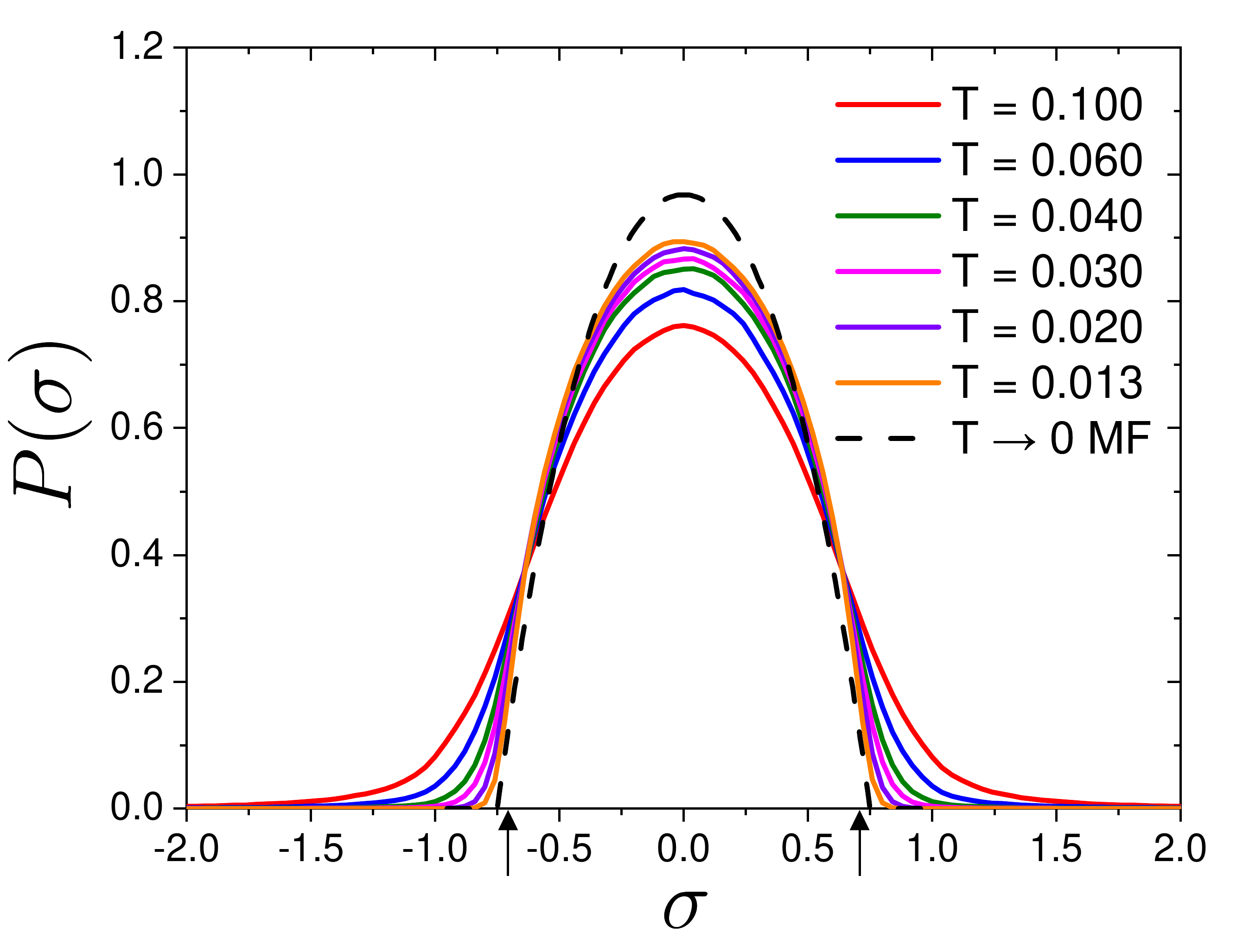}
\caption{Probability distribution function of the local stress, $P(\sigma)$. The dashed curve indicates the solution of the MF theory at $T \to 0$. The vertical arrows indicate the location of $\bar \sigma^{\rm act}$ and $-\bar \sigma^{\rm act}$ extracted from $\Delta E(\overline \sigma^{\rm act})$ with the activation energy barrier in Fig.~\ref{fig:dynamics}(a).
Recall that $\sigma_c=1$.
%\MO{we keep this value here $\bar \sigma = 0.70735$}
}
\label{fig:stress_distribution}
\end{figure}

We now offer a theoretical explanation for the phenomenological behavior presented above. Our starting point is the Kinetic Theory of plastic flow developed in Ref.~\cite{bocquet2009kinetic}. Using translation invariance, one finds that the kinetic equations of Ref.~\cite{bocquet2009kinetic} boils down in our case to the following H\'ebraud-Lequeux-like model~\cite{hebraud1998mode,agoritsas2015relevance,popovic2020thermally} for the probability distribution $P(\sigma, t)$ of the local stress $\sigma$ at a given site (see SI):
\begin{eqnarray}
\frac{\partial P(\sigma, t)}{\partial t} = D(t) \frac{\partial^2 P(\sigma, t)}{\partial \sigma^2} &-& \nu(\sigma, \sigma_c) P(\sigma, t) \nonumber \\ 
&+& \Gamma(t)y(\sigma),
\label{eq:HL_dynamics-text}
\end{eqnarray}
where the three terms on the RHS in Eq.~(\ref{eq:HL_dynamics-text}) respectively correspond (from left to right) to (i) the redistribution of the stress due to elastic interactions, (ii) a loss term due to rearrangements that change the local value of $\sigma$, and (iii) a gain term due to rearrangements in which the local value of the stress becomes equal to $\sigma$ after stress drops that take place with rate $\Gamma(t)= \int_{-\infty}^{\infty} \mathrm{d} \sigma \ \nu(\sigma, \sigma_c) P(\sigma, t)$. 
The strength of the first term, $D(t)$, is related to the total relaxation rate by $D(t)=\alpha \Gamma(t)$ \cite{bocquet2009kinetic}. In our case, in which the Eshelby orientations are randomly oriented, $\alpha\simeq 0.110$  (see SI), whereas $\nu(\sigma, \sigma_c)$ reads: 
\begin{eqnarray}
 \nu(\sigma, \sigma_c) = \frac{1}{\tau_0} \theta(|\sigma|-\sigma_c) + \frac{1}{\tau_0} e^{-\frac{\Delta E(\sigma)}{T}} \theta(\sigma_c-|\sigma|).
 \label{eq:nu_text}
\end{eqnarray}
The first term on the RHS in Eq.~(\ref{eq:nu_text}) is due to spontaneous relaxation when the system is locally unstable (beyond $\sigma_c$ or below $-\sigma_c$), whereas the second one is due to thermally activated relaxation~\cite{popovic2020thermally}. 
Stationary dynamics is described by a time-independent solution of Eq.~(\ref{eq:HL_dynamics-text}). Thus from now on, we drop the time index $t$.
As explained in the SI, we find that in the small temperature limit the stationary $P(\sigma)$ has a symmetric bell shape and is non-zero for $-\overline \sigma <\sigma<\overline \sigma$ with $\overline \sigma $ strictly less than $\sigma_c$. Its analytic expression is reported in the SI. We plot it in Fig. \ref{fig:stress_distribution} as a dashed (black) line and compare it to
$P(\sigma)$ obtained from numerical simulations of the two-dimensional model at different temperatures. The agreement is very good. Figure~\ref{fig:stress_distribution} numerically confirms that the support of  $\lim_{T\rightarrow 0}P(\sigma)$ is within the interval $[ -\overline \sigma,\overline \sigma]$ with $\overline \sigma<\sigma_c$, and show a numerical value of $\overline \sigma$ very close to the one we computed analytically. The stress $\overline \sigma$ defines the smallest typical energy barrier $\Delta E(\overline \sigma)$. As shown in the SI, 
the latter determines the relaxation rate $\Gamma$ for $T\rightarrow 0$: 
\begin{equation}
    \Gamma\simeq \frac{1}{\tau_0} e^{-\frac{\Delta E(\overline \sigma)}{T}},
\end{equation}
where $1/\tau_0$ is the rate of plastic event.
The above results lead to a scenario in which there is a spatially heterogeneous distribution of local stresses given by $P(\sigma)$. This leads to a distribution of energy barriers in the system. The sites having the smallest barriers, corresponding to $|\sigma|\simeq\overline \sigma$, are the ones triggering rearrangements and controlling the relaxation time at small temperatures. Our numerical findings presented before fully agrees with this picture: indeed, the activation energy associated with the Arrhenius behavior in Fig.~\ref{fig:dynamics}(b) corresponds to a stress $\overline \sigma^{\rm act}$ through the relation, $\tau_{\alpha} \sim 1/\Gamma \sim e^{\Delta E(\overline \sigma^{\rm act})/T}$, which is highlighted by the vertical arrow in  Fig.~\ref{fig:stress_distribution} and is identical (or very close) to the edge $\overline \sigma$ of the support of the stress distribution.
The effect of elasticity-induced facilitation on the relaxation time-scale, shown in Fig.~\ref{fig:dynamics}(b), is correctly reproduced by our analysis of Eq.~(\ref{eq:HL_dynamics-text}) which predicts $\tau_{\alpha} \sim 1/\Gamma \sim e^{\Delta E(0)/T}$ in absence of stress redistribution, hence a larger barrier $\Delta E(0)>\Delta E(\overline \sigma^{\rm act})$ (see SI).

The above scenario also offers an explanation for the development of DH. 
Once a site with $|\sigma|\simeq\overline \sigma$ triggers a local rearrangement, stress is redistributed around and can lead to subsequent relaxations. This is how dynamic facilitation takes place. 
A simple (upper bound) argument allows us to rationalize the existence of larger dynamic heterogeneities and a growing length-scale at a small temperature: in order to induce dynamical correlations, the stress redistribution received by a given site should at least change the local barrier by a factor larger than $T$ because only in this case the local relaxation time can change considerably. Since the stress redistribution is of order $1/r^d$ at a distance $r$ in $d$ spatial dimensions, this simple argument suggests that sites at linear distance $\ell<\xi(T)$, where $1/\xi^d(T) \sim T$, are dynamically correlated (this is actually only an upper bound on $\xi(T)$; developing  a complete theoretical argument is left for future work). 
The spatial relaxation pattern at $t \simeq \tau_\alpha$ is therefore formed by dynamically correlated regions of size $\xi^d$, which as explained in Ref.~\cite{toninelli2005dynamical}, implies a peak of $\chi_4\sim \xi^d\sim \frac{1}{T}\sim \log \tau_\alpha$. This conjectured behavior of $\chi_4$ is in qualitative agreement with our numerical findings, and it indeed leads to bending in the log-log plot of $\chi_4$ versus $\tau_\alpha$  (see also SI). 
The mechanism described above for DH is different from the ones at play in  KCMs~\cite{garrahan2011kinetically} and argued to hold for the Random First Order Transition \cite{lubchenko2007theory,biroli2012random}. 
Although EPM-Q shares some similarities with Kinetically Constrained Models, relaxation in EPM-Q is due to a combination of activation and elasticity instead of sub-diffusion of conserved rare defects \cite{garrahan2011kinetically}. In this new mechanism, avalanches of motion have a finite size and appears intermittently \footnote{Note that although the EPM-Q suggests a mechanism in which there is first a local activated relaxation and then a subsequent avalanche of motion, in a more refined model, verifying time-reversibility, also the reverse process would be possible. Therefore, one should think in terms of finite size avalanches resulting from elasticity and activation, and not in terms of an initiator event and a subsequent avalanche.}.  
Determining from atomistic simulations of glass-forming liquids which one of these scenarios holds is a very interesting open challenge, see e.g., Refs.~\cite{candelier2010spatiotemporal,keys2011excitations,scalliet2022thirty}.

In summary, we have shown that the simple EPM-Q model, which encodes  "elasticity-induced facilitation", is able to reproduce the salient features associated with the growth of dynamical correlation in glass-forming liquids. Some other important facts are instead missed, but as we argue below more realistic versions of the model should be able to capture them. There are, in fact, several research directions that stem from our work and are worth pursuing. 

For instance, DH is not only associated with heterogeneity in space but also with stretched exponential relaxation~\cite{ediger1996supercooled}. In the EPM-Q presented in this study, there is no heterogeneity (or disorder) in the solid state, as all the parameters associated with local relaxation events (threshold stress $\sigma_c$, the constants $z_0$ and $K$ in Eq.~(\ref{eq:local-t})) are the same on all sites. In a more realistic version, which takes into account the disorder of the amorphous solid~\cite{agoritsas2015relevance}, they should be random variables to be redrawn after a local relaxation. This would lead to a more heterogeneous distribution of barriers and hence of relaxation times, providing a possible mechanism for stretched relaxation. Furthermore, a more realistic model should also take into account that local stress relaxation is not instantaneous ~\cite{martens2012spontaneous}, and that a complete description should be tensorial instead of scalar~\cite{nicolas2014universal,budrikis2017universal}.

Another important generalization concerns the behavior of the relaxation time and the nature of the "local relaxation event". Even if the associated dynamical process were truly local and non-cooperative, the typical value of the local energy barriers could depend on the temperature. In fact, it is known that
the elastic constants increase when decreasing temperature. As argued in Refs.~\cite{dyre2006colloquium,nemilov2006interrelation,rouxel2011thermodynamics,mirigian2013unified,kapteijns2021does}, this leads to a mean activation energy for the structural relaxation
that increases approaching the glass transition. Combining these insights with our model (and hence introducing a realistic $T$ dependence of $K$ and $\sigma_c$) is certainly worth future studies, as it is a simple way to describe DH and super-Arrhenius behavior at the same time. Another possibility is that the single (mesoscopic) site relaxation process of our model could actually correspond to a cooperative many-particle rearrangement as the one envisioned to take place in Random First Order Transition theory. Within this perspective, the local degrees of freedom in our model correspond to what has been called {\it activons} in Ref.~\cite{biroli2022rfot} and are represented as local traps in Ref.~\cite{scalliet2021excess}. It would be very interesting to generalize our model to combine the Random First Order Transition scenario for cooperative relaxation with elasticity-induced facilitation. 

Finally, facilitation leads to avalanches of mobile particles, which are the building blocks of DH. EPMs have been fruitfully used to develop a scaling theory of avalanches in sheared amorphous solids~\cite{lin2014scaling}. The EPM-Q model paves the way for an analogous study of the avalanches of motion in super-cooled liquids~\cite{avalanche_inprogress}.

\begin{acknowledgments}
{\it Acknowledgments:}
We thank E. Agoritsas, L. Berthier, E. Bertin, J-P. Bouchaud, M. Cates, D. Frenkel, R. Jack, C. Liu, R. Mari, K. Martens, S. Rossi, S. Patinet, C. Scalliet, G. Tarjus, H. Yoshino, F. van Wijland, M. Wyart, and F. Zamponi for insightful discussions. This work was supported by grants from the Simons Foundation (\#454935 Giulio Biroli).
\end{acknowledgments}

\bibliography{apssamp.bib}

%merlin.mbs apsrev4-1.bst 2010-07-25 4.21a (PWD, AO, DPC) hacked
%Control: key (0)
%Control: author (0) dotless jnrlst
%Control: editor formatted (1) identically to author
%Control: production of article title (0) allowed
%Control: page (1) range
%Control: year (0) verbatim
%Control: production of eprint (0) enabled
\begin{thebibliography}{71}%
\makeatletter
\providecommand \@ifxundefined [1]{%
 \@ifx{#1\undefined}
}%
\providecommand \@ifnum [1]{%
 \ifnum #1\expandafter \@firstoftwo
 \else \expandafter \@secondoftwo
 \fi
}%
\providecommand \@ifx [1]{%
 \ifx #1\expandafter \@firstoftwo
 \else \expandafter \@secondoftwo
 \fi
}%
\providecommand \natexlab [1]{#1}%
\providecommand \enquote  [1]{``#1''}%
\providecommand \bibnamefont  [1]{#1}%
\providecommand \bibfnamefont [1]{#1}%
\providecommand \citenamefont [1]{#1}%
\providecommand \href@noop [0]{\@secondoftwo}%
\providecommand \href [0]{\begingroup \@sanitize@url \@href}%
\providecommand \@href[1]{\@@startlink{#1}\@@href}%
\providecommand \@@href[1]{\endgroup#1\@@endlink}%
\providecommand \@sanitize@url [0]{\catcode `\\12\catcode `\$12\catcode
  `\&12\catcode `\#12\catcode `\^12\catcode `\_12\catcode `\%12\relax}%
\providecommand \@@startlink[1]{}%
\providecommand \@@endlink[0]{}%
\providecommand \url  [0]{\begingroup\@sanitize@url \@url }%
\providecommand \@url [1]{\endgroup\@href {#1}{\urlprefix }}%
\providecommand \urlprefix  [0]{URL }%
\providecommand \Eprint [0]{\href }%
\providecommand \doibase [0]{http://dx.doi.org/}%
\providecommand \selectlanguage [0]{\@gobble}%
\providecommand \bibinfo  [0]{\@secondoftwo}%
\providecommand \bibfield  [0]{\@secondoftwo}%
\providecommand \translation [1]{[#1]}%
\providecommand \BibitemOpen [0]{}%
\providecommand \bibitemStop [0]{}%
\providecommand \bibitemNoStop [0]{.\EOS\space}%
\providecommand \EOS [0]{\spacefactor3000\relax}%
\providecommand \BibitemShut  [1]{\csname bibitem#1\endcsname}%
\let\auto@bib@innerbib\@empty
%</preamble>
\bibitem [{\citenamefont {Berthier}\ and\ \citenamefont
  {Biroli}(2011)}]{berthier2011theoretical}%
  \BibitemOpen
  \bibfield  {author} {\bibinfo {author} {\bibfnamefont {Ludovic}\ \bibnamefont
  {Berthier}}\ and\ \bibinfo {author} {\bibfnamefont {Giulio}\ \bibnamefont
  {Biroli}},\ }\bibfield  {title} {\enquote {\bibinfo {title} {Theoretical
  perspective on the glass transition and amorphous materials},}\ }\href@noop
  {} {\bibfield  {journal} {\bibinfo  {journal} {Reviews of modern physics}\
  }\textbf {\bibinfo {volume} {83}},\ \bibinfo {pages} {587} (\bibinfo {year}
  {2011})}\BibitemShut {NoStop}%
\bibitem [{\citenamefont {Berthier}\ \emph {et~al.}(2011)\citenamefont
  {Berthier}, \citenamefont {Biroli}, \citenamefont {Bouchaud}, \citenamefont
  {Cipelletti},\ and\ \citenamefont {van Saarloos}}]{berthier2011dynamical}%
  \BibitemOpen
  \bibfield  {author} {\bibinfo {author} {\bibfnamefont {Ludovic}\ \bibnamefont
  {Berthier}}, \bibinfo {author} {\bibfnamefont {Giulio}\ \bibnamefont
  {Biroli}}, \bibinfo {author} {\bibfnamefont {Jean-Philippe}\ \bibnamefont
  {Bouchaud}}, \bibinfo {author} {\bibfnamefont {Luca}\ \bibnamefont
  {Cipelletti}}, \ and\ \bibinfo {author} {\bibfnamefont {Wim}\ \bibnamefont
  {van Saarloos}},\ }\href@noop {} {\emph {\bibinfo {title} {Dynamical
  heterogeneities in glasses, colloids, and granular media}}},\ Vol.\ \bibinfo
  {volume} {150}\ (\bibinfo  {publisher} {OUP Oxford},\ \bibinfo {year}
  {2011})\BibitemShut {NoStop}%
\bibitem [{\citenamefont {Chandler}\ and\ \citenamefont
  {Garrahan}(2010)}]{chandler2010dynamics}%
  \BibitemOpen
  \bibfield  {author} {\bibinfo {author} {\bibfnamefont {David}\ \bibnamefont
  {Chandler}}\ and\ \bibinfo {author} {\bibfnamefont {Juan~P}\ \bibnamefont
  {Garrahan}},\ }\bibfield  {title} {\enquote {\bibinfo {title} {Dynamics on
  the way to forming glass: Bubbles in space-time},}\ }\href@noop {} {\bibfield
   {journal} {\bibinfo  {journal} {Annual review of physical chemistry}\
  }\textbf {\bibinfo {volume} {61}},\ \bibinfo {pages} {191--217} (\bibinfo
  {year} {2010})}\BibitemShut {NoStop}%
\bibitem [{\citenamefont {Bhattacharyya}\ \emph {et~al.}(2008)\citenamefont
  {Bhattacharyya}, \citenamefont {Bagchi},\ and\ \citenamefont
  {Wolynes}}]{bhattacharyya2008facilitation}%
  \BibitemOpen
  \bibfield  {author} {\bibinfo {author} {\bibfnamefont {Sarika~Maitra}\
  \bibnamefont {Bhattacharyya}}, \bibinfo {author} {\bibfnamefont {Biman}\
  \bibnamefont {Bagchi}}, \ and\ \bibinfo {author} {\bibfnamefont {Peter~G}\
  \bibnamefont {Wolynes}},\ }\bibfield  {title} {\enquote {\bibinfo {title}
  {Facilitation, complexity growth, mode coupling, and activated dynamics in
  supercooled liquids},}\ }\href@noop {} {\bibfield  {journal} {\bibinfo
  {journal} {Proceedings of the National Academy of Sciences}\ }\textbf
  {\bibinfo {volume} {105}},\ \bibinfo {pages} {16077--16082} (\bibinfo {year}
  {2008})}\BibitemShut {NoStop}%
\bibitem [{\citenamefont {Biroli}\ and\ \citenamefont
  {Bouchaud}(2022)}]{biroli2022rfot}%
  \BibitemOpen
  \bibfield  {author} {\bibinfo {author} {\bibfnamefont {Giulio}\ \bibnamefont
  {Biroli}}\ and\ \bibinfo {author} {\bibfnamefont {Jean-Philippe}\
  \bibnamefont {Bouchaud}},\ }\bibfield  {title} {\enquote {\bibinfo {title}
  {The rfot theory of glasses: Recent progress and open issues},}\ }\href@noop
  {} {\bibfield  {journal} {\bibinfo  {journal} {arXiv preprint
  arXiv:2208.05866}\ } (\bibinfo {year} {2022})}\BibitemShut {NoStop}%
\bibitem [{\citenamefont {Garrahan}\ \emph {et~al.}(2011)\citenamefont
  {Garrahan}, \citenamefont {Sollich},\ and\ \citenamefont
  {Toninelli}}]{garrahan2011kinetically}%
  \BibitemOpen
  \bibfield  {author} {\bibinfo {author} {\bibfnamefont {Juan~P}\ \bibnamefont
  {Garrahan}}, \bibinfo {author} {\bibfnamefont {Peter}\ \bibnamefont
  {Sollich}}, \ and\ \bibinfo {author} {\bibfnamefont {Cristina}\ \bibnamefont
  {Toninelli}},\ }\bibfield  {title} {\enquote {\bibinfo {title} {Kinetically
  constrained models},}\ }\href@noop {} {\bibfield  {journal} {\bibinfo
  {journal} {Dynamical heterogeneities in glasses, colloids, and granular
  media}\ }\textbf {\bibinfo {volume} {150}},\ \bibinfo {pages} {111--137}
  (\bibinfo {year} {2011})}\BibitemShut {NoStop}%
\bibitem [{\citenamefont {Keys}\ \emph {et~al.}(2011)\citenamefont {Keys},
  \citenamefont {Hedges}, \citenamefont {Garrahan}, \citenamefont {Glotzer},\
  and\ \citenamefont {Chandler}}]{keys2011excitations}%
  \BibitemOpen
  \bibfield  {author} {\bibinfo {author} {\bibfnamefont {Aaron~S}\ \bibnamefont
  {Keys}}, \bibinfo {author} {\bibfnamefont {Lester~O}\ \bibnamefont {Hedges}},
  \bibinfo {author} {\bibfnamefont {Juan~P}\ \bibnamefont {Garrahan}}, \bibinfo
  {author} {\bibfnamefont {Sharon~C}\ \bibnamefont {Glotzer}}, \ and\ \bibinfo
  {author} {\bibfnamefont {David}\ \bibnamefont {Chandler}},\ }\bibfield
  {title} {\enquote {\bibinfo {title} {Excitations are localized and relaxation
  is hierarchical in glass-forming liquids},}\ }\href@noop {} {\bibfield
  {journal} {\bibinfo  {journal} {Physical Review X}\ }\textbf {\bibinfo
  {volume} {1}},\ \bibinfo {pages} {021013} (\bibinfo {year}
  {2011})}\BibitemShut {NoStop}%
\bibitem [{\citenamefont {Ritort}\ and\ \citenamefont
  {Sollich}(2003)}]{ritort2003glassy}%
  \BibitemOpen
  \bibfield  {author} {\bibinfo {author} {\bibfnamefont {Felix}\ \bibnamefont
  {Ritort}}\ and\ \bibinfo {author} {\bibfnamefont {Peter}\ \bibnamefont
  {Sollich}},\ }\bibfield  {title} {\enquote {\bibinfo {title} {Glassy dynamics
  of kinetically constrained models},}\ }\href@noop {} {\bibfield  {journal}
  {\bibinfo  {journal} {Advances in physics}\ }\textbf {\bibinfo {volume}
  {52}},\ \bibinfo {pages} {219--342} (\bibinfo {year} {2003})}\BibitemShut
  {NoStop}%
\bibitem [{\citenamefont {Dyre}(2006)}]{dyre2006colloquium}%
  \BibitemOpen
  \bibfield  {author} {\bibinfo {author} {\bibfnamefont {Jeppe~C}\ \bibnamefont
  {Dyre}},\ }\bibfield  {title} {\enquote {\bibinfo {title} {Colloquium: The
  glass transition and elastic models of glass-forming liquids},}\ }\href@noop
  {} {\bibfield  {journal} {\bibinfo  {journal} {Reviews of modern physics}\
  }\textbf {\bibinfo {volume} {78}},\ \bibinfo {pages} {953} (\bibinfo {year}
  {2006})}\BibitemShut {NoStop}%
\bibitem [{\citenamefont {Lema{\^\i}tre}(2014)}]{lemaitre2014structural}%
  \BibitemOpen
  \bibfield  {author} {\bibinfo {author} {\bibfnamefont {Ana{\"e}l}\
  \bibnamefont {Lema{\^\i}tre}},\ }\bibfield  {title} {\enquote {\bibinfo
  {title} {Structural relaxation is a scale-free process},}\ }\href@noop {}
  {\bibfield  {journal} {\bibinfo  {journal} {Physical review letters}\
  }\textbf {\bibinfo {volume} {113}},\ \bibinfo {pages} {245702} (\bibinfo
  {year} {2014})}\BibitemShut {NoStop}%
\bibitem [{\citenamefont {Chacko}\ \emph {et~al.}(2021)\citenamefont {Chacko},
  \citenamefont {Landes}, \citenamefont {Biroli}, \citenamefont {Dauchot},
  \citenamefont {Liu},\ and\ \citenamefont
  {Reichman}}]{chacko2021elastoplasticity}%
  \BibitemOpen
  \bibfield  {author} {\bibinfo {author} {\bibfnamefont {Rahul~N}\ \bibnamefont
  {Chacko}}, \bibinfo {author} {\bibfnamefont {Fran{\c{c}}ois~P}\ \bibnamefont
  {Landes}}, \bibinfo {author} {\bibfnamefont {Giulio}\ \bibnamefont {Biroli}},
  \bibinfo {author} {\bibfnamefont {Olivier}\ \bibnamefont {Dauchot}}, \bibinfo
  {author} {\bibfnamefont {Andrea~J}\ \bibnamefont {Liu}}, \ and\ \bibinfo
  {author} {\bibfnamefont {David~R}\ \bibnamefont {Reichman}},\ }\bibfield
  {title} {\enquote {\bibinfo {title} {Elastoplasticity mediates dynamical
  heterogeneity below the mode coupling temperature},}\ }\href@noop {}
  {\bibfield  {journal} {\bibinfo  {journal} {Physical Review Letters}\
  }\textbf {\bibinfo {volume} {127}},\ \bibinfo {pages} {048002} (\bibinfo
  {year} {2021})}\BibitemShut {NoStop}%
\bibitem [{\citenamefont {Hasyim}\ and\ \citenamefont
  {Mandadapu}(2021)}]{hasyim2021theory}%
  \BibitemOpen
  \bibfield  {author} {\bibinfo {author} {\bibfnamefont {Muhammad~R}\
  \bibnamefont {Hasyim}}\ and\ \bibinfo {author} {\bibfnamefont {Kranthi~K}\
  \bibnamefont {Mandadapu}},\ }\bibfield  {title} {\enquote {\bibinfo {title}
  {A theory of localized excitations in supercooled liquids},}\ }\href@noop {}
  {\bibfield  {journal} {\bibinfo  {journal} {The Journal of chemical physics}\
  }\textbf {\bibinfo {volume} {155}},\ \bibinfo {pages} {044504} (\bibinfo
  {year} {2021})}\BibitemShut {NoStop}%
\bibitem [{\citenamefont {Schoenholz}\ \emph {et~al.}(2016)\citenamefont
  {Schoenholz}, \citenamefont {Cubuk}, \citenamefont {Sussman}, \citenamefont
  {Kaxiras},\ and\ \citenamefont {Liu}}]{schoenholz2016structural}%
  \BibitemOpen
  \bibfield  {author} {\bibinfo {author} {\bibfnamefont {Samuel~S}\
  \bibnamefont {Schoenholz}}, \bibinfo {author} {\bibfnamefont {Ekin~D}\
  \bibnamefont {Cubuk}}, \bibinfo {author} {\bibfnamefont {Daniel~M}\
  \bibnamefont {Sussman}}, \bibinfo {author} {\bibfnamefont {Efthimios}\
  \bibnamefont {Kaxiras}}, \ and\ \bibinfo {author} {\bibfnamefont {Andrea~J}\
  \bibnamefont {Liu}},\ }\bibfield  {title} {\enquote {\bibinfo {title} {A
  structural approach to relaxation in glassy liquids},}\ }\href@noop {}
  {\bibfield  {journal} {\bibinfo  {journal} {Nature Physics}\ }\textbf
  {\bibinfo {volume} {12}},\ \bibinfo {pages} {469--471} (\bibinfo {year}
  {2016})}\BibitemShut {NoStop}%
\bibitem [{\citenamefont {Tah}\ \emph {et~al.}(2022)\citenamefont {Tah},
  \citenamefont {Ridout},\ and\ \citenamefont {Liu}}]{tah2022fragility}%
  \BibitemOpen
  \bibfield  {author} {\bibinfo {author} {\bibfnamefont {Indrajit}\
  \bibnamefont {Tah}}, \bibinfo {author} {\bibfnamefont {Sean~A}\ \bibnamefont
  {Ridout}}, \ and\ \bibinfo {author} {\bibfnamefont {Andrea~J}\ \bibnamefont
  {Liu}},\ }\bibfield  {title} {\enquote {\bibinfo {title} {Fragility in glassy
  liquids: A structural approach based on machine learning},}\ }\href@noop {}
  {\bibfield  {journal} {\bibinfo  {journal} {arXiv preprint arXiv:2205.07187}\
  } (\bibinfo {year} {2022})}\BibitemShut {NoStop}%
\bibitem [{\citenamefont {Candelier}\ \emph {et~al.}(2010)\citenamefont
  {Candelier}, \citenamefont {Widmer-Cooper}, \citenamefont {Kummerfeld},
  \citenamefont {Dauchot}, \citenamefont {Biroli}, \citenamefont {Harrowell},\
  and\ \citenamefont {Reichman}}]{candelier2010spatiotemporal}%
  \BibitemOpen
  \bibfield  {author} {\bibinfo {author} {\bibfnamefont {Rapha{\"e}l}\
  \bibnamefont {Candelier}}, \bibinfo {author} {\bibfnamefont {Asaph}\
  \bibnamefont {Widmer-Cooper}}, \bibinfo {author} {\bibfnamefont {Jonathan~K}\
  \bibnamefont {Kummerfeld}}, \bibinfo {author} {\bibfnamefont {Olivier}\
  \bibnamefont {Dauchot}}, \bibinfo {author} {\bibfnamefont {Giulio}\
  \bibnamefont {Biroli}}, \bibinfo {author} {\bibfnamefont {Peter}\
  \bibnamefont {Harrowell}}, \ and\ \bibinfo {author} {\bibfnamefont {David~R}\
  \bibnamefont {Reichman}},\ }\bibfield  {title} {\enquote {\bibinfo {title}
  {Spatiotemporal hierarchy of relaxation events, dynamical heterogeneities,
  and structural reorganization in a supercooled liquid},}\ }\href@noop {}
  {\bibfield  {journal} {\bibinfo  {journal} {Physical review letters}\
  }\textbf {\bibinfo {volume} {105}},\ \bibinfo {pages} {135702} (\bibinfo
  {year} {2010})}\BibitemShut {NoStop}%
\bibitem [{\citenamefont {Guiselin}\ \emph {et~al.}(2021)\citenamefont
  {Guiselin}, \citenamefont {Scalliet},\ and\ \citenamefont
  {Berthier}}]{guiselin2021microscopic}%
  \BibitemOpen
  \bibfield  {author} {\bibinfo {author} {\bibfnamefont {Benjamin}\
  \bibnamefont {Guiselin}}, \bibinfo {author} {\bibfnamefont {Camille}\
  \bibnamefont {Scalliet}}, \ and\ \bibinfo {author} {\bibfnamefont {Ludovic}\
  \bibnamefont {Berthier}},\ }\bibfield  {title} {\enquote {\bibinfo {title}
  {Microscopic origin of excess wings in relaxation spectra of deeply
  supercooled liquids},}\ }\href@noop {} {\bibfield  {journal} {\bibinfo
  {journal} {Nature Physics (in press) arXiv preprint arXiv:2103.01569}\ }
  (\bibinfo {year} {2021})}\BibitemShut {NoStop}%
\bibitem [{\citenamefont {Scalliet}\ \emph {et~al.}(2022)\citenamefont
  {Scalliet}, \citenamefont {Guiselin},\ and\ \citenamefont
  {Berthier}}]{scalliet2022thirty}%
  \BibitemOpen
  \bibfield  {author} {\bibinfo {author} {\bibfnamefont {Camille}\ \bibnamefont
  {Scalliet}}, \bibinfo {author} {\bibfnamefont {Benjamin}\ \bibnamefont
  {Guiselin}}, \ and\ \bibinfo {author} {\bibfnamefont {Ludovic}\ \bibnamefont
  {Berthier}},\ }\bibfield  {title} {\enquote {\bibinfo {title} {Thirty
  milliseconds in the life of a supercooled liquid},}\ }\href@noop {}
  {\bibfield  {journal} {\bibinfo  {journal} {arXiv preprint arXiv:2207.00491}\
  } (\bibinfo {year} {2022})}\BibitemShut {NoStop}%
\bibitem [{\citenamefont {Menon}\ \emph {et~al.}(1992)\citenamefont {Menon},
  \citenamefont {O'Brien}, \citenamefont {Dixon}, \citenamefont {Wu},
  \citenamefont {Nagel}, \citenamefont {Williams},\ and\ \citenamefont
  {Carini}}]{menon1992wide}%
  \BibitemOpen
  \bibfield  {author} {\bibinfo {author} {\bibfnamefont {N}~\bibnamefont
  {Menon}}, \bibinfo {author} {\bibfnamefont {Kevin~P}\ \bibnamefont
  {O'Brien}}, \bibinfo {author} {\bibfnamefont {Paul~K}\ \bibnamefont {Dixon}},
  \bibinfo {author} {\bibfnamefont {Lei}\ \bibnamefont {Wu}}, \bibinfo {author}
  {\bibfnamefont {Sidney~R}\ \bibnamefont {Nagel}}, \bibinfo {author}
  {\bibfnamefont {Bruce~D}\ \bibnamefont {Williams}}, \ and\ \bibinfo {author}
  {\bibfnamefont {John~P}\ \bibnamefont {Carini}},\ }\bibfield  {title}
  {\enquote {\bibinfo {title} {Wide-frequency dielectric susceptibility
  measurements in glycerol},}\ }\href@noop {} {\bibfield  {journal} {\bibinfo
  {journal} {Journal of non-crystalline solids}\ }\textbf {\bibinfo {volume}
  {141}},\ \bibinfo {pages} {61--65} (\bibinfo {year} {1992})}\BibitemShut
  {NoStop}%
\bibitem [{\citenamefont {Bouchaud}\ \emph {et~al.}(1995)\citenamefont
  {Bouchaud}, \citenamefont {Comtet},\ and\ \citenamefont
  {Monthus}}]{bouchaud1995dynamical}%
  \BibitemOpen
  \bibfield  {author} {\bibinfo {author} {\bibfnamefont {Jean-Philippe}\
  \bibnamefont {Bouchaud}}, \bibinfo {author} {\bibfnamefont {Alain}\
  \bibnamefont {Comtet}}, \ and\ \bibinfo {author} {\bibfnamefont {C{\'e}cile}\
  \bibnamefont {Monthus}},\ }\bibfield  {title} {\enquote {\bibinfo {title} {On
  a dynamical model of glasses},}\ }\href@noop {} {\bibfield  {journal}
  {\bibinfo  {journal} {Journal de Physique I}\ }\textbf {\bibinfo {volume}
  {5}},\ \bibinfo {pages} {1521--1526} (\bibinfo {year} {1995})}\BibitemShut
  {NoStop}%
\bibitem [{\citenamefont {Scalliet}\ \emph {et~al.}(2021)\citenamefont
  {Scalliet}, \citenamefont {Guiselin},\ and\ \citenamefont
  {Berthier}}]{scalliet2021excess}%
  \BibitemOpen
  \bibfield  {author} {\bibinfo {author} {\bibfnamefont {Camille}\ \bibnamefont
  {Scalliet}}, \bibinfo {author} {\bibfnamefont {Benjamin}\ \bibnamefont
  {Guiselin}}, \ and\ \bibinfo {author} {\bibfnamefont {Ludovic}\ \bibnamefont
  {Berthier}},\ }\bibfield  {title} {\enquote {\bibinfo {title} {Excess wings
  and asymmetric relaxation spectra in a facilitated trap model},}\ }\href@noop
  {} {\bibfield  {journal} {\bibinfo  {journal} {The Journal of Chemical
  Physics}\ }\textbf {\bibinfo {volume} {155}},\ \bibinfo {pages} {064505}
  (\bibinfo {year} {2021})}\BibitemShut {NoStop}%
\bibitem [{\citenamefont {Lema{\^\i}tre}(2015)}]{lemaitre2015tensorial}%
  \BibitemOpen
  \bibfield  {author} {\bibinfo {author} {\bibfnamefont {Ana{\"e}l}\
  \bibnamefont {Lema{\^\i}tre}},\ }\bibfield  {title} {\enquote {\bibinfo
  {title} {Tensorial analysis of eshelby stresses in 3d supercooled liquids},}\
  }\href@noop {} {\bibfield  {journal} {\bibinfo  {journal} {The Journal of
  chemical physics}\ }\textbf {\bibinfo {volume} {143}},\ \bibinfo {pages}
  {164515} (\bibinfo {year} {2015})}\BibitemShut {NoStop}%
\bibitem [{\citenamefont {Chowdhury}\ \emph {et~al.}(2016)\citenamefont
  {Chowdhury}, \citenamefont {Abraham}, \citenamefont {Hudson},\ and\
  \citenamefont {Harrowell}}]{chowdhury2016long}%
  \BibitemOpen
  \bibfield  {author} {\bibinfo {author} {\bibfnamefont {Sadrul}\ \bibnamefont
  {Chowdhury}}, \bibinfo {author} {\bibfnamefont {Sneha}\ \bibnamefont
  {Abraham}}, \bibinfo {author} {\bibfnamefont {Toby}\ \bibnamefont {Hudson}},
  \ and\ \bibinfo {author} {\bibfnamefont {Peter}\ \bibnamefont {Harrowell}},\
  }\bibfield  {title} {\enquote {\bibinfo {title} {Long range stress
  correlations in the inherent structures of liquids at rest},}\ }\href@noop {}
  {\bibfield  {journal} {\bibinfo  {journal} {The Journal of chemical physics}\
  }\textbf {\bibinfo {volume} {144}},\ \bibinfo {pages} {124508} (\bibinfo
  {year} {2016})}\BibitemShut {NoStop}%
\bibitem [{\citenamefont {Tong}\ \emph {et~al.}(2020)\citenamefont {Tong},
  \citenamefont {Sengupta},\ and\ \citenamefont {Tanaka}}]{tong2020emergent}%
  \BibitemOpen
  \bibfield  {author} {\bibinfo {author} {\bibfnamefont {Hua}\ \bibnamefont
  {Tong}}, \bibinfo {author} {\bibfnamefont {Shiladitya}\ \bibnamefont
  {Sengupta}}, \ and\ \bibinfo {author} {\bibfnamefont {Hajime}\ \bibnamefont
  {Tanaka}},\ }\bibfield  {title} {\enquote {\bibinfo {title} {Emergent
  solidity of amorphous materials as a consequence of mechanical
  self-organisation},}\ }\href@noop {} {\bibfield  {journal} {\bibinfo
  {journal} {Nature communications}\ }\textbf {\bibinfo {volume} {11}},\
  \bibinfo {pages} {1--10} (\bibinfo {year} {2020})}\BibitemShut {NoStop}%
\bibitem [{\citenamefont {Nishikawa}\ \emph {et~al.}(2022)\citenamefont
  {Nishikawa}, \citenamefont {Ozawa}, \citenamefont {Ikeda}, \citenamefont
  {Chaudhuri},\ and\ \citenamefont {Berthier}}]{nishikawa2022relaxation}%
  \BibitemOpen
  \bibfield  {author} {\bibinfo {author} {\bibfnamefont {Yoshihiko}\
  \bibnamefont {Nishikawa}}, \bibinfo {author} {\bibfnamefont {Misaki}\
  \bibnamefont {Ozawa}}, \bibinfo {author} {\bibfnamefont {Atsushi}\
  \bibnamefont {Ikeda}}, \bibinfo {author} {\bibfnamefont {Pinaki}\
  \bibnamefont {Chaudhuri}}, \ and\ \bibinfo {author} {\bibfnamefont {Ludovic}\
  \bibnamefont {Berthier}},\ }\bibfield  {title} {\enquote {\bibinfo {title}
  {Relaxation dynamics in the energy landscape of glass-forming liquids},}\
  }\href@noop {} {\bibfield  {journal} {\bibinfo  {journal} {Physical Review
  X}\ }\textbf {\bibinfo {volume} {12}},\ \bibinfo {pages} {021001} (\bibinfo
  {year} {2022})}\BibitemShut {NoStop}%
\bibitem [{\citenamefont {Wu}\ \emph {et~al.}(2015)\citenamefont {Wu},
  \citenamefont {Iwashita},\ and\ \citenamefont {Egami}}]{wu2015anisotropic}%
  \BibitemOpen
  \bibfield  {author} {\bibinfo {author} {\bibfnamefont {Bin}\ \bibnamefont
  {Wu}}, \bibinfo {author} {\bibfnamefont {Takuya}\ \bibnamefont {Iwashita}}, \
  and\ \bibinfo {author} {\bibfnamefont {Takeshi}\ \bibnamefont {Egami}},\
  }\bibfield  {title} {\enquote {\bibinfo {title} {Anisotropic stress
  correlations in two-dimensional liquids},}\ }\href@noop {} {\bibfield
  {journal} {\bibinfo  {journal} {Physical Review E}\ }\textbf {\bibinfo
  {volume} {91}},\ \bibinfo {pages} {032301} (\bibinfo {year}
  {2015})}\BibitemShut {NoStop}%
\bibitem [{\citenamefont {Maier}\ \emph {et~al.}(2017)\citenamefont {Maier},
  \citenamefont {Zippelius},\ and\ \citenamefont {Fuchs}}]{maier2017emergence}%
  \BibitemOpen
  \bibfield  {author} {\bibinfo {author} {\bibfnamefont {Manuel}\ \bibnamefont
  {Maier}}, \bibinfo {author} {\bibfnamefont {Annette}\ \bibnamefont
  {Zippelius}}, \ and\ \bibinfo {author} {\bibfnamefont {Matthias}\
  \bibnamefont {Fuchs}},\ }\bibfield  {title} {\enquote {\bibinfo {title}
  {Emergence of long-ranged stress correlations at the liquid to glass
  transition},}\ }\href@noop {} {\bibfield  {journal} {\bibinfo  {journal}
  {Physical review letters}\ }\textbf {\bibinfo {volume} {119}},\ \bibinfo
  {pages} {265701} (\bibinfo {year} {2017})}\BibitemShut {NoStop}%
\bibitem [{\citenamefont {Steffen}\ \emph {et~al.}(2022)\citenamefont
  {Steffen}, \citenamefont {Schneider}, \citenamefont {M{\"u}ller},\ and\
  \citenamefont {Rottler}}]{steffen2022molecular}%
  \BibitemOpen
  \bibfield  {author} {\bibinfo {author} {\bibfnamefont {David}\ \bibnamefont
  {Steffen}}, \bibinfo {author} {\bibfnamefont {Ludwig}\ \bibnamefont
  {Schneider}}, \bibinfo {author} {\bibfnamefont {Marcus}\ \bibnamefont
  {M{\"u}ller}}, \ and\ \bibinfo {author} {\bibfnamefont {J{\"o}rg}\
  \bibnamefont {Rottler}},\ }\bibfield  {title} {\enquote {\bibinfo {title}
  {Molecular simulations and hydrodynamic theory of nonlocal shear-stress
  correlations in supercooled fluids},}\ }\href@noop {} {\bibfield  {journal}
  {\bibinfo  {journal} {The Journal of Chemical Physics}\ }\textbf {\bibinfo
  {volume} {157}},\ \bibinfo {pages} {064501} (\bibinfo {year}
  {2022})}\BibitemShut {NoStop}%
\bibitem [{\citenamefont {Lerbinger}\ \emph {et~al.}(2021)\citenamefont
  {Lerbinger}, \citenamefont {Barbot}, \citenamefont {Vandembroucq},\ and\
  \citenamefont {Patinet}}]{lerbinger2021relevance}%
  \BibitemOpen
  \bibfield  {author} {\bibinfo {author} {\bibfnamefont {Matthias}\
  \bibnamefont {Lerbinger}}, \bibinfo {author} {\bibfnamefont {Armand}\
  \bibnamefont {Barbot}}, \bibinfo {author} {\bibfnamefont {Damien}\
  \bibnamefont {Vandembroucq}}, \ and\ \bibinfo {author} {\bibfnamefont
  {Sylvain}\ \bibnamefont {Patinet}},\ }\bibfield  {title} {\enquote {\bibinfo
  {title} {On the relevance of shear transformations in the relaxation of
  supercooled liquids},}\ }\href@noop {} {\bibfield  {journal} {\bibinfo
  {journal} {arXiv preprint arXiv:2109.12639}\ } (\bibinfo {year}
  {2021})}\BibitemShut {NoStop}%
\bibitem [{\citenamefont {Li}\ \emph {et~al.}(2022)\citenamefont {Li},
  \citenamefont {Yao},\ and\ \citenamefont {Ciamarra}}]{li2022local}%
  \BibitemOpen
  \bibfield  {author} {\bibinfo {author} {\bibfnamefont {Yan-Wei}\ \bibnamefont
  {Li}}, \bibinfo {author} {\bibfnamefont {Yugui}\ \bibnamefont {Yao}}, \ and\
  \bibinfo {author} {\bibfnamefont {Massimo~Pica}\ \bibnamefont {Ciamarra}},\
  }\bibfield  {title} {\enquote {\bibinfo {title} {Local plastic response and
  slow heterogeneous dynamics of supercooled liquids},}\ }\href@noop {}
  {\bibfield  {journal} {\bibinfo  {journal} {Physical Review Letters}\
  }\textbf {\bibinfo {volume} {128}},\ \bibinfo {pages} {258001} (\bibinfo
  {year} {2022})}\BibitemShut {NoStop}%
\bibitem [{\citenamefont {Nicolas}\ \emph {et~al.}(2018)\citenamefont
  {Nicolas}, \citenamefont {Ferrero}, \citenamefont {Martens},\ and\
  \citenamefont {Barrat}}]{nicolas2018deformation}%
  \BibitemOpen
  \bibfield  {author} {\bibinfo {author} {\bibfnamefont {Alexandre}\
  \bibnamefont {Nicolas}}, \bibinfo {author} {\bibfnamefont {Ezequiel~E}\
  \bibnamefont {Ferrero}}, \bibinfo {author} {\bibfnamefont {Kirsten}\
  \bibnamefont {Martens}}, \ and\ \bibinfo {author} {\bibfnamefont
  {Jean-Louis}\ \bibnamefont {Barrat}},\ }\bibfield  {title} {\enquote
  {\bibinfo {title} {Deformation and flow of amorphous solids: Insights from
  elastoplastic models},}\ }\href@noop {} {\bibfield  {journal} {\bibinfo
  {journal} {Reviews of Modern Physics}\ }\textbf {\bibinfo {volume} {90}},\
  \bibinfo {pages} {045006} (\bibinfo {year} {2018})}\BibitemShut {NoStop}%
\bibitem [{\citenamefont {Lin}\ \emph {et~al.}(2014)\citenamefont {Lin},
  \citenamefont {Lerner}, \citenamefont {Rosso},\ and\ \citenamefont
  {Wyart}}]{lin2014scaling}%
  \BibitemOpen
  \bibfield  {author} {\bibinfo {author} {\bibfnamefont {Jie}\ \bibnamefont
  {Lin}}, \bibinfo {author} {\bibfnamefont {Edan}\ \bibnamefont {Lerner}},
  \bibinfo {author} {\bibfnamefont {Alberto}\ \bibnamefont {Rosso}}, \ and\
  \bibinfo {author} {\bibfnamefont {Matthieu}\ \bibnamefont {Wyart}},\
  }\bibfield  {title} {\enquote {\bibinfo {title} {Scaling description of the
  yielding transition in soft amorphous solids at zero temperature},}\
  }\href@noop {} {\bibfield  {journal} {\bibinfo  {journal} {Proceedings of the
  National Academy of Sciences}\ }\textbf {\bibinfo {volume} {111}},\ \bibinfo
  {pages} {14382--14387} (\bibinfo {year} {2014})}\BibitemShut {NoStop}%
\bibitem [{\citenamefont {Bulatov}\ and\ \citenamefont
  {Argon}(1994)}]{bulatov1994stochastic}%
  \BibitemOpen
  \bibfield  {author} {\bibinfo {author} {\bibfnamefont {VV}~\bibnamefont
  {Bulatov}}\ and\ \bibinfo {author} {\bibfnamefont {AS}~\bibnamefont
  {Argon}},\ }\bibfield  {title} {\enquote {\bibinfo {title} {A stochastic
  model for continuum elasto-plastic behavior. ii. a study of the glass
  transition and structural relaxation},}\ }\href@noop {} {\bibfield  {journal}
  {\bibinfo  {journal} {Modelling and Simulation in Materials Science and
  Engineering}\ }\textbf {\bibinfo {volume} {2}},\ \bibinfo {pages} {185}
  (\bibinfo {year} {1994})}\BibitemShut {NoStop}%
\bibitem [{\citenamefont {Ferrero}\ \emph {et~al.}(2014)\citenamefont
  {Ferrero}, \citenamefont {Martens},\ and\ \citenamefont
  {Barrat}}]{ferrero2014relaxation}%
  \BibitemOpen
  \bibfield  {author} {\bibinfo {author} {\bibfnamefont {Ezequiel~E}\
  \bibnamefont {Ferrero}}, \bibinfo {author} {\bibfnamefont {Kirsten}\
  \bibnamefont {Martens}}, \ and\ \bibinfo {author} {\bibfnamefont
  {Jean-Louis}\ \bibnamefont {Barrat}},\ }\bibfield  {title} {\enquote
  {\bibinfo {title} {Relaxation in yield stress systems through elastically
  interacting activated events},}\ }\href@noop {} {\bibfield  {journal}
  {\bibinfo  {journal} {Physical review letters}\ }\textbf {\bibinfo {volume}
  {113}},\ \bibinfo {pages} {248301} (\bibinfo {year} {2014})}\BibitemShut
  {NoStop}%
\bibitem [{\citenamefont {Popovi{\'c}}\ \emph {et~al.}(2020)\citenamefont
  {Popovi{\'c}}, \citenamefont {de~Geus}, \citenamefont {Ji},\ and\
  \citenamefont {Wyart}}]{popovic2020thermally}%
  \BibitemOpen
  \bibfield  {author} {\bibinfo {author} {\bibfnamefont {Marko}\ \bibnamefont
  {Popovi{\'c}}}, \bibinfo {author} {\bibfnamefont {Tom~WJ}\ \bibnamefont
  {de~Geus}}, \bibinfo {author} {\bibfnamefont {Wencheng}\ \bibnamefont {Ji}},
  \ and\ \bibinfo {author} {\bibfnamefont {Matthieu}\ \bibnamefont {Wyart}},\
  }\bibfield  {title} {\enquote {\bibinfo {title} {Thermally activated flow in
  models of amorphous solids},}\ }\href@noop {} {\bibfield  {journal} {\bibinfo
   {journal} {arXiv preprint arXiv:2009.04963}\ } (\bibinfo {year}
  {2020})}\BibitemShut {NoStop}%
\bibitem [{\citenamefont {Ferrero}\ \emph {et~al.}(2021)\citenamefont
  {Ferrero}, \citenamefont {Kolton},\ and\ \citenamefont
  {Jagla}}]{ferrero2021yielding}%
  \BibitemOpen
  \bibfield  {author} {\bibinfo {author} {\bibfnamefont {Ezequiel~E}\
  \bibnamefont {Ferrero}}, \bibinfo {author} {\bibfnamefont {Alejandro~B}\
  \bibnamefont {Kolton}}, \ and\ \bibinfo {author} {\bibfnamefont {Eduardo~A}\
  \bibnamefont {Jagla}},\ }\bibfield  {title} {\enquote {\bibinfo {title}
  {Yielding of amorphous solids at finite temperatures},}\ }\href@noop {}
  {\bibfield  {journal} {\bibinfo  {journal} {Physical Review Materials}\
  }\textbf {\bibinfo {volume} {5}},\ \bibinfo {pages} {115602} (\bibinfo {year}
  {2021})}\BibitemShut {NoStop}%
\bibitem [{\citenamefont {Picard}\ \emph {et~al.}(2004)\citenamefont {Picard},
  \citenamefont {Ajdari}, \citenamefont {Lequeux},\ and\ \citenamefont
  {Bocquet}}]{picard2004elastic}%
  \BibitemOpen
  \bibfield  {author} {\bibinfo {author} {\bibfnamefont {Guillemette}\
  \bibnamefont {Picard}}, \bibinfo {author} {\bibfnamefont {Armand}\
  \bibnamefont {Ajdari}}, \bibinfo {author} {\bibfnamefont {Fran{\c{c}}ois}\
  \bibnamefont {Lequeux}}, \ and\ \bibinfo {author} {\bibfnamefont
  {Lyd{\'e}ric}\ \bibnamefont {Bocquet}},\ }\bibfield  {title} {\enquote
  {\bibinfo {title} {Elastic consequences of a single plastic event: A step
  towards the microscopic modeling of the flow of yield stress fluids},}\
  }\href@noop {} {\bibfield  {journal} {\bibinfo  {journal} {The European
  Physical Journal E}\ }\textbf {\bibinfo {volume} {15}},\ \bibinfo {pages}
  {371--381} (\bibinfo {year} {2004})}\BibitemShut {NoStop}%
\bibitem [{\citenamefont {Barbot}\ \emph {et~al.}(2018)\citenamefont {Barbot},
  \citenamefont {Lerbinger}, \citenamefont {Hernandez-Garcia}, \citenamefont
  {Garc{\'\i}a-Garc{\'\i}a}, \citenamefont {Falk}, \citenamefont
  {Vandembroucq},\ and\ \citenamefont {Patinet}}]{barbot2018local}%
  \BibitemOpen
  \bibfield  {author} {\bibinfo {author} {\bibfnamefont {Armand}\ \bibnamefont
  {Barbot}}, \bibinfo {author} {\bibfnamefont {Matthias}\ \bibnamefont
  {Lerbinger}}, \bibinfo {author} {\bibfnamefont {Anier}\ \bibnamefont
  {Hernandez-Garcia}}, \bibinfo {author} {\bibfnamefont {Reinaldo}\
  \bibnamefont {Garc{\'\i}a-Garc{\'\i}a}}, \bibinfo {author} {\bibfnamefont
  {Michael~L}\ \bibnamefont {Falk}}, \bibinfo {author} {\bibfnamefont {Damien}\
  \bibnamefont {Vandembroucq}}, \ and\ \bibinfo {author} {\bibfnamefont
  {Sylvain}\ \bibnamefont {Patinet}},\ }\bibfield  {title} {\enquote {\bibinfo
  {title} {Local yield stress statistics in model amorphous solids},}\
  }\href@noop {} {\bibfield  {journal} {\bibinfo  {journal} {Physical Review
  E}\ }\textbf {\bibinfo {volume} {97}},\ \bibinfo {pages} {033001} (\bibinfo
  {year} {2018})}\BibitemShut {NoStop}%
\bibitem [{\citenamefont {Maloney}\ and\ \citenamefont
  {Lacks}(2006)}]{maloney2006energy}%
  \BibitemOpen
  \bibfield  {author} {\bibinfo {author} {\bibfnamefont {Craig~E}\ \bibnamefont
  {Maloney}}\ and\ \bibinfo {author} {\bibfnamefont {Daniel~J}\ \bibnamefont
  {Lacks}},\ }\bibfield  {title} {\enquote {\bibinfo {title} {Energy barrier
  scalings in driven systems},}\ }\href@noop {} {\bibfield  {journal} {\bibinfo
   {journal} {Physical Review E}\ }\textbf {\bibinfo {volume} {73}},\ \bibinfo
  {pages} {061106} (\bibinfo {year} {2006})}\BibitemShut {NoStop}%
\bibitem [{\citenamefont {Fan}\ \emph {et~al.}(2014)\citenamefont {Fan},
  \citenamefont {Iwashita},\ and\ \citenamefont {Egami}}]{fan2014thermally}%
  \BibitemOpen
  \bibfield  {author} {\bibinfo {author} {\bibfnamefont {Yue}\ \bibnamefont
  {Fan}}, \bibinfo {author} {\bibfnamefont {Takuya}\ \bibnamefont {Iwashita}},
  \ and\ \bibinfo {author} {\bibfnamefont {Takeshi}\ \bibnamefont {Egami}},\
  }\bibfield  {title} {\enquote {\bibinfo {title} {How thermally activated
  deformation starts in metallic glass},}\ }\href@noop {} {\bibfield  {journal}
  {\bibinfo  {journal} {Nature communications}\ }\textbf {\bibinfo {volume}
  {5}},\ \bibinfo {pages} {1--7} (\bibinfo {year} {2014})}\BibitemShut
  {NoStop}%
\bibitem [{Note1()}]{Note1}%
  \BibitemOpen
  \bibinfo {note} {Note that the dynamics of EPM-Q does not verify
  time-reversal symmetry (or detailed balance). This is not necessarily a major
  concern as no spontaneous activity is created at small temperature. Moreover,
  the phenomenology of DH is observed also for non time-reversible systems,
  e.g. granular media \cite {berthier2011dynamical}. However, it is certainly a
  property that would be worth restoring in a more complete version of the
  model}\BibitemShut {NoStop}%
\bibitem [{\citenamefont {Kob}\ and\ \citenamefont
  {Andersen}(1995)}]{kob1995testing}%
  \BibitemOpen
  \bibfield  {author} {\bibinfo {author} {\bibfnamefont {Walter}\ \bibnamefont
  {Kob}}\ and\ \bibinfo {author} {\bibfnamefont {Hans~C}\ \bibnamefont
  {Andersen}},\ }\bibfield  {title} {\enquote {\bibinfo {title} {Testing
  mode-coupling theory for a supercooled binary lennard-jones mixture. ii.
  intermediate scattering function and dynamic susceptibility},}\ }\href@noop
  {} {\bibfield  {journal} {\bibinfo  {journal} {Physical Review E}\ }\textbf
  {\bibinfo {volume} {52}},\ \bibinfo {pages} {4134} (\bibinfo {year}
  {1995})}\BibitemShut {NoStop}%
\bibitem [{\citenamefont {Whitelam}\ \emph {et~al.}(2005)\citenamefont
  {Whitelam}, \citenamefont {Berthier},\ and\ \citenamefont
  {Garrahan}}]{whitelam2005renormalization}%
  \BibitemOpen
  \bibfield  {author} {\bibinfo {author} {\bibfnamefont {Stephen}\ \bibnamefont
  {Whitelam}}, \bibinfo {author} {\bibfnamefont {Ludovic}\ \bibnamefont
  {Berthier}}, \ and\ \bibinfo {author} {\bibfnamefont {Juan~P}\ \bibnamefont
  {Garrahan}},\ }\bibfield  {title} {\enquote {\bibinfo {title}
  {Renormalization group study of a kinetically constrained model for strong
  glasses},}\ }\href@noop {} {\bibfield  {journal} {\bibinfo  {journal}
  {Physical Review E}\ }\textbf {\bibinfo {volume} {71}},\ \bibinfo {pages}
  {026128} (\bibinfo {year} {2005})}\BibitemShut {NoStop}%
\bibitem [{\citenamefont {Donati}\ \emph {et~al.}(2002)\citenamefont {Donati},
  \citenamefont {Franz}, \citenamefont {Glotzer},\ and\ \citenamefont
  {Parisi}}]{donati2002theory}%
  \BibitemOpen
  \bibfield  {author} {\bibinfo {author} {\bibfnamefont {Claudio}\ \bibnamefont
  {Donati}}, \bibinfo {author} {\bibfnamefont {Silvio}\ \bibnamefont {Franz}},
  \bibinfo {author} {\bibfnamefont {Sharon~C}\ \bibnamefont {Glotzer}}, \ and\
  \bibinfo {author} {\bibfnamefont {Giorgio}\ \bibnamefont {Parisi}},\
  }\bibfield  {title} {\enquote {\bibinfo {title} {Theory of non-linear
  susceptibility and correlation length in glasses and liquids},}\ }\href@noop
  {} {\bibfield  {journal} {\bibinfo  {journal} {Journal of non-crystalline
  solids}\ }\textbf {\bibinfo {volume} {307}},\ \bibinfo {pages} {215--224}
  (\bibinfo {year} {2002})}\BibitemShut {NoStop}%
\bibitem [{\citenamefont {Dalle-Ferrier}\ \emph {et~al.}(2007)\citenamefont
  {Dalle-Ferrier}, \citenamefont {Thibierge}, \citenamefont {Alba-Simionesco},
  \citenamefont {Berthier}, \citenamefont {Biroli}, \citenamefont {Bouchaud},
  \citenamefont {Ladieu}, \citenamefont {L’H{\^o}te},\ and\ \citenamefont
  {Tarjus}}]{dalle2007spatial}%
  \BibitemOpen
  \bibfield  {author} {\bibinfo {author} {\bibfnamefont {C{\'e}cile}\
  \bibnamefont {Dalle-Ferrier}}, \bibinfo {author} {\bibfnamefont {Caroline}\
  \bibnamefont {Thibierge}}, \bibinfo {author} {\bibfnamefont {Christiane}\
  \bibnamefont {Alba-Simionesco}}, \bibinfo {author} {\bibfnamefont {Ludovic}\
  \bibnamefont {Berthier}}, \bibinfo {author} {\bibfnamefont {Giulio}\
  \bibnamefont {Biroli}}, \bibinfo {author} {\bibfnamefont {J-P}\ \bibnamefont
  {Bouchaud}}, \bibinfo {author} {\bibfnamefont {Fran{\c{c}}ois}\ \bibnamefont
  {Ladieu}}, \bibinfo {author} {\bibfnamefont {Denis}\ \bibnamefont
  {L’H{\^o}te}}, \ and\ \bibinfo {author} {\bibfnamefont {Gilles}\
  \bibnamefont {Tarjus}},\ }\bibfield  {title} {\enquote {\bibinfo {title}
  {Spatial correlations in the dynamics of glassforming liquids: Experimental
  determination of their temperature dependence},}\ }\href@noop {} {\bibfield
  {journal} {\bibinfo  {journal} {Physical Review E}\ }\textbf {\bibinfo
  {volume} {76}},\ \bibinfo {pages} {041510} (\bibinfo {year}
  {2007})}\BibitemShut {NoStop}%
\bibitem [{\citenamefont {Lubchenko}\ and\ \citenamefont
  {Wolynes}(2007)}]{lubchenko2007theory}%
  \BibitemOpen
  \bibfield  {author} {\bibinfo {author} {\bibfnamefont {Vassiliy}\
  \bibnamefont {Lubchenko}}\ and\ \bibinfo {author} {\bibfnamefont {Peter~G}\
  \bibnamefont {Wolynes}},\ }\bibfield  {title} {\enquote {\bibinfo {title}
  {Theory of structural glasses and supercooled liquids},}\ }\href@noop {}
  {\bibfield  {journal} {\bibinfo  {journal} {Annu. Rev. Phys. Chem.}\ }\textbf
  {\bibinfo {volume} {58}},\ \bibinfo {pages} {235--266} (\bibinfo {year}
  {2007})}\BibitemShut {NoStop}%
\bibitem [{\citenamefont {Biroli}\ and\ \citenamefont
  {Bouchaud}(2012)}]{biroli2012random}%
  \BibitemOpen
  \bibfield  {author} {\bibinfo {author} {\bibfnamefont {Giulio}\ \bibnamefont
  {Biroli}}\ and\ \bibinfo {author} {\bibfnamefont {Jean-Philippe}\
  \bibnamefont {Bouchaud}},\ }\bibfield  {title} {\enquote {\bibinfo {title}
  {The random first-order transition theory of glasses: A critical
  assessment},}\ }\href@noop {} {\bibfield  {journal} {\bibinfo  {journal}
  {Structural Glasses and Supercooled Liquids: Theory, Experiment, and
  Applications}\ ,\ \bibinfo {pages} {31--113}} (\bibinfo {year}
  {2012})}\BibitemShut {NoStop}%
\bibitem [{\citenamefont {Bocquet}\ \emph {et~al.}(2009)\citenamefont
  {Bocquet}, \citenamefont {Colin},\ and\ \citenamefont
  {Ajdari}}]{bocquet2009kinetic}%
  \BibitemOpen
  \bibfield  {author} {\bibinfo {author} {\bibfnamefont {Lyd{\'e}ric}\
  \bibnamefont {Bocquet}}, \bibinfo {author} {\bibfnamefont {Annie}\
  \bibnamefont {Colin}}, \ and\ \bibinfo {author} {\bibfnamefont {Armand}\
  \bibnamefont {Ajdari}},\ }\bibfield  {title} {\enquote {\bibinfo {title}
  {Kinetic theory of plastic flow in soft glassy materials},}\ }\href@noop {}
  {\bibfield  {journal} {\bibinfo  {journal} {Physical review letters}\
  }\textbf {\bibinfo {volume} {103}},\ \bibinfo {pages} {036001} (\bibinfo
  {year} {2009})}\BibitemShut {NoStop}%
\bibitem [{\citenamefont {H{\'e}braud}\ and\ \citenamefont
  {Lequeux}(1998)}]{hebraud1998mode}%
  \BibitemOpen
  \bibfield  {author} {\bibinfo {author} {\bibfnamefont {Pascal}\ \bibnamefont
  {H{\'e}braud}}\ and\ \bibinfo {author} {\bibfnamefont {Fran{\c{c}}ois}\
  \bibnamefont {Lequeux}},\ }\bibfield  {title} {\enquote {\bibinfo {title}
  {Mode-coupling theory for the pasty rheology of soft glassy materials},}\
  }\href@noop {} {\bibfield  {journal} {\bibinfo  {journal} {Physical review
  letters}\ }\textbf {\bibinfo {volume} {81}},\ \bibinfo {pages} {2934}
  (\bibinfo {year} {1998})}\BibitemShut {NoStop}%
\bibitem [{\citenamefont {Agoritsas}\ \emph {et~al.}(2015)\citenamefont
  {Agoritsas}, \citenamefont {Bertin}, \citenamefont {Martens},\ and\
  \citenamefont {Barrat}}]{agoritsas2015relevance}%
  \BibitemOpen
  \bibfield  {author} {\bibinfo {author} {\bibfnamefont {Elisabeth}\
  \bibnamefont {Agoritsas}}, \bibinfo {author} {\bibfnamefont {Eric}\
  \bibnamefont {Bertin}}, \bibinfo {author} {\bibfnamefont {Kirsten}\
  \bibnamefont {Martens}}, \ and\ \bibinfo {author} {\bibfnamefont
  {Jean-Louis}\ \bibnamefont {Barrat}},\ }\bibfield  {title} {\enquote
  {\bibinfo {title} {On the relevance of disorder in athermal amorphous
  materials under shear},}\ }\href@noop {} {\bibfield  {journal} {\bibinfo
  {journal} {The European Physical Journal E}\ }\textbf {\bibinfo {volume}
  {38}},\ \bibinfo {pages} {1--22} (\bibinfo {year} {2015})}\BibitemShut
  {NoStop}%
\bibitem [{\citenamefont {Toninelli}\ \emph {et~al.}(2005)\citenamefont
  {Toninelli}, \citenamefont {Wyart}, \citenamefont {Berthier}, \citenamefont
  {Biroli},\ and\ \citenamefont {Bouchaud}}]{toninelli2005dynamical}%
  \BibitemOpen
  \bibfield  {author} {\bibinfo {author} {\bibfnamefont {Cristina}\
  \bibnamefont {Toninelli}}, \bibinfo {author} {\bibfnamefont {Matthieu}\
  \bibnamefont {Wyart}}, \bibinfo {author} {\bibfnamefont {Ludovic}\
  \bibnamefont {Berthier}}, \bibinfo {author} {\bibfnamefont {Giulio}\
  \bibnamefont {Biroli}}, \ and\ \bibinfo {author} {\bibfnamefont
  {Jean-Philippe}\ \bibnamefont {Bouchaud}},\ }\bibfield  {title} {\enquote
  {\bibinfo {title} {Dynamical susceptibility of glass formers: Contrasting the
  predictions of theoretical scenarios},}\ }\href@noop {} {\bibfield  {journal}
  {\bibinfo  {journal} {Physical Review E}\ }\textbf {\bibinfo {volume} {71}},\
  \bibinfo {pages} {041505} (\bibinfo {year} {2005})}\BibitemShut {NoStop}%
\bibitem [{Note2()}]{Note2}%
  \BibitemOpen
  \bibinfo {note} {Note that although the EPM-Q suggests a mechanism in which
  there is first a local activated relaxation and then a subsequent avalanche
  of motion, in a more refined model, verifying time-reversibility, also the
  reverse process would be possible. Therefore, one should think in terms of
  finite size avalanches resulting from elasticity and activation, and not in
  terms of an initiator event and a subsequent avalanche.}\BibitemShut {Stop}%
\bibitem [{\citenamefont {Ediger}\ \emph {et~al.}(1996)\citenamefont {Ediger},
  \citenamefont {Angell},\ and\ \citenamefont {Nagel}}]{ediger1996supercooled}%
  \BibitemOpen
  \bibfield  {author} {\bibinfo {author} {\bibfnamefont {Mark~D}\ \bibnamefont
  {Ediger}}, \bibinfo {author} {\bibfnamefont {C~Austen}\ \bibnamefont
  {Angell}}, \ and\ \bibinfo {author} {\bibfnamefont {Sidney~R}\ \bibnamefont
  {Nagel}},\ }\bibfield  {title} {\enquote {\bibinfo {title} {Supercooled
  liquids and glasses},}\ }\href@noop {} {\bibfield  {journal} {\bibinfo
  {journal} {The journal of physical chemistry}\ }\textbf {\bibinfo {volume}
  {100}},\ \bibinfo {pages} {13200--13212} (\bibinfo {year}
  {1996})}\BibitemShut {NoStop}%
\bibitem [{\citenamefont {Martens}\ \emph {et~al.}(2012)\citenamefont
  {Martens}, \citenamefont {Bocquet},\ and\ \citenamefont
  {Barrat}}]{martens2012spontaneous}%
  \BibitemOpen
  \bibfield  {author} {\bibinfo {author} {\bibfnamefont {Kirsten}\ \bibnamefont
  {Martens}}, \bibinfo {author} {\bibfnamefont {Lyd{\'e}ric}\ \bibnamefont
  {Bocquet}}, \ and\ \bibinfo {author} {\bibfnamefont {Jean-Louis}\
  \bibnamefont {Barrat}},\ }\bibfield  {title} {\enquote {\bibinfo {title}
  {Spontaneous formation of permanent shear bands in a mesoscopic model of
  flowing disordered matter},}\ }\href@noop {} {\bibfield  {journal} {\bibinfo
  {journal} {Soft Matter}\ }\textbf {\bibinfo {volume} {8}},\ \bibinfo {pages}
  {4197--4205} (\bibinfo {year} {2012})}\BibitemShut {NoStop}%
\bibitem [{\citenamefont {Nicolas}\ \emph {et~al.}(2014)\citenamefont
  {Nicolas}, \citenamefont {Martens}, \citenamefont {Bocquet},\ and\
  \citenamefont {Barrat}}]{nicolas2014universal}%
  \BibitemOpen
  \bibfield  {author} {\bibinfo {author} {\bibfnamefont {Alexandre}\
  \bibnamefont {Nicolas}}, \bibinfo {author} {\bibfnamefont {Kirsten}\
  \bibnamefont {Martens}}, \bibinfo {author} {\bibfnamefont {Lyd{\'e}ric}\
  \bibnamefont {Bocquet}}, \ and\ \bibinfo {author} {\bibfnamefont
  {Jean-Louis}\ \bibnamefont {Barrat}},\ }\bibfield  {title} {\enquote
  {\bibinfo {title} {Universal and non-universal features in coarse-grained
  models of flow in disordered solids},}\ }\href@noop {} {\bibfield  {journal}
  {\bibinfo  {journal} {Soft matter}\ }\textbf {\bibinfo {volume} {10}},\
  \bibinfo {pages} {4648--4661} (\bibinfo {year} {2014})}\BibitemShut {NoStop}%
\bibitem [{\citenamefont {Budrikis}\ \emph {et~al.}(2017)\citenamefont
  {Budrikis}, \citenamefont {Castellanos}, \citenamefont {Sandfeld},
  \citenamefont {Zaiser},\ and\ \citenamefont
  {Zapperi}}]{budrikis2017universal}%
  \BibitemOpen
  \bibfield  {author} {\bibinfo {author} {\bibfnamefont {Zoe}\ \bibnamefont
  {Budrikis}}, \bibinfo {author} {\bibfnamefont {David~Fernandez}\ \bibnamefont
  {Castellanos}}, \bibinfo {author} {\bibfnamefont {Stefan}\ \bibnamefont
  {Sandfeld}}, \bibinfo {author} {\bibfnamefont {Michael}\ \bibnamefont
  {Zaiser}}, \ and\ \bibinfo {author} {\bibfnamefont {Stefano}\ \bibnamefont
  {Zapperi}},\ }\bibfield  {title} {\enquote {\bibinfo {title} {Universal
  features of amorphous plasticity},}\ }\href@noop {} {\bibfield  {journal}
  {\bibinfo  {journal} {Nature communications}\ }\textbf {\bibinfo {volume}
  {8}},\ \bibinfo {pages} {1--10} (\bibinfo {year} {2017})}\BibitemShut
  {NoStop}%
\bibitem [{\citenamefont {Nemilov}(2006)}]{nemilov2006interrelation}%
  \BibitemOpen
  \bibfield  {author} {\bibinfo {author} {\bibfnamefont {SV}~\bibnamefont
  {Nemilov}},\ }\bibfield  {title} {\enquote {\bibinfo {title} {Interrelation
  between shear modulus and the molecular parameters of viscous flow for glass
  forming liquids},}\ }\href@noop {} {\bibfield  {journal} {\bibinfo  {journal}
  {Journal of non-crystalline solids}\ }\textbf {\bibinfo {volume} {352}},\
  \bibinfo {pages} {2715--2725} (\bibinfo {year} {2006})}\BibitemShut {NoStop}%
\bibitem [{\citenamefont {Rouxel}(2011)}]{rouxel2011thermodynamics}%
  \BibitemOpen
  \bibfield  {author} {\bibinfo {author} {\bibfnamefont {Tanguy}\ \bibnamefont
  {Rouxel}},\ }\bibfield  {title} {\enquote {\bibinfo {title} {Thermodynamics
  of viscous flow and elasticity of glass forming liquids in the glass
  transition range},}\ }\href@noop {} {\bibfield  {journal} {\bibinfo
  {journal} {The Journal of chemical physics}\ }\textbf {\bibinfo {volume}
  {135}},\ \bibinfo {pages} {184501} (\bibinfo {year} {2011})}\BibitemShut
  {NoStop}%
\bibitem [{\citenamefont {Mirigian}\ and\ \citenamefont
  {Schweizer}(2013)}]{mirigian2013unified}%
  \BibitemOpen
  \bibfield  {author} {\bibinfo {author} {\bibfnamefont {Stephen}\ \bibnamefont
  {Mirigian}}\ and\ \bibinfo {author} {\bibfnamefont {Kenneth~S}\ \bibnamefont
  {Schweizer}},\ }\bibfield  {title} {\enquote {\bibinfo {title} {Unified
  theory of activated relaxation in liquids over 14 decades in time},}\
  }\href@noop {} {\bibfield  {journal} {\bibinfo  {journal} {The Journal of
  Physical Chemistry Letters}\ }\textbf {\bibinfo {volume} {4}},\ \bibinfo
  {pages} {3648--3653} (\bibinfo {year} {2013})}\BibitemShut {NoStop}%
\bibitem [{\citenamefont {Kapteijns}\ \emph {et~al.}(2021)\citenamefont
  {Kapteijns}, \citenamefont {Richard}, \citenamefont {Bouchbinder},
  \citenamefont {Schr{\o}der}, \citenamefont {Dyre},\ and\ \citenamefont
  {Lerner}}]{kapteijns2021does}%
  \BibitemOpen
  \bibfield  {author} {\bibinfo {author} {\bibfnamefont {Geert}\ \bibnamefont
  {Kapteijns}}, \bibinfo {author} {\bibfnamefont {David}\ \bibnamefont
  {Richard}}, \bibinfo {author} {\bibfnamefont {Eran}\ \bibnamefont
  {Bouchbinder}}, \bibinfo {author} {\bibfnamefont {Thomas~B}\ \bibnamefont
  {Schr{\o}der}}, \bibinfo {author} {\bibfnamefont {Jeppe~C}\ \bibnamefont
  {Dyre}}, \ and\ \bibinfo {author} {\bibfnamefont {Edan}\ \bibnamefont
  {Lerner}},\ }\bibfield  {title} {\enquote {\bibinfo {title} {Does mesoscopic
  elasticity control viscous slowing down in glassforming liquids?}}\
  }\href@noop {} {\bibfield  {journal} {\bibinfo  {journal} {The Journal of
  chemical physics}\ }\textbf {\bibinfo {volume} {155}},\ \bibinfo {pages}
  {074502} (\bibinfo {year} {2021})}\BibitemShut {NoStop}%
\bibitem [{\citenamefont {Biroli}\ \emph {et~al.}(2022)\citenamefont {Biroli},
  \citenamefont {Ozawa}, \citenamefont {Popovi\ifmmode~\acute{c}\else
  \'{c}\fi{}}, \citenamefont {Tahaei},\ and\ \citenamefont
  {Wyart}}]{avalanche_inprogress}%
  \BibitemOpen
  \bibfield  {author} {\bibinfo {author} {\bibfnamefont {Giulio}\ \bibnamefont
  {Biroli}}, \bibinfo {author} {\bibfnamefont {Misaki}\ \bibnamefont {Ozawa}},
  \bibinfo {author} {\bibfnamefont {Marko}\ \bibnamefont
  {Popovi\ifmmode~\acute{c}\else \'{c}\fi{}}}, \bibinfo {author} {\bibfnamefont
  {Ali}\ \bibnamefont {Tahaei}}, \ and\ \bibinfo {author} {\bibfnamefont
  {Matthieu}\ \bibnamefont {Wyart}},\ }\href@noop {} {\bibfield  {journal}
  {\bibinfo  {journal} {Work in progress}\ } (\bibinfo {year}
  {2022})}\BibitemShut {NoStop}%
\bibitem [{\citenamefont {Berthier}\ and\ \citenamefont
  {Kob}(2007)}]{berthier2007monte}%
  \BibitemOpen
  \bibfield  {author} {\bibinfo {author} {\bibfnamefont {Ludovic}\ \bibnamefont
  {Berthier}}\ and\ \bibinfo {author} {\bibfnamefont {Walter}\ \bibnamefont
  {Kob}},\ }\bibfield  {title} {\enquote {\bibinfo {title} {The monte carlo
  dynamics of a binary lennard-jones glass-forming mixture},}\ }\href@noop {}
  {\bibfield  {journal} {\bibinfo  {journal} {Journal of Physics: Condensed
  Matter}\ }\textbf {\bibinfo {volume} {19}},\ \bibinfo {pages} {205130}
  (\bibinfo {year} {2007})}\BibitemShut {NoStop}%
\bibitem [{\citenamefont {Krauth}(2006)}]{krauth2006statistical}%
  \BibitemOpen
  \bibfield  {author} {\bibinfo {author} {\bibfnamefont {Werner}\ \bibnamefont
  {Krauth}},\ }\href@noop {} {\emph {\bibinfo {title} {Statistical mechanics:
  algorithms and computations}}},\ Vol.~\bibinfo {volume} {13}\ (\bibinfo
  {publisher} {OUP Oxford},\ \bibinfo {year} {2006})\BibitemShut {NoStop}%
\bibitem [{\citenamefont {Maloney}\ and\ \citenamefont
  {Lemaitre}(2006)}]{maloney2006amorphous}%
  \BibitemOpen
  \bibfield  {author} {\bibinfo {author} {\bibfnamefont {Craig~E}\ \bibnamefont
  {Maloney}}\ and\ \bibinfo {author} {\bibfnamefont {Anael}\ \bibnamefont
  {Lemaitre}},\ }\bibfield  {title} {\enquote {\bibinfo {title} {Amorphous
  systems in athermal, quasistatic shear},}\ }\href@noop {} {\bibfield
  {journal} {\bibinfo  {journal} {Physical Review E}\ }\textbf {\bibinfo
  {volume} {74}},\ \bibinfo {pages} {016118} (\bibinfo {year}
  {2006})}\BibitemShut {NoStop}%
\bibitem [{\citenamefont {Dasgupta}\ \emph {et~al.}(2013)\citenamefont
  {Dasgupta}, \citenamefont {Hentschel},\ and\ \citenamefont
  {Procaccia}}]{dasgupta2013yield}%
  \BibitemOpen
  \bibfield  {author} {\bibinfo {author} {\bibfnamefont {Ratul}\ \bibnamefont
  {Dasgupta}}, \bibinfo {author} {\bibfnamefont {H~George~E}\ \bibnamefont
  {Hentschel}}, \ and\ \bibinfo {author} {\bibfnamefont {Itamar}\ \bibnamefont
  {Procaccia}},\ }\bibfield  {title} {\enquote {\bibinfo {title} {Yield strain
  in shear banding amorphous solids},}\ }\href@noop {} {\bibfield  {journal}
  {\bibinfo  {journal} {Physical Review E}\ }\textbf {\bibinfo {volume} {87}},\
  \bibinfo {pages} {022810} (\bibinfo {year} {2013})}\BibitemShut {NoStop}%
\bibitem [{\citenamefont {Jagla}(2020)}]{jagla2020tensorial}%
  \BibitemOpen
  \bibfield  {author} {\bibinfo {author} {\bibfnamefont {Eduardo~Alberto}\
  \bibnamefont {Jagla}},\ }\bibfield  {title} {\enquote {\bibinfo {title}
  {Tensorial description of the plasticity of amorphous composites},}\
  }\href@noop {} {\bibfield  {journal} {\bibinfo  {journal} {Physical Review
  E}\ }\textbf {\bibinfo {volume} {101}},\ \bibinfo {pages} {043004} (\bibinfo
  {year} {2020})}\BibitemShut {NoStop}%
\bibitem [{\citenamefont {Cao}\ \emph {et~al.}(2018)\citenamefont {Cao},
  \citenamefont {Nicolas}, \citenamefont {Trimcev},\ and\ \citenamefont
  {Rosso}}]{cao2018soft}%
  \BibitemOpen
  \bibfield  {author} {\bibinfo {author} {\bibfnamefont {Xiangyu}\ \bibnamefont
  {Cao}}, \bibinfo {author} {\bibfnamefont {Alexandre}\ \bibnamefont
  {Nicolas}}, \bibinfo {author} {\bibfnamefont {Denny}\ \bibnamefont
  {Trimcev}}, \ and\ \bibinfo {author} {\bibfnamefont {Alberto}\ \bibnamefont
  {Rosso}},\ }\bibfield  {title} {\enquote {\bibinfo {title} {Soft modes and
  strain redistribution in continuous models of amorphous plasticity: the
  eshelby paradigm, and beyond?}}\ }\href@noop {} {\bibfield  {journal}
  {\bibinfo  {journal} {Soft matter}\ }\textbf {\bibinfo {volume} {14}},\
  \bibinfo {pages} {3640--3651} (\bibinfo {year} {2018})}\BibitemShut {NoStop}%
\bibitem [{\citenamefont {Ferrero}\ and\ \citenamefont
  {Jagla}(2019)}]{ferrero2019criticality}%
  \BibitemOpen
  \bibfield  {author} {\bibinfo {author} {\bibfnamefont {Ezequiel~E}\
  \bibnamefont {Ferrero}}\ and\ \bibinfo {author} {\bibfnamefont {Eduardo~A}\
  \bibnamefont {Jagla}},\ }\bibfield  {title} {\enquote {\bibinfo {title}
  {Criticality in elastoplastic models of amorphous solids with
  stress-dependent yielding rates},}\ }\href@noop {} {\bibfield  {journal}
  {\bibinfo  {journal} {Soft matter}\ }\textbf {\bibinfo {volume} {15}},\
  \bibinfo {pages} {9041--9055} (\bibinfo {year} {2019})}\BibitemShut {NoStop}%
\bibitem [{\citenamefont {Popovi{\'c}}\ \emph {et~al.}(2018)\citenamefont
  {Popovi{\'c}}, \citenamefont {de~Geus},\ and\ \citenamefont
  {Wyart}}]{popovic2018elastoplastic}%
  \BibitemOpen
  \bibfield  {author} {\bibinfo {author} {\bibfnamefont {Marko}\ \bibnamefont
  {Popovi{\'c}}}, \bibinfo {author} {\bibfnamefont {Tom~WJ}\ \bibnamefont
  {de~Geus}}, \ and\ \bibinfo {author} {\bibfnamefont {Matthieu}\ \bibnamefont
  {Wyart}},\ }\bibfield  {title} {\enquote {\bibinfo {title} {Elastoplastic
  description of sudden failure in athermal amorphous materials during
  quasistatic loading},}\ }\href@noop {} {\bibfield  {journal} {\bibinfo
  {journal} {Physical Review E}\ }\textbf {\bibinfo {volume} {98}},\ \bibinfo
  {pages} {040901} (\bibinfo {year} {2018})}\BibitemShut {NoStop}%
\bibitem [{\citenamefont {Rossi}\ \emph {et~al.}(2022)\citenamefont {Rossi},
  \citenamefont {Biroli}, \citenamefont {Ozawa}, \citenamefont {Tarjus},\ and\
  \citenamefont {Zamponi}}]{rossi2022finite}%
  \BibitemOpen
  \bibfield  {author} {\bibinfo {author} {\bibfnamefont {Saverio}\ \bibnamefont
  {Rossi}}, \bibinfo {author} {\bibfnamefont {Giulio}\ \bibnamefont {Biroli}},
  \bibinfo {author} {\bibfnamefont {Misaki}\ \bibnamefont {Ozawa}}, \bibinfo
  {author} {\bibfnamefont {Gilles}\ \bibnamefont {Tarjus}}, \ and\ \bibinfo
  {author} {\bibfnamefont {Francesco}\ \bibnamefont {Zamponi}},\ }\bibfield
  {title} {\enquote {\bibinfo {title} {Finite-disorder critical point in the
  yielding transition of elasto-plastic models},}\ }\href@noop {} {\bibfield
  {journal} {\bibinfo  {journal} {arXiv preprint arXiv:2204.10683}\ } (\bibinfo
  {year} {2022})}\BibitemShut {NoStop}%
\bibitem [{\citenamefont {Arceri}\ \emph {et~al.}(2020)\citenamefont {Arceri},
  \citenamefont {Landes}, \citenamefont {Berthier},\ and\ \citenamefont
  {Biroli}}]{arceri2020glasses}%
  \BibitemOpen
  \bibfield  {author} {\bibinfo {author} {\bibfnamefont {Francesco}\
  \bibnamefont {Arceri}}, \bibinfo {author} {\bibfnamefont {Fran{\c{c}}ois~P}\
  \bibnamefont {Landes}}, \bibinfo {author} {\bibfnamefont {Ludovic}\
  \bibnamefont {Berthier}}, \ and\ \bibinfo {author} {\bibfnamefont {Giulio}\
  \bibnamefont {Biroli}},\ }\bibfield  {title} {\enquote {\bibinfo {title}
  {Glasses and aging: a statistical mechanics perspective},}\ }\href@noop {}
  {\bibfield  {journal} {\bibinfo  {journal} {arXiv preprint arXiv:2006.09725}\
  } (\bibinfo {year} {2020})}\BibitemShut {NoStop}%
\bibitem [{\citenamefont {Monthus}\ and\ \citenamefont
  {Bouchaud}(1996)}]{monthus1996models}%
  \BibitemOpen
  \bibfield  {author} {\bibinfo {author} {\bibfnamefont {C{\'e}cile}\
  \bibnamefont {Monthus}}\ and\ \bibinfo {author} {\bibfnamefont
  {Jean-Philippe}\ \bibnamefont {Bouchaud}},\ }\bibfield  {title} {\enquote
  {\bibinfo {title} {Models of traps and glass phenomenology},}\ }\href@noop {}
  {\bibfield  {journal} {\bibinfo  {journal} {Journal of Physics A:
  Mathematical and General}\ }\textbf {\bibinfo {volume} {29}},\ \bibinfo
  {pages} {3847} (\bibinfo {year} {1996})}\BibitemShut {NoStop}%
\end{thebibliography}%

\appendix

\section{Elastoplastic model}

We consider a two-dimensional lattice whose linear box length is $L$ using the lattice constant as the unit of length. We employ a simple scalar description of an elastoplastic model. For each site, we assign local shear stress $\sigma_i$ and local plastic strain $\gamma^{\rm p}_i$ at a position ${\bf r}_i$.

The dynamical rule for our simulation model is inspired by the Monte-Carlo dynamics in molecular simulations~\cite{berthier2007monte}. 
We pick a site, say $i$, up randomly among $L^2$ sites. 
If $\sigma_i$ is greater (or lower) than or equal to a threshold $\sigma_c>0$ (or $-\sigma_c<0$), namely,
$|\sigma_i| \geq \sigma_c$, this site shows a plastic event: $\sigma_i \to \sigma_i - \delta \sigma_i$, where $\delta \sigma_i$ is the local stress drop.
In this paper, we use an uniform threshold, $\sigma_c=1$.
Instead, if $|\sigma_i| < \sigma_c$, with probability $e^{-\Delta E(\sigma_i)/T}$, where $\Delta E(\sigma_i)$ is a local energy barrier and $T$ is the temperature, this site shows a plastic event: $\sigma_i \to \sigma_i - \delta \sigma_i$. This corresponds to a plastic rearragement induced by a local thermal activation. We employ $\Delta E(\sigma_i)=(\sigma_c-|\sigma_i|)^a$ with $a=3/2$~\cite{maloney2006energy}.
This specific form of the local energy barrier is suggested by molecular simulation studies~\cite{fan2014thermally,lerbinger2021relevance} and previous elastoplastic models under shear~\cite{popovic2020thermally,ferrero2021yielding}.
The stress drop $\delta \sigma_i$ associated with a plastic event is a stochastic variable. In this paper, we use $\delta \sigma_i=(z+|\sigma_i|-\sigma_c){\rm sgn}(\sigma_i)$, where ${\rm sgn}(x)$ is the sign function and $z >0$ is a random number drawn by an exponential distribution, $p(z)=\frac{1}{z_0} e^{-z/z_0}$. $z_0$ is the mean value and we set $z_0=1$. This exponential distribution would be realistic according to molecular simulations in Ref.~\cite{barbot2018local}.
Due to the local stress drop, local plastic strain is updated as $\gamma^{\rm p}_i \to \gamma^{\rm p}_i + \delta \sigma_i/\mu$, where $\mu$ is the local shear modulus and we set $\mu=1$.

A local plastic event at site $i$ influences all other sites ($\forall j \neq i$) as 
\begin{equation}
    \sigma_j \to \sigma_j + {G}^{\psi_i}_{{\bf r}_{ji}} \ \delta \sigma_i,
\end{equation}
where ${\bf r}_{ji}={\bf r}_{j}-{\bf r}_{i}$ and $\psi_i \in [0, \pi/2)$ is a random orientation of the Eshelby kernel ${G}^{\psi}_{\bf r}$. Numerical implementation of ${G}^{\psi}_{\bf r}$ will be described below.
Similar to the Monte-Carlo dynamics, we repeat the above attempt $L^2$ times, which corresponds to unit time.

For the initial condition, we draw the local stress $\sigma_i$ $(\forall i)$ by the Gaussian distribution with zero mean and the standard deviation $R=1$, and we set $\gamma^{\rm p}_i=0$ $(\forall i)$.
To study dynamical properties at the steady-state, we monitor the waiting time dependence of correlation functions (see below), and we report various observables only at the steady-state, discarding the initial transient part.

We study $L=64$ and $L=128$ systems. We report mainly the results from $L=64$ unless otherwise stated.
For various observables such as correlation functions, we average over 10-40 and 5 independent trajectories for $L=64$ and $128$, respectively.

Note that our numerical implementation still has some room for improvement in terms of efficiency. Further optimization of simulation codes, such as incorporating the Faster-than-the-clock algorithm~\cite{krauth2006statistical} and efficient implementation of the Eshelby kernel, would enable us to study lower temperatures and larger system sizes, which will allow us to perform detailed finite-size scaling, etc. We leave such detailed studies for future investigation.

\section{Rotation of Eshelby kernel}

In supercooled liquids without external field, the Eshelby propagation are oriented in any direction~\cite{lemaitre2014structural,wu2015anisotropic}, unlike amorphous solids under shear where the Eshelbies tend to orient along the direction of shear~\cite{maloney2006amorphous,dasgupta2013yield}. We incorporate this feature in our elastoplastic modeling while keeping the scalar description with the shear stress component.
The method we introduce here is an approximated way compared with fully tensorial description~\cite{nicolas2014universal,budrikis2017universal,jagla2020tensorial}.
Yet it captures essential features of elasticity, e.g., long-range and anisotropic interactions.

In two-dimensional continuous medium, the Eshelby kernel in real space, $G({\bf r})$, for the shear stress component is given by 
\begin{equation}
    G({\bf r}) = \frac{\cos(4 \theta)}{\pi r^2},
\end{equation}
where $\theta$ is the angle between ${\bf r}=(x,y)$ and the $x$-axis, and $r=|{\bf r}|$.
Now we rotate this kernel randomly with an angle $\psi \in [0, \pi/2)$:
\begin{equation}
    G^{\psi}({\bf r}) = \frac{\cos(4 (\theta-\psi))}{\pi r^2}.
    \label{eq:rot_eshelby_2D_real}
\end{equation}
Note that due to symmetry it is enough to restrict $\psi$ up to $\pi/2$.
We then perform the Fourier transform~\cite{cao2018soft} of Eq.~(\ref{eq:rot_eshelby_2D_real}):
\begin{equation}
    \mathcal{F}\left[ \frac{\cos(4 (\theta-\psi))}{\pi r^2} \right] = \frac{1}{2} \cos(4(\phi-\psi)),
\end{equation}
where $\phi$ is the angle of the spherical coordinates in the Fourier space, namely, ${\bf q} =(q_x,q_y)=\left(q \cos(\phi), q \sin(\phi) \right)$, where $q=|{\bf q}|$.
This means that the rotation with the angle $\psi$ in real space corresponds to the rotation $\psi$ in the Fourier space.

As usual, we numerically implement the Eshelby kernel by using the expression in the Fourier space. First, we consider the standard case, $\psi=0$.
In the continuous space, the Fourier transform of the kernel $\widehat G({\bf q})$ is given by
\begin{equation}
\widehat G({\bf q}) = -\frac{4 q_x^2 q_y^2}{(q_x^2+q_y^2)^2}.
\end{equation}
Since we study a finite discrete lattice system, we employ the Laplacian correction method~\cite{ferrero2019criticality}:
\begin{equation}
q_{\alpha}^2 \to 2-2\cos(q_\alpha),
\end{equation}
where $q_\alpha=\frac{2\pi n_\alpha}{L}$ ($\alpha=x, y$), with $n_\alpha=-L/2+1,\dots,L/2$ for a system of the linear box length $L$ (even number).
Thus we get the kernel $\widehat G_{\bf q}$ in discrete Fourier space:
\begin{equation}
\widehat G_{\bf q} = -\frac{4 \left(1-\cos(q_x)\right)\left(1-\cos(q_y)\right)}{(2-\cos(q_x)-\cos(q_y))^2}.
\end{equation}

In order to take into account the rotation of the kernel, we rotate the axes in the Fourier space by $\psi$,
\begin{eqnarray}
q^{\psi}_x &=&  \cos(\psi) q_x + \sin(\psi) q_y, \\
q^{\psi}_y &=& - \sin(\psi) q_x + \cos(\psi) q_y,
\end{eqnarray}
and we define
\begin{equation}
    \widehat G^{\psi}_{\bf q} = -\frac{4 \left(1-\cos(q^{\psi}_x)\right)\left(1-\cos(q^{\psi}_y)\right)}{(2-\cos(q^{\psi}_x)-\cos(q^{\psi}_y))^2}.
    \label{eq:kernel_rotation_fourier_2D}
\end{equation}
We then perform the inverse of the discrete Fourier transform to get the real space kernel on the lattice:
\begin{equation}
    G^{\psi}_{\bf r} = \frac{1}{L^2} \sum_{\bf q} \widehat G^{\psi}_{\bf q} \ e^{i {\bf q} \cdot {\bf r}}.
\end{equation}

Next, we assign $G^{\psi}_{\bf r=0}$ and $\widehat G^{\psi}_{\bf q=0}$ in order to fulfill desired physical conditions in isotropic liquids.
In this situation, the total stress is conserved. 
To model this situation we impose $G^{\psi}_{{\bf r}=0}=-1$ and $\widehat G^{\psi}_{\bf q=0}=0$ numerically.
The original kernel in Eq.~(\ref{eq:kernel_rotation_fourier_2D}) is only defined on ${\bf q \neq 0}$.
Thus we introduce new kernel $\widehat G'^{\psi}_{\bf q}$ for the numerical simulation:
\begin{equation}
  \widehat G'^{\psi}_{\bf q} = \begin{cases}
    \frac{\widehat G^{\psi}_{\bf q}}{\mathcal{N}} & ({\bf q \neq 0}) \\
    0 & ({\bf q=0})
  \end{cases} 
  \label{eq:kernel_final}
\end{equation}
where $\mathcal{N} = - \frac{1}{L^2}\sum_{{\bf q \neq 0}}\widehat{G}^{\psi}_{\bf q}>0$.
One can easily see that this new kernel $\widehat{G}'^{\psi}_{\bf q}$ satisfies ${G}'^{\psi}_{\bf r=0} = \frac{1}{L^2}\sum_{{\bf q}}\widehat{G}'^{\psi}_{\bf q}=-1$~\cite{popovic2018elastoplastic,rossi2022finite}.

\begin{figure}
\includegraphics[width=0.32\columnwidth]{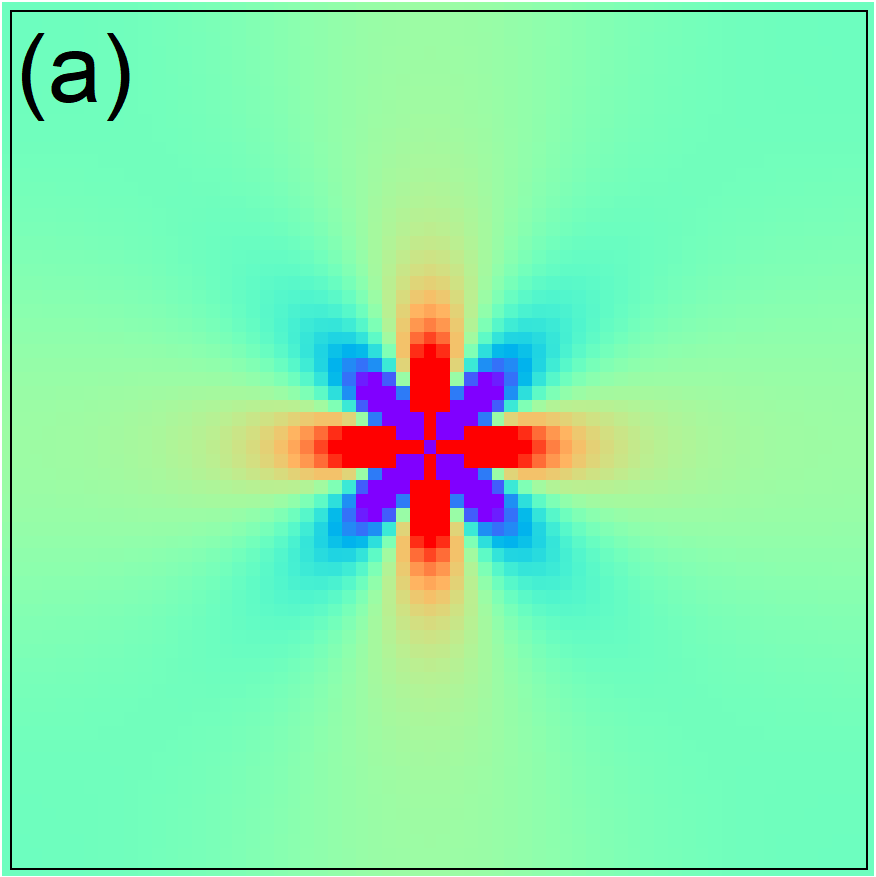}
\includegraphics[width=0.32\columnwidth]{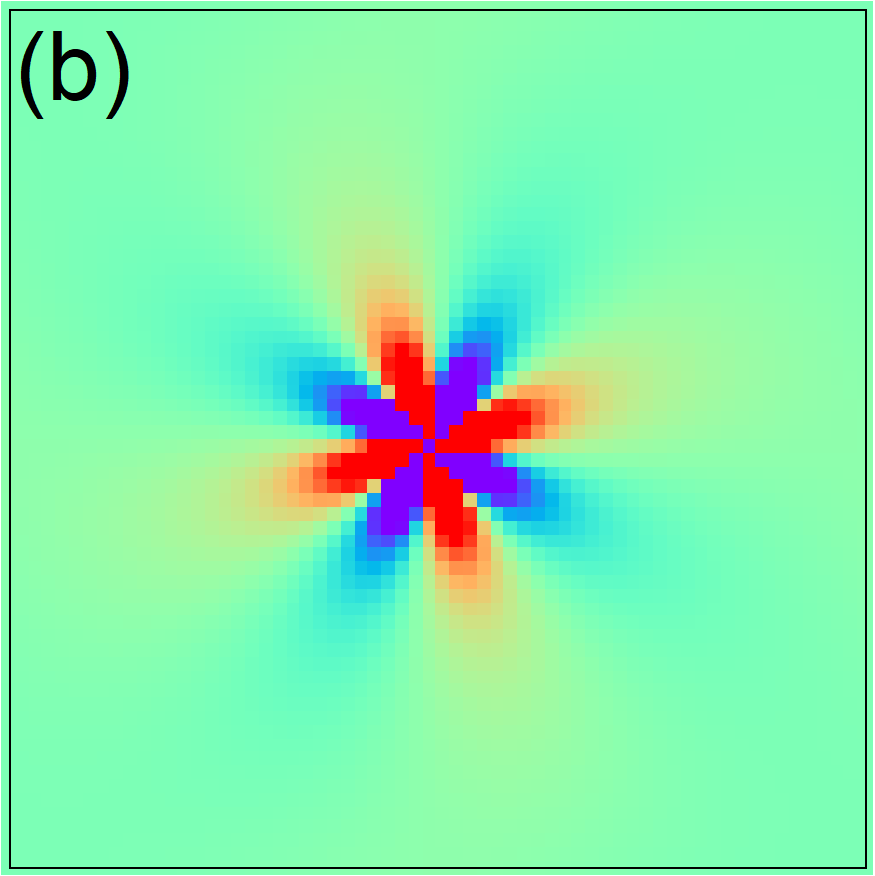}
\includegraphics[width=0.32\columnwidth]{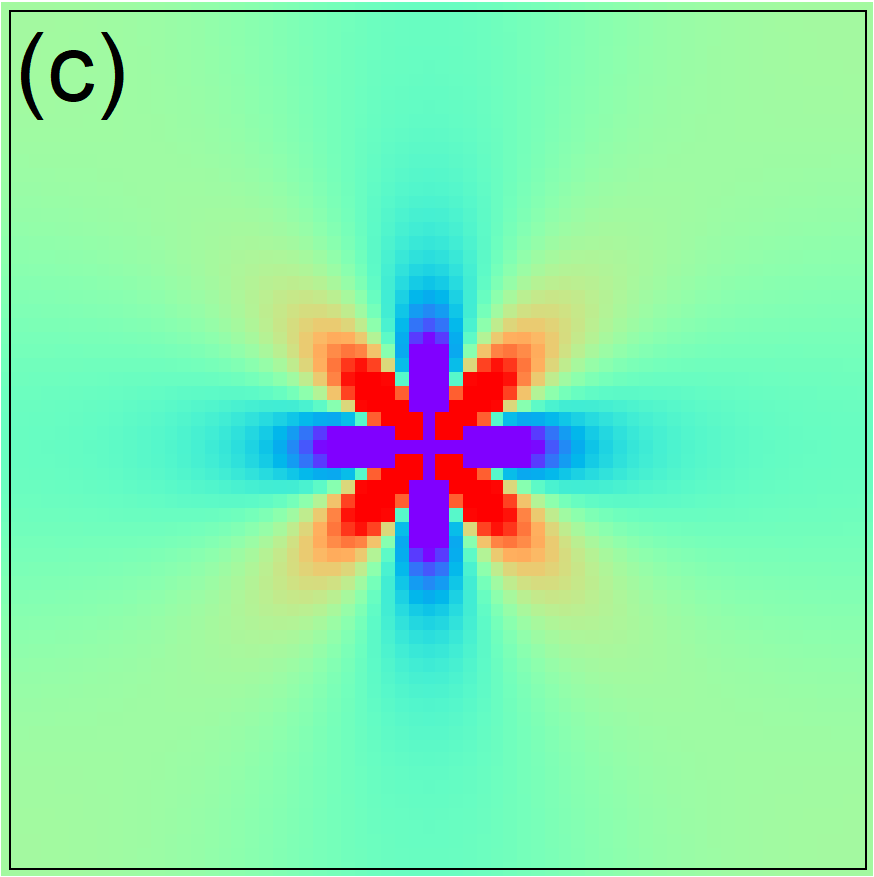}
\caption{Eshelby interaction kernel $G'^{\psi}_{\bf r}$ for $\psi=0$ (a), $\psi=0.3$ (b), and $\psi=\pi/4$ (c). Red and blue colors correspond to positive and negative values.
%System size is $L=64$.
}
\label{fig:Eshelby_rotation}
\end{figure}

We realized that a naive numerical implementation of the above method shows numerical instability. In particular, we found strong oscillations in $G'^{\psi}_{\bf r}$ along the $x$- and $y$-axes, except $\psi=0$ and $\pi/4$ (not shown). In order to avoid this instability, we employ the fine grid method used in Ref.~\cite{nicolas2014universal}.
In this method, we prepare a finer lattice whose size is $2L \times 2L$. We associate one site in the original $L \times L$ lattice with four sites in the finer lattice. The inverse of the discrete Fourier transform is performed on the finer lattice, and we average over four sites to get the value for the corresponding site on the original lattice.
Figure~\ref{fig:Eshelby_rotation} shows ${G}'^{\psi}_{\bf r}$ for $\psi=0, 0.3$ and $\pi/4$, obtained by the fine grid method, removing the numerical instability observed in the naive implementation.

\section{Time correlation functions}

\begin{figure}
\includegraphics[width=0.95\columnwidth]{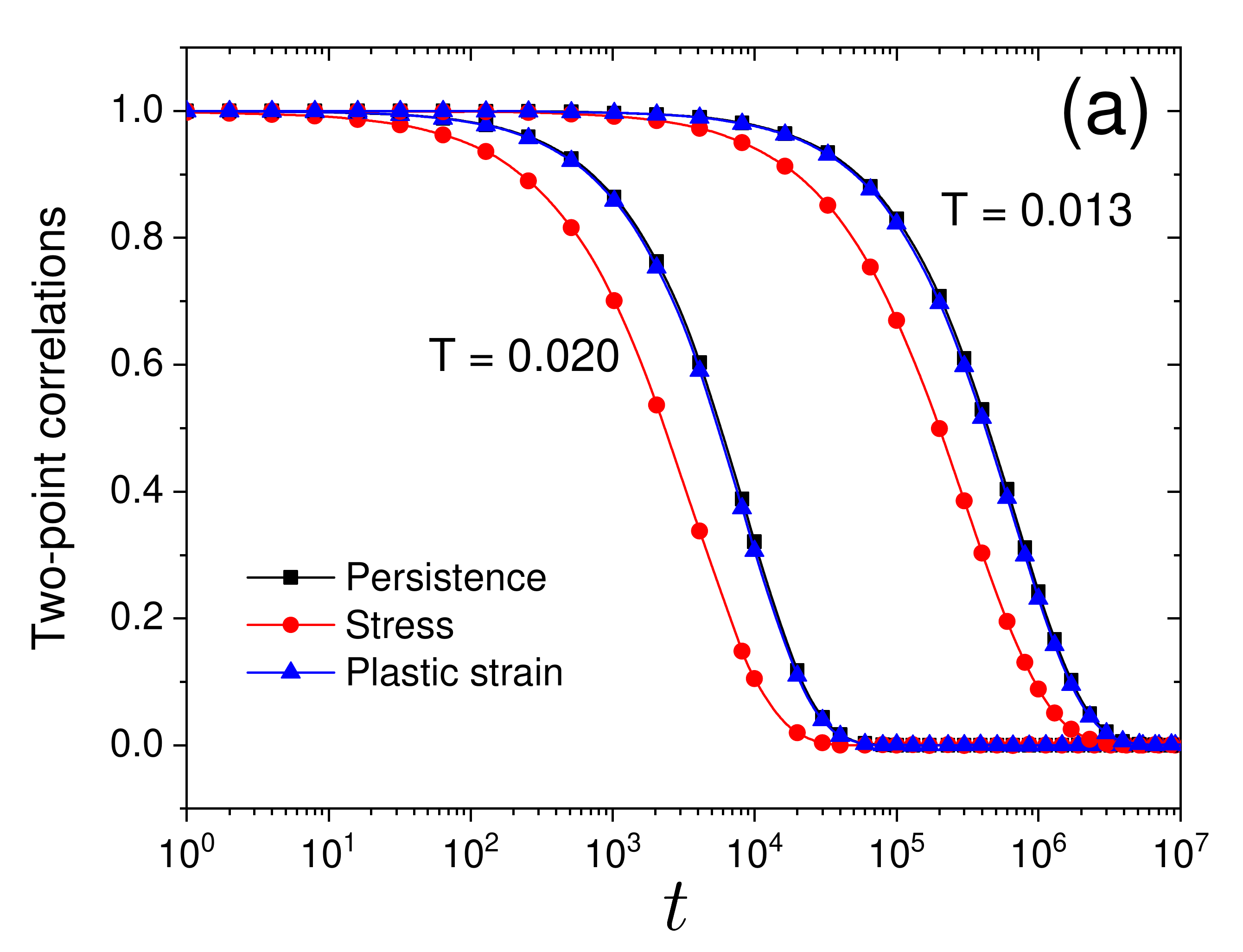}
\includegraphics[width=0.95\columnwidth]{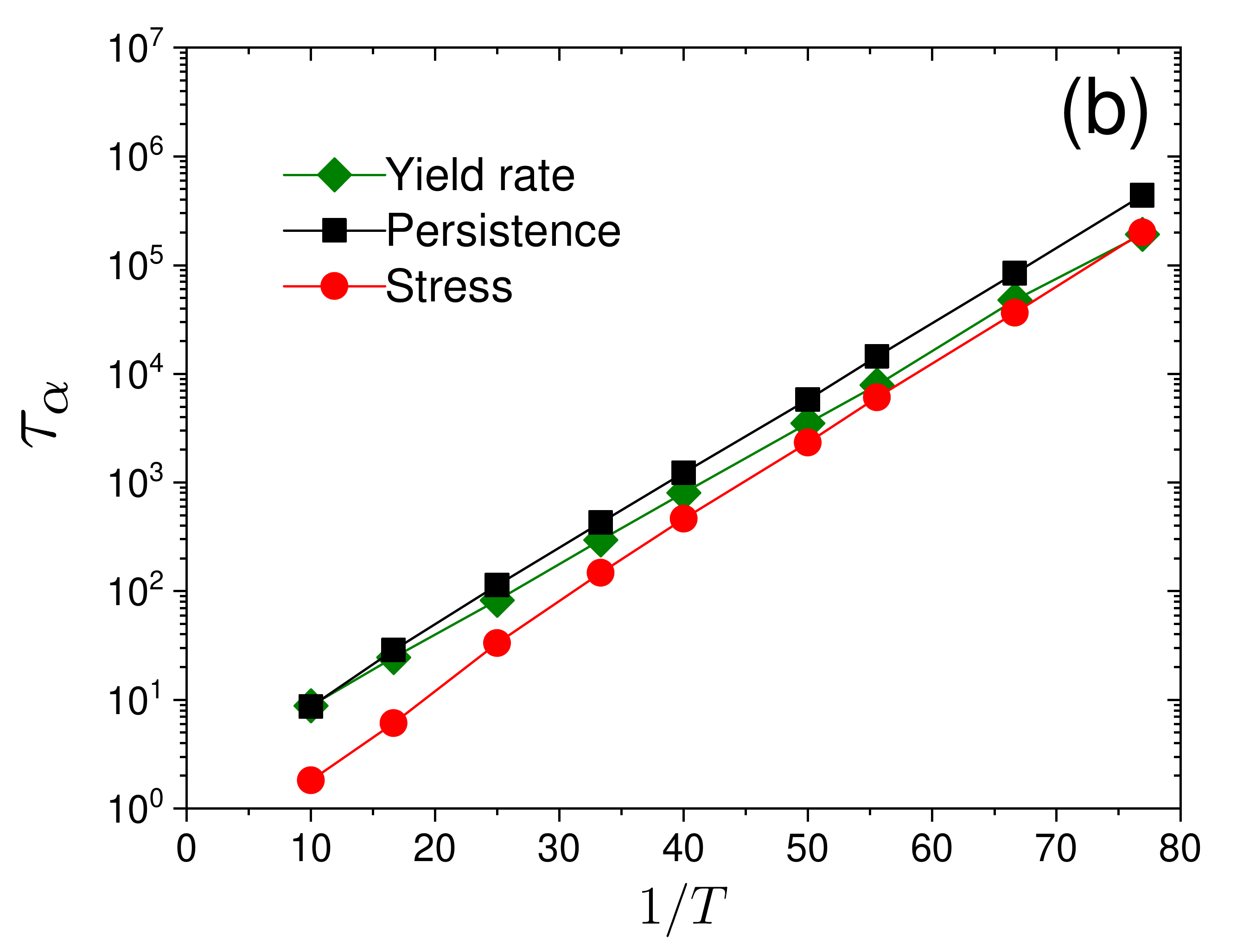}
\caption{(a): Three different two-point time correlation functions, $\langle P(t) \rangle$ (Persistence), $C_{\rm s}(t)$ (Stress), and $C_{\rm p}(t)$ (Plastic strain), for $T=0.020$ (left) and $T=0.013$ (right). (b): Relaxation time $\tau_{\alpha}$ obtained from $\tau_\alpha=1/\langle \Gamma \rangle$ (Yield rate), $\langle P(\tau_\alpha) \rangle=1/2$ (Persistence), and $C_{\rm s}(\tau_\alpha)=1/2$ (Stress). }
\label{fig:correlations}
\end{figure}

To study glassy slow dynamics of the model, we measure several time correlation functions.

First, we consider the following two-point time correlation functions.
The persistence correlation function, $\langle P(t) \rangle$, has been used widely in the context of kinetically constrained models~\cite{garrahan2011kinetically}. $P(t)$ is defined by $P(t) = \frac{1}{L^2}\sum_i p_i(t)$,
where
$p_i(t)=1$ if the site $i$ did not show a plastic event until time $t$ from $t=0$, and $p_i(t)=0$ otherwise.
$\langle \cdots \rangle$ denotes the time average at the stationary state.
We also measure the local stress auto-correlation function, $C_{\rm s}(t)$, given by 
\begin{equation}
    C_{\rm s}(t) = \frac{\sum_i \langle \sigma_i(t) \sigma_i(0) \rangle}{\sum_i \langle \sigma_i^2(0) \rangle}.
\end{equation}
Besides, we compute a correlation function for the local plastic strain, $C_{\rm p}(t)$, which is defined by
\begin{equation}
    C_{\rm p}(t) = \frac{1}{L^2} \sum_i \left\langle \cos\left(k\{\gamma_i^{\rm p}(t)-\gamma_i^{\rm p}(0)\}\right) \right\rangle,
\end{equation}
where $k$ is the wave number and we set $k=2\pi$. $C_{\rm p}(t)$ is an analog of the self-intermediate scattering function~\cite{kob1995testing}.

Figure~\ref{fig:correlations}(a) presents $\langle P(t) \rangle$, $C_{\rm s}(t)$, and $C_{\rm p}(t)$, for $T=0.20$ and $T=0.013$.
We find that $\langle P(t) \rangle$ and $C_{\rm p}(t)$ decay the same way in the temperature range we studied ($T=0.100-0.013$), while $C_{\rm s}(t)$ tend to decay slightly faster.
This discrepancy would come from the fact that $C_{\rm s}(t)$ contains stress fluctuations from both plastic events and stress variations in an elastic state (due to the stress redistribution), while the decay of $\langle P(t) \rangle$ and $C_{\rm p}(t)$ originates solely from the plastic activities. 

We then define the relaxation timescale $\tau_{\alpha}$ at which these correlation functions decay to $1/2$ (e.g. $\langle P(\tau_\alpha) \rangle=1/2$).
Figure~\ref{fig:correlations}(b) shows the obtained $\tau_\alpha$ versus $1/T$ plot, together with $1/\langle \Gamma \rangle$, where $\langle \Gamma \rangle$ is the yield rate.
All definitions of $\tau_\alpha$ follow the Arrhenius law with the essentially same activation energy barrier. We do not plot $\tau_\alpha$ from $C_{\rm p}(t)$, as it follows the same way as the one from $\langle P(t) \rangle$.
To conclude, the Arrhenius temperature dependence presented in Fig.~\ref{fig:dynamics}(b) in the main text is a robust feature irrespective of the definition of the timescale.

\begin{figure}
\includegraphics[width=0.95\columnwidth]{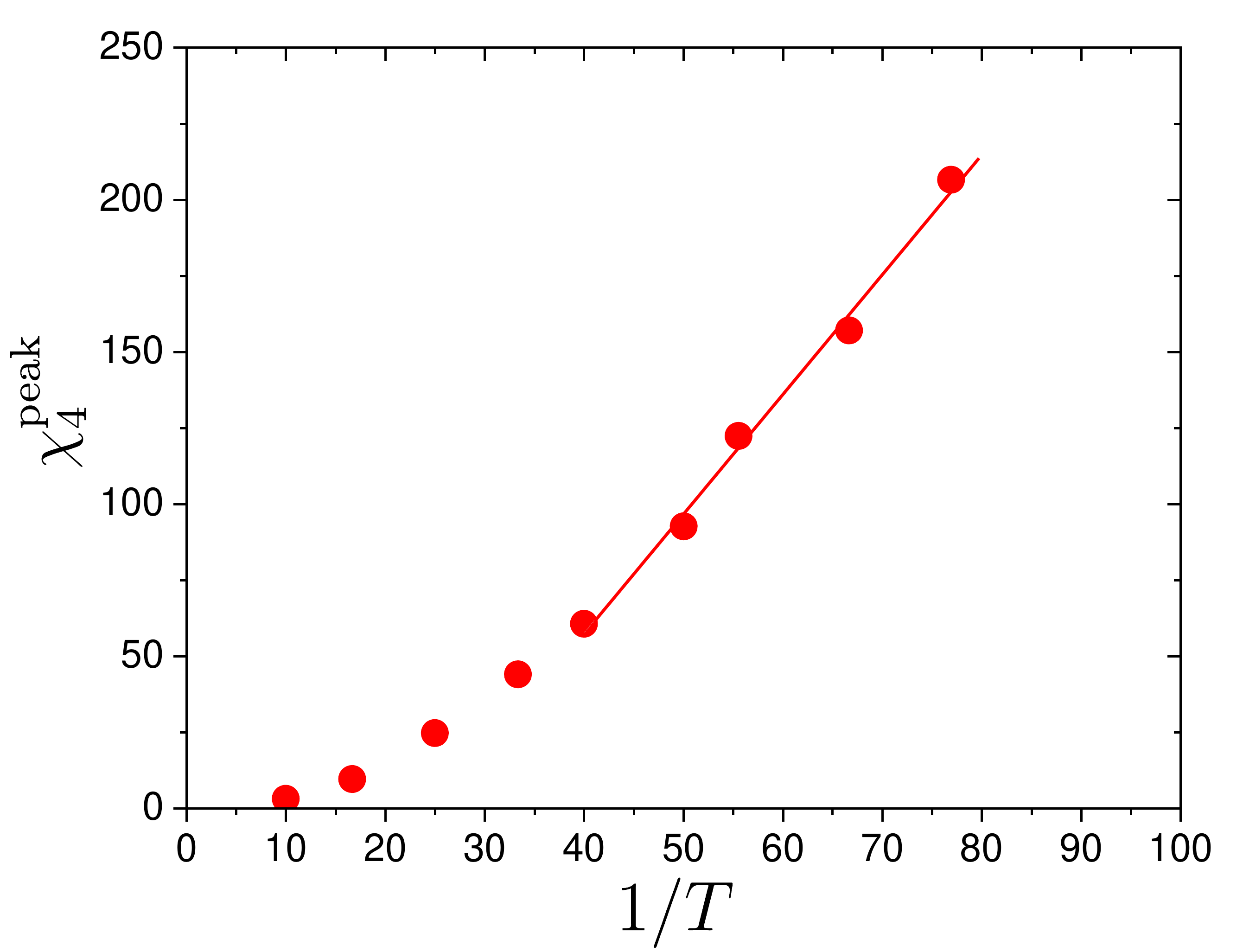}
\caption{
The peak of $\chi_4$ as a function of $1/T$. The solid line corresponds to a linear fit at lower temperatures.}
\label{fig:chi4_linear}
\end{figure}

In addition to the two-point correlation functions, we measure a four-point correlations function~\cite{donati2002theory}, $\chi_4(t)$, defined by
\begin{equation}
 \chi_4(t) = L^2\left( \langle P^2(t)\rangle - \langle P(t) \rangle^2 \right).   
\end{equation}
The time-temperature evolution of $\chi_4(t)$ is presented in Fig.~\ref{fig:chi4}(a) in the main text. It has a peak around the timescale of $\tau_\alpha$ and the peak grows with decreasing temperature, reflecting the growth of dynamically correlated plastic activities. 
Figure~\ref{fig:chi4_linear} shows the peak, $\chi_4^{\rm peak}$, plotted as a function of the inverse temperature. 
As argued in the main text, $\chi_4^{\rm peak} \sim 1/T$ is observed at lower temperatures.

\section{Absence of non-stationary effect}

\begin{figure}
\includegraphics[width=0.95\columnwidth]{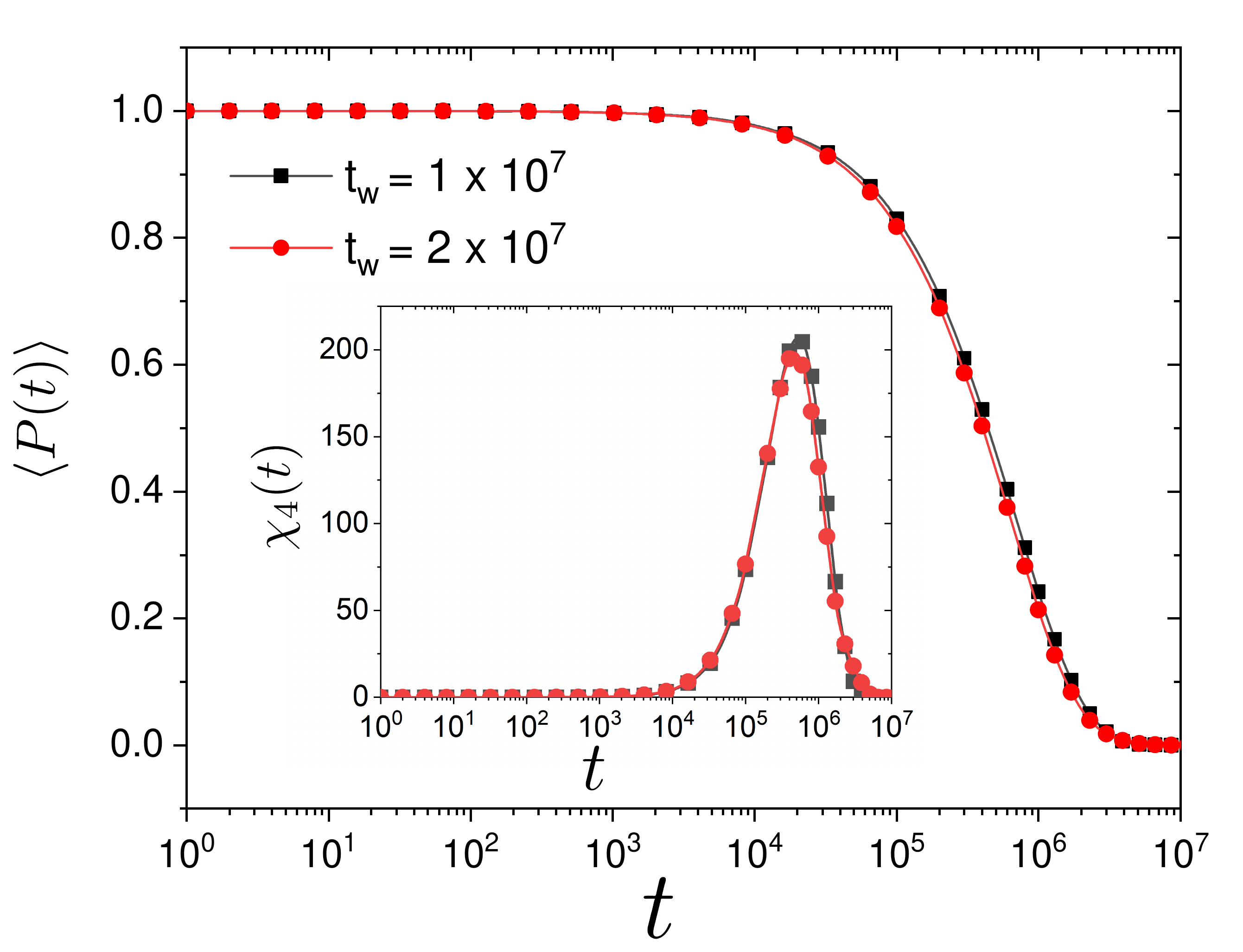}
\caption{Waiting time $t_{\rm w}$ dependence of $\langle P(t) \rangle$ for $T=0.013$. Inset: The corresponding $\chi_4(t)$.}
\label{fig:absence_aging}
\end{figure}

Glassy systems show the waiting time $t_{\rm w}$ dependence~\cite{arceri2020glasses}. Yet, in this paper, we focus on equilibrium, steady-state dynamics, where non-stationary effect is absent. 
To check its absence, we show the $t_{\rm w}$ dependence of $\langle P(t) \rangle$ and associated $\chi_4(t)$ in Fig.~\ref{fig:absence_aging}.
We compute $\langle P(t) \rangle$ and $\chi_4(t)$ within the time intervals, $[1\times 10^7, 2\times 10^7]$ (denoted as $t_{\rm w}=1\times 10^7$) and $[2\times 10^7, 3\times 10^7]$ ($t_{\rm w}=2\times 10^7$), starting from the initial condition given by the Gaussian stress distribution. 
We confirm that the data for $t_{\rm w}=1\times 10^7$ and $t_{\rm w}=2\times 10^7$ match very well within our numerical accuracy, demonstrating that non-stationary effect is indeed absent.

\section{Finite size effects}

\begin{figure}
\includegraphics[width=0.95\columnwidth]{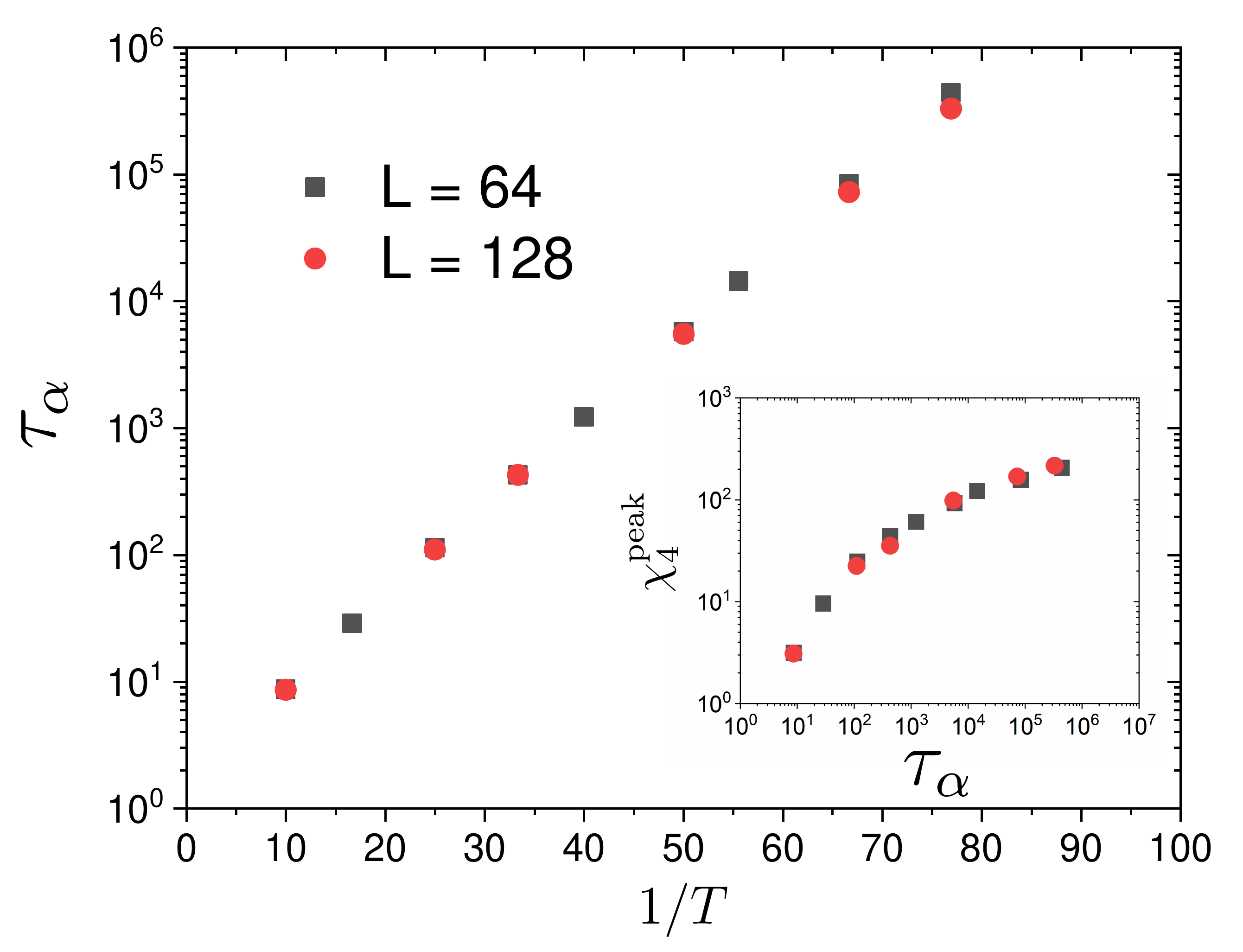}
\caption{System size dependence of $\tau_{\alpha}$ obtained from $\langle P(\tau_\alpha) \rangle=1/2$ for $L=64$ and $L=128$. Inset: The corresponding $\chi_4^{\rm peak}$ versus $\tau_\alpha$ plot.}
\label{fig:finite_size}
\end{figure}

One might naively speculate that the concave curvature presented in the $\chi_4^{\rm peak}$ versus $\tau_\alpha$ plot in Fig.~\ref{fig:chi4}(b) in the main text originates from a finite size effect in dynamics. 
To assess this issue, we compare results from two system sizes, $L=64$ and $L=128$, in terms of $\tau_\alpha$ and $\chi_4^{\rm peak}$ in Fig.~\ref{fig:finite_size}.
We confirm that $\tau_\alpha$ as well as $\chi_4^{\rm peak}$ do not show a noticeable finite size effect within our numerical accuracy, which allows us to conclude that the curvature in Fig.~\ref{fig:chi4}(b) in the main text is a real physical feature caused by the mechanism of elastic interactions.

\section{Dynamic facilitation}

\begin{figure}
\includegraphics[width=0.48\columnwidth]{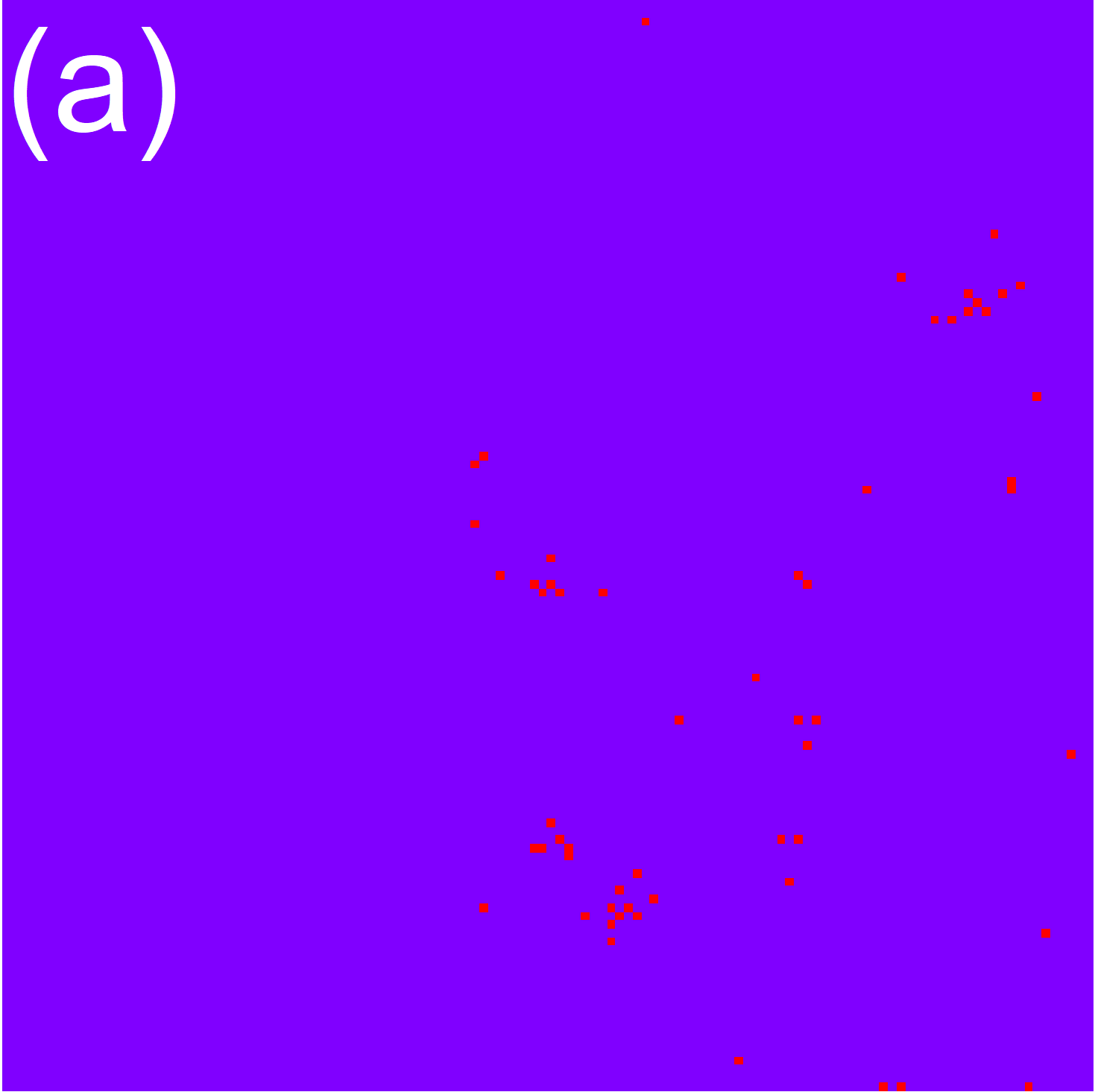}
\includegraphics[width=0.48\columnwidth]{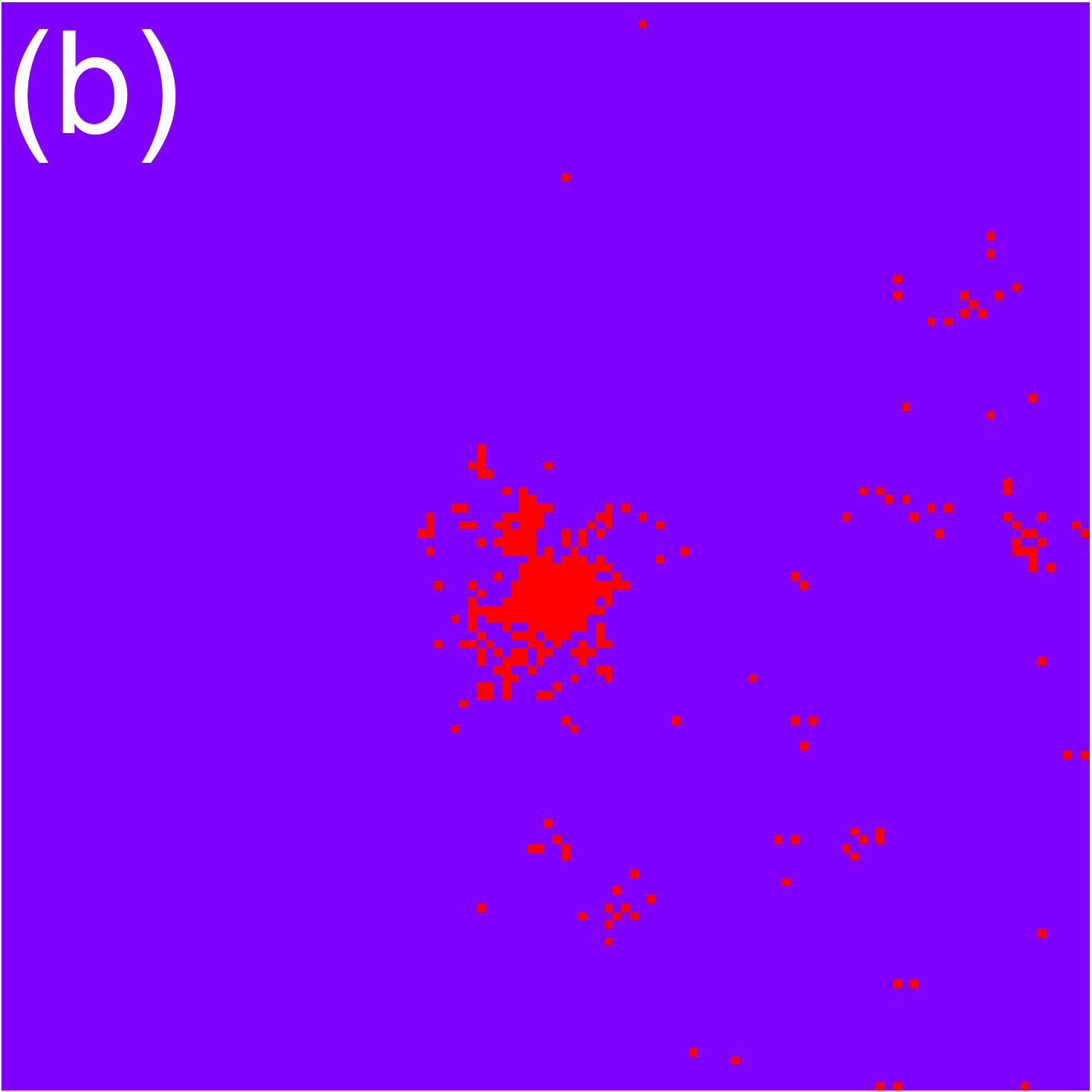}
\includegraphics[width=0.48\columnwidth]{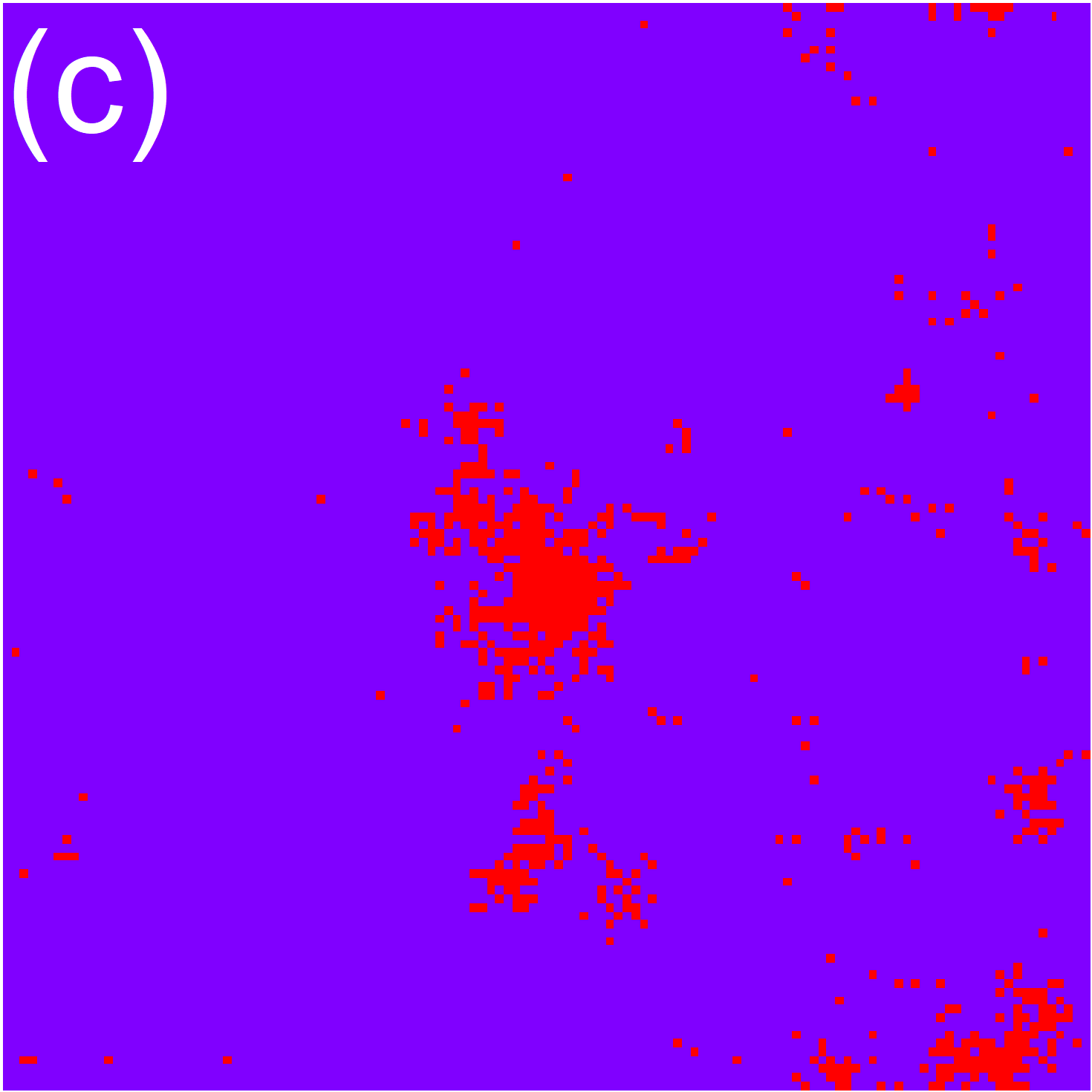}
\includegraphics[width=0.48\columnwidth]{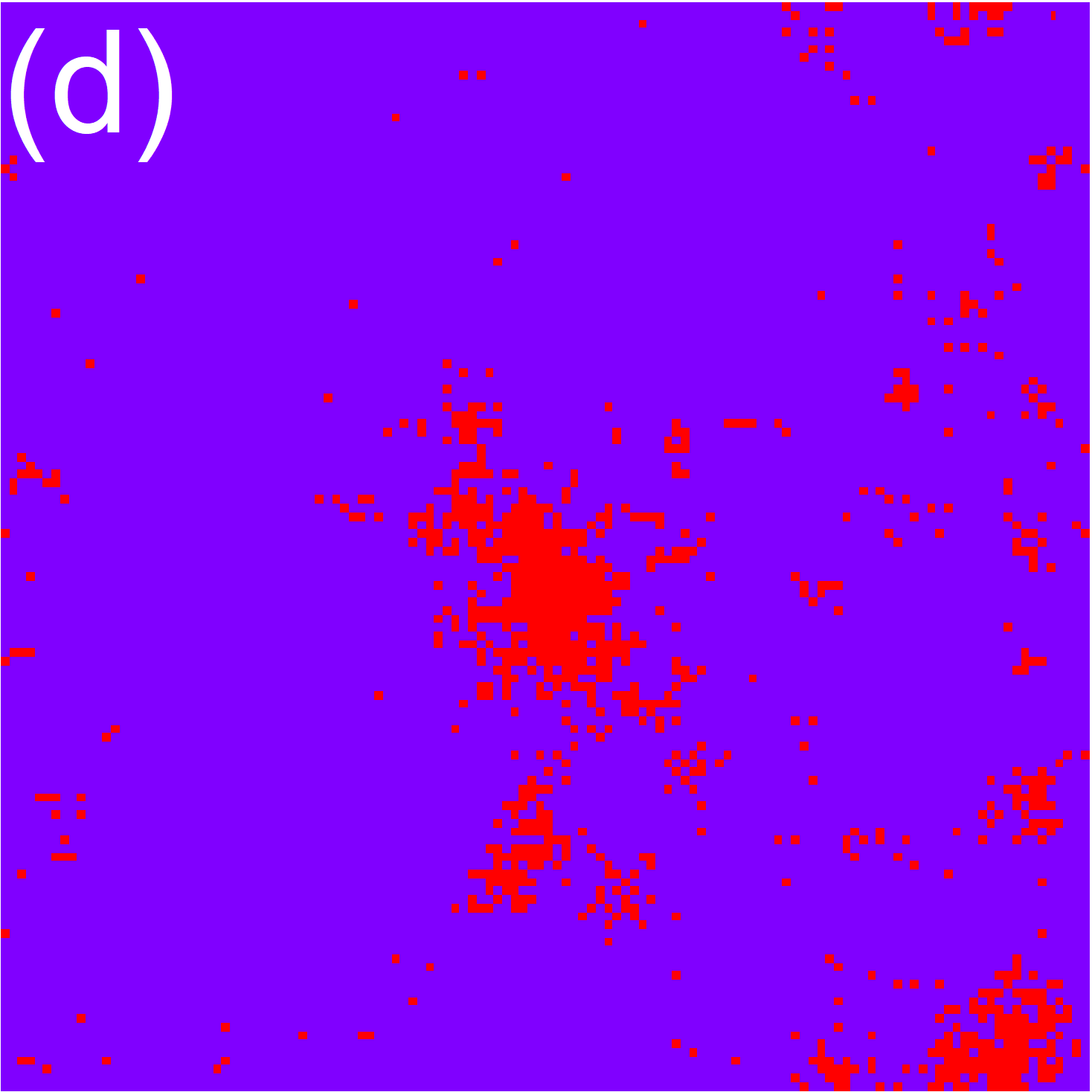}
\includegraphics[width=0.48\columnwidth]{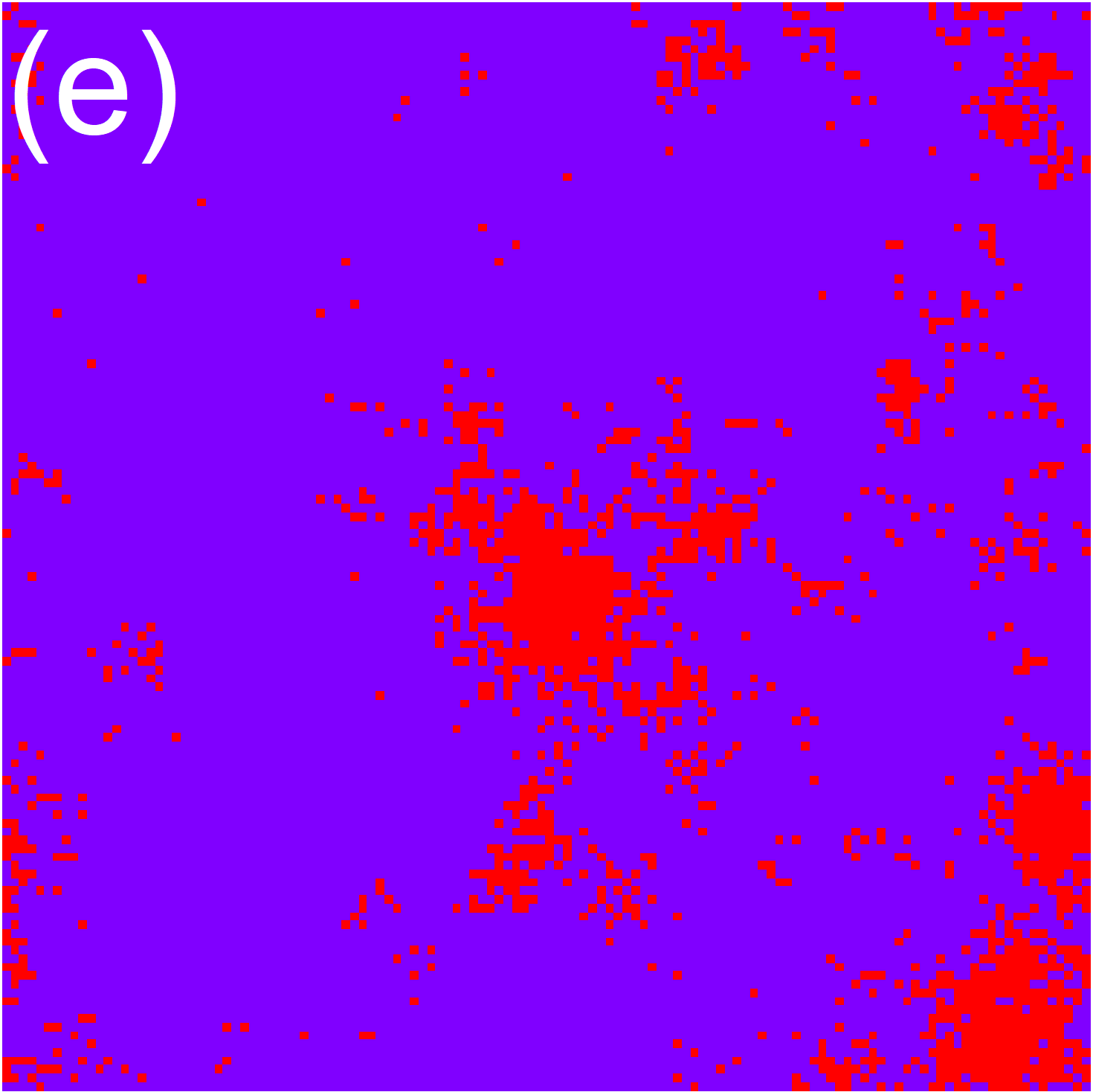}
\includegraphics[width=0.48\columnwidth]{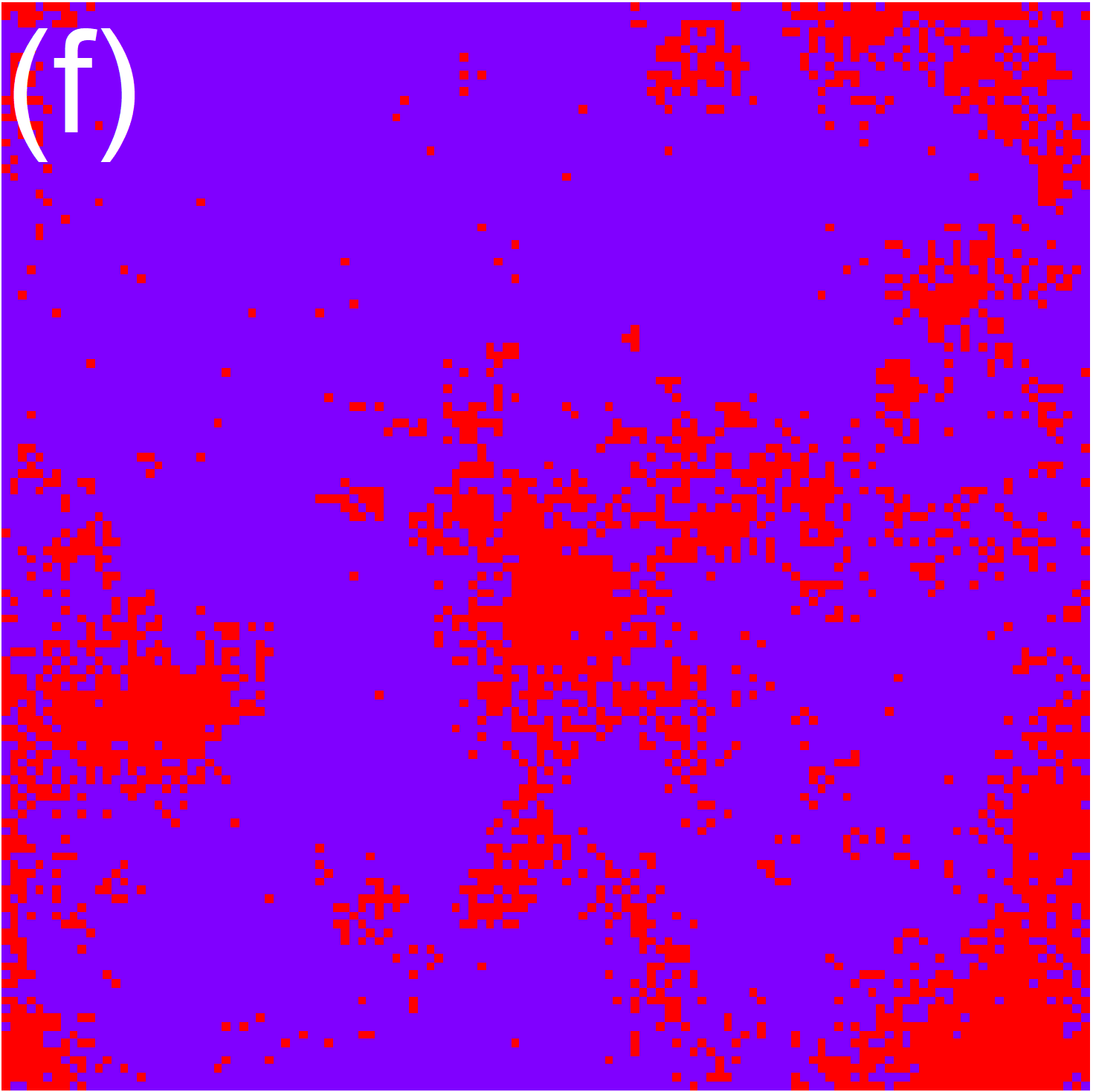}
\includegraphics[width=0.48\columnwidth]{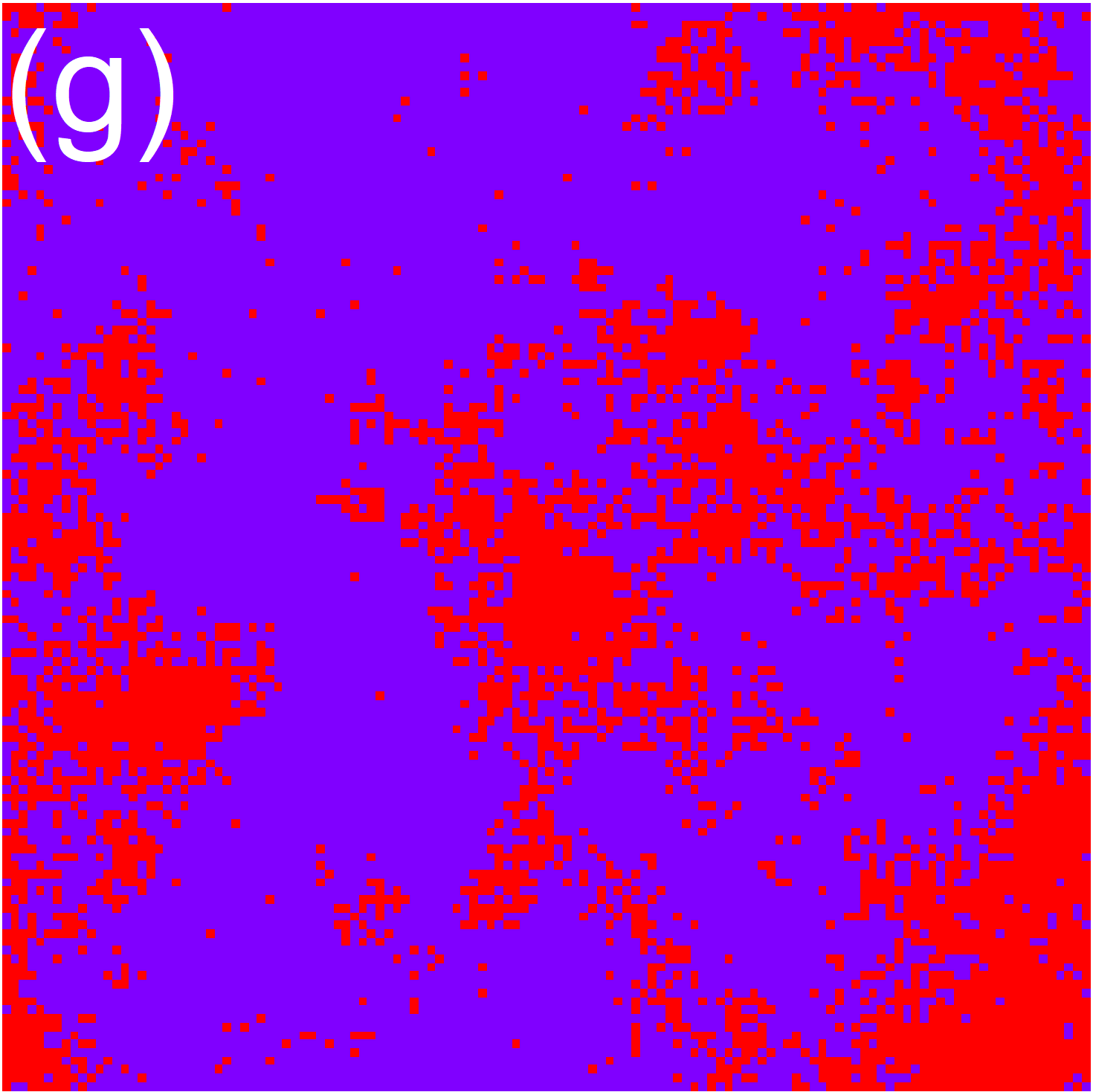}
\includegraphics[width=0.48\columnwidth]{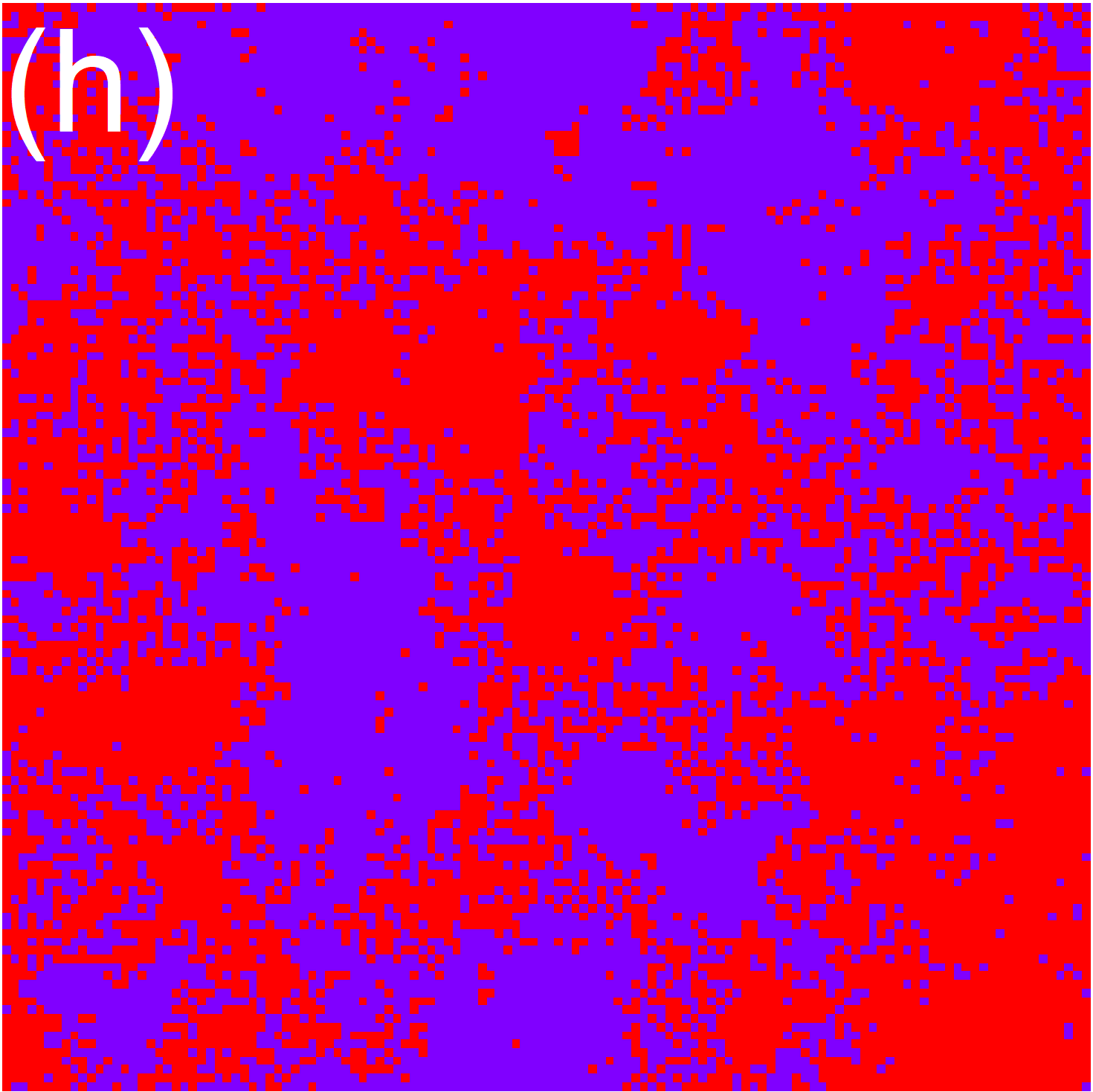}
\caption{
Time evolution of snapshots for local persistence, $p_i(t)$, for $T=0.013$ and $L=128$.
Red and blue sites correspond to mobile ($p_i(t)=0$) and immobile ($p_i(t)=1$) sites, respectively.  
The snapshot at $t=0$ corresponds to $p_i(0)=1$ ($\forall i$) (not shown).
(a): $t= 2048$.
(b): $t= 4096$.
(c): $t= 8192$.
(d): $t= 16384$.
(e): $t= 32768$.
(f): $t= 65536$.
(g): $t= 1 \times 10^5$.
(h): $t= 2 \times 10^5$.
}
\label{fig:dynamical_facilitation}
\end{figure}

The data for $\chi_4$ presented in Fig.~\ref{fig:chi4} in the main text demonstrate the growth of dynamic heterogeneity with time and temperature. To get more insight into the physical mechanism, in Fig.~\ref{fig:dynamical_facilitation}, we show a sequence of local persistence maps (or mobility maps), characterizing the time evolution of dynamic heterogeneity. The time intervals are selected in a nearly logarithmic way. This visualization reveals the following phenomenology: A localized mobile site ($p_i(t)=0$, red) induces other mobile sites in the neighbor region. These induced mobile sites again induce other mobile sites in nearby regions, showing a cascade or propagation of rearrangements, that is, the so-called dynamic facilitation~\cite{chandler2010dynamics,keys2011excitations}. 
These plots show striking similarity with a recent extensive molecular simulation study at very low temperature~\cite{guiselin2021microscopic}.
On top of that, our results based on the elastoplastic modeling provide evidence that the elastic interaction causes dynamic facilitation.  

\section{Kinetic elastoplastic theory (KEP)}

%Mean-field model provides us with the relation between $\alpha$ and $\overline \sigma$ in Eq.~(\ref{eq:alpha_MF}) and plotted in Fig.~\ref{fig:MF}(b).
%Yet the value of $\alpha$ which originates from the magnitude of elastic interaction cannot be determined within the mean-field theory.
In the following, we apply the theory developed in Ref.~\cite{bocquet2009kinetic} to the EPM-Q model. This allows us to obtain the value of $\alpha$ associated with the magnitude of elastic interactions, and it is the starting point of our analytical treatment. 

In the approach~\cite{bocquet2009kinetic}, one considers the dynamical evolution of the probability distribution for each site $i$,
\begin{eqnarray}
\frac{\partial P_i(\sigma, t)}{\partial t} &=&  - \nu(\sigma, \sigma_c) P_i(\sigma, t) + \Gamma_i(t)y(\sigma) + \mathcal{L}(P,P), \nonumber \\
\mathcal{L}(P,P) &=& \sum_{j \neq i} \Gamma_j(t) \left( P_i(\sigma+\Delta \sigma_i, t) -P_i(\sigma,t)\right),
\label{eq:KEP}
\end{eqnarray}
with $\Delta \sigma_i = G_{{\bf r}_{ij}}^{\psi_j}\delta \sigma_j$, where ${\bf r}_{ij}={\bf r}_i-{\bf r}_j$ and $\psi_j$ is the rotation angle of the Eshelby kernel centered at the site $j$. 
The term $\mathcal{L}(P,P)$ corresponds to contributions of elastic propagation from plastic events at other sites.
$\Gamma_i(t)$ corresponds to the rate of relaxations taking place at site $i$, and can be obtained self-consistently from $P_i(\sigma, t)$ (see below and Ref.~\cite{bocquet2009kinetic}).

One further assumes that $\Delta \sigma_i$ is small and hence $\mathcal{L}(P,P)$ can be expanded up to the second order:
\begin{eqnarray}
\mathcal{L}(P,P) &\simeq& \left( \sum_{j \neq i} \Gamma_j(t) G_{{\bf r}_{ij}}^{\psi_j}\delta \sigma_j \right) \frac{\partial P_i(\sigma, t)}{\partial \sigma} \nonumber \\
&\qquad& + \left( \frac{1}{2} \sum_{j \neq i} \Gamma_j(t) \left(G_{{\bf r}_{ij}}^{\psi_j}\delta \sigma_j \right)^2 \right) \frac{\partial^2 P_i(\sigma, t)}{\partial \sigma^2}. \nonumber \\
\label{eq:LPP_expansion}
\end{eqnarray}
The coefficient of the first term of RHS in Eq.~(\ref{eq:LPP_expansion}) vanishes in isotropic fluids because $\delta \sigma_j$ can be positive and negative in a symmetric way. 
Instead, the second term, together with Eq.~(\ref{eq:KEP}) provides an effective diffusion constant, leading to the equation: 
\begin{eqnarray}
\frac{\partial P_i(\sigma, t)}{\partial t} &=&  - \nu(\sigma, \sigma_c)  P_i(\sigma, t) + \Gamma_i(t)y(\sigma) +\nonumber \\
&&+D_i(t)\frac{\partial^2 P_i(\sigma,t)}{\partial \sigma^2},
\label{eq:KEP-dif}
\end{eqnarray}
where
\begin{equation}
    D_i(t) = \frac{1}{2} \sum_{j \neq i} \Gamma_j(t) \left(G_{{\bf r}_{ij}}^{\psi_j}\delta \sigma_j\right)^2.
\end{equation}
Considering a uniform situation, where $D_i(t) \to D(t)$ and $\Gamma_j(t) \to \Gamma(t)$ one obtains (skipping now the spatial index $i$)
\begin{eqnarray}
\frac{\partial P(\sigma, t)}{\partial t} = D(t) \frac{\partial^2 P(\sigma, t)}{\partial \sigma^2} &-& \nu(\sigma, \sigma_c) P(\sigma, t) \nonumber \\ 
&+& \Gamma(t)y(\sigma),
\label{eq:HL_dynamics}
\end{eqnarray}
where $D(t) = \alpha \Gamma(t)$ with 
\begin{equation}
\alpha = \frac{1}{2} \sum_{j \neq i} \left(G_{{\bf r}_{ij}}^{\psi_j} \delta \sigma_j\right)^2.     
\end{equation}
We approximate this further as follows (similarly to Ref.~\cite{bocquet2009kinetic}):
\begin{equation}
\alpha \simeq \frac{\sigma_0^2}{2} \sum_{j \neq i} \left(G_{{\bf r}_{ij}}^{\psi_j} \right)^2 = \frac{\sigma_0^2}{\pi} \int_0^{\frac{\pi}{2}} \mathrm{d} \psi \sum_{j \neq i} \left(G_{{\bf r}_{ij}}^{\psi}\right)^2.
\label{eq:alpha_approx}
\end{equation}
We evaluate numerically Eq.~(\ref{eq:alpha_approx}) by using the Eshelby kernel in Eq.~(\ref{eq:kernel_final}), which provides us with $\alpha=0.110$.
%As we shall show, this corresponds to $\overline \sigma= 0.749$ by using Eq.~(\ref{eq:alpha_MF}). These values are reported in Fig.~\ref{fig:MF} and used in Fig.~\ref{fig:stress_distribution} of the main text. 

\section{Analytical theory}

In the following, we first obtain the steady state equation, and then present its solution in the regime of low temperatures. We then discuss elasticity-induced facilitation within the mean-field theory.

\subsection{Steady state equation}
Equation~(\ref{eq:HL_dynamics}) resembles the one of Hébraud-Lequeux-like mean-field models~\cite{hebraud1998mode,agoritsas2015relevance}.
Since $P(\sigma, t)$ and $y(\sigma)$ are normalized, $\int_{-\infty}^{\infty} \mathrm{d} \sigma \ P(\sigma, t)=1$ and $\int_{-\infty}^{\infty} \mathrm{d} \sigma \ y(\sigma)=1$, by integrating Eq.~(\ref{eq:HL_dynamics}) over $\sigma$ one obtains the expression for the yield rate,
\begin{equation}
    \Gamma(t) = \int_{-\infty}^{\infty} \mathrm{d} \sigma \ \nu(\sigma, \sigma_c) P(\sigma,t).
    \label{eq:yield_rate}
\end{equation}
For the original Hébraud-Lequeux model at zero temperature~\cite{hebraud1998mode}, $\nu(\sigma, \sigma_c)=\frac{1}{\tau_0}\theta(|\sigma|-\sigma_c)$ and $y(\sigma)=\delta(\sigma)$ were employed, where $1/\tau_0$ is the rate of plastic event and $\delta(x)$ is the Dirac's delta function.
Following the numerical simulations of the two-dimensional model described above, we consider thermal activation~\cite{popovic2020thermally} and exponential local stress drop, given by
\begin{eqnarray}
 \nu(\sigma, \sigma_c) &=& \frac{1}{\tau_0} \theta(|\sigma|-\sigma_c) + \frac{1}{\tau_0} e^{-\frac{\Delta E(\sigma)}{T}} \theta(\sigma_c-|\sigma|), \\
 y(\sigma) &=& \frac{1}{\mathcal{N}} e^{(|\sigma|-\sigma_c)/\sigma_0} \theta(\sigma_c-|\sigma|),
 \label{eq:def_y}
\end{eqnarray}
where $\Delta E(\sigma)=(\sigma_c-|\sigma|)^a$ is the local energy barrier and $\mathcal{N}=2 \sigma_0(1-e^{-\sigma_c/\sigma_0})$ is the normalization factor.

We then consider the steady state solution with $\lim_{t \to \infty} P(\sigma, t)=P(\sigma)$, $\lim_{t \to \infty} \Gamma(t)=\Gamma$, and $\lim_{t \to \infty} D(t)=D$.
Therefore, Eq.~(\ref{eq:HL_dynamics}) for the steady state becomes
\begin{equation}
\alpha \Gamma \frac{\partial^2 P(\sigma)}{\partial \sigma^2} - \nu(\sigma, \sigma_c)P(\sigma) + \Gamma y(\sigma) = 0.
\label{eq:MF_SS}
\end{equation}
Since $P(\sigma)$ is symmetric in thermal equilibrium, we consider only the $\sigma \geq 0$ case.

\subsection{Low $T$ solution}

We consider Eq.~(\ref{eq:MF_SS}) at very low temperature. We assume that $\Gamma \simeq \frac{1}{\tau_0}e^{-\overline E/T}$ for $T\rightarrow 0$, where $\overline E$ is an energy barrier, and then check a posteriori the validity of this assumption (and also determine $\overline{E}$).
For the analysis of  Eq.~(\ref{eq:MF_SS}) it is useful to define a stress value $\overline{\sigma}$ as the stress associated with $\overline{E}$ through the equation $\overline{E}=\Delta E( \overline{\sigma})$.
There are three different regimes to study separately. 

\vspace{0.2cm}
{\it Regime I:} $0 \leq \sigma <\overline{\sigma}$. 
In this case 
%we assume that the local stress $\sigma$ is localized below some $\overline \sigma$, namely, $\sigma \lesssim \overline \sigma < \sigma_c$ and $P(\sigma)$ is nearly supported by $\overline \sigma$.
%Thus, the integration for the yield rate in Eq.~(\ref{eq:yield_rate}) is dominated by the value at $\overline \sigma$ and can be evaluated as $\Gamma \sim \nu(\overline \sigma, \sigma_c)=\frac{1}{\tau_0}e^{-\Delta E(\overline \sigma)/T}$.
$\nu(\sigma, \sigma_c)/\Gamma \simeq e^{-(\Delta E(\sigma)-\Delta E(\overline \sigma))/T} \to 0$ when $T \to 0$, which means that the second term in Eq.~(\ref{eq:MF_SS}) can be neglected at very low temperature.
We thus get
\begin{equation}
    \alpha \frac{\partial^2 P(\sigma)}{\partial \sigma^2} +  \frac{e^{(\sigma-\sigma_c)/\sigma_0}}{\mathcal{N}} = 0.
    \label{eq:MF_final}
\end{equation}
We solve Eq.~(\ref{eq:MF_final}) at $T \to 0$ with the boundary conditions, $\lim_{\sigma \to \overline \sigma} P(\sigma)=0$ and $\lim_{\sigma \to 0} \frac{\partial P(\sigma)}{\partial \sigma}=0$. 
We note that unlike the standard Hébraud-Lequeux model with $y(\sigma)=\delta(\sigma)$, we assume that $P(\sigma)$ is smooth at $\sigma=0$ because our $y(\sigma)$ in Eq.~(\ref{eq:def_y}) does not involve a singurarity at $\sigma=0$.

We then obtain the expression for $P(\sigma)$:
\begin{eqnarray}
P(\sigma) &=& \frac{\sigma_0^2}{\alpha \mathcal{N}} \left( e^{(\overline \sigma -\sigma_c)/\sigma_0} - e^{(\sigma -\sigma_c)/\sigma_0} \right) \nonumber \\
&\qquad& \qquad \qquad \qquad + \ \frac{\sigma_0 e^{-\sigma_c/\sigma_0}}{\alpha \mathcal{N}} (\sigma-\overline \sigma).
\label{eq:P_MF}
\end{eqnarray}
$P(\sigma)$ with specific values of parameters (see below) is plotted in Fig.~\ref{fig:MF}(a).

With the normalization, $\frac{1}{2} = \int_0^{\overline \sigma} \mathrm{d} \sigma \ P(\sigma)$, 
we get
\begin{equation}
\frac{1}{2} = \frac{\sigma_0^2 e^{(\overline \sigma-\sigma_c)/\sigma_0}}{\alpha \mathcal{N}}(\overline \sigma - \sigma_0) + \frac{\sigma_0 e^{-\sigma_c/\sigma_0}}{\alpha \mathcal{N}}\left(\sigma_0^2 -  \frac{\overline \sigma^2}{2}\right), \nonumber
\end{equation}
which determines the relation between $\overline \sigma$ and $\alpha$, as given by
\begin{equation}
\alpha = \frac{\sigma_0(\overline \sigma - \sigma_0) e^{(\overline \sigma - \sigma_c)/\sigma_0} + (\sigma_0^2 - \overline \sigma^2/2) e^{-\sigma_c/\sigma_0}}{1-e^{-\sigma_c/\sigma_0}}.
\label{eq:alpha_MF}
\end{equation}
This solution with $\sigma_c=\sigma_0=1$ is shown in Fig.~\ref{fig:MF}(b).
Using the value of $\alpha$ determined previously ($\alpha=0.110$), this gives $\overline \sigma= 0.749$. These values are reported in Fig.~\ref{fig:MF} and used in Fig.~\ref{fig:stress_distribution} of the main text. 
\begin{figure}
\includegraphics[width=0.95\columnwidth]{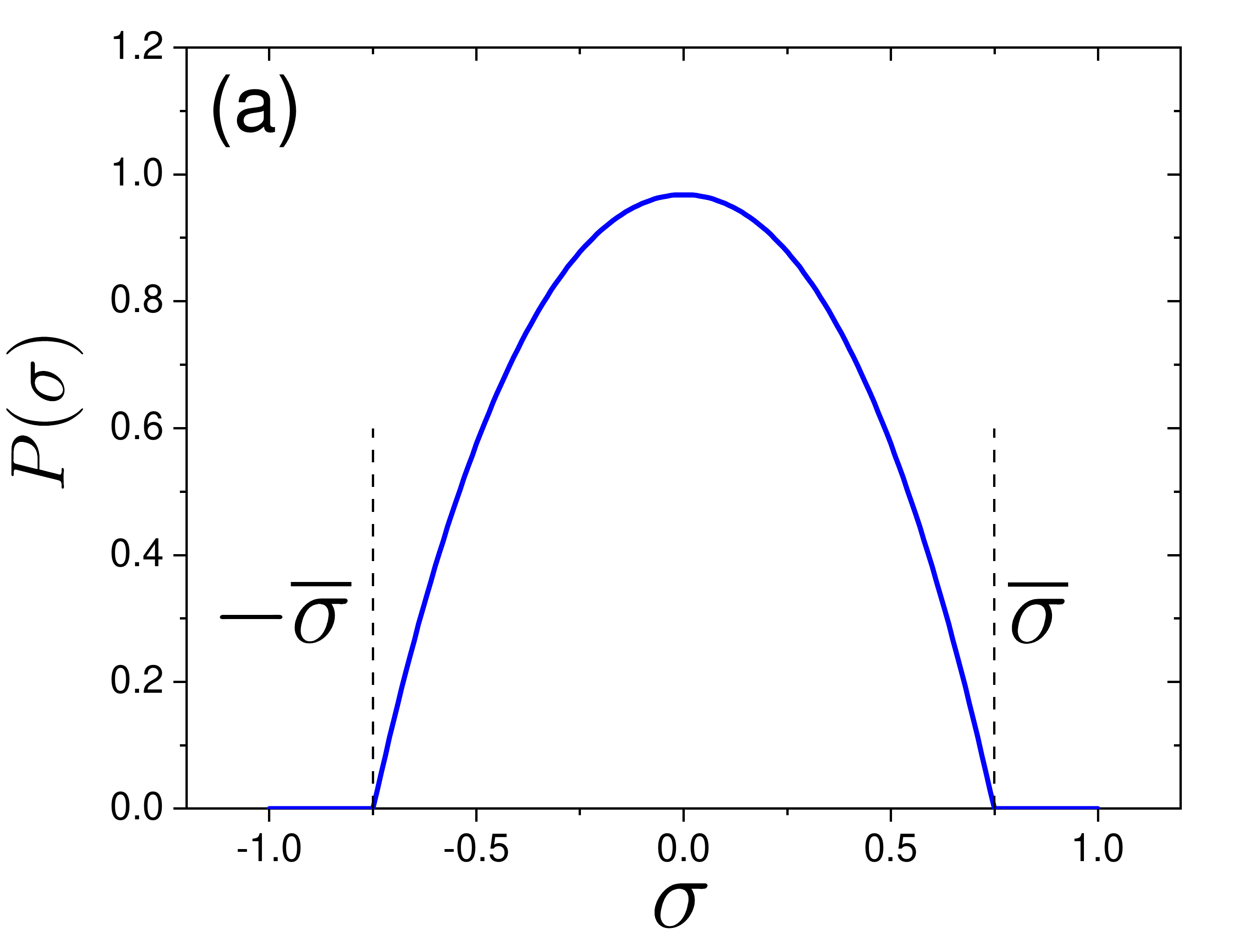}
\includegraphics[width=0.95\columnwidth]{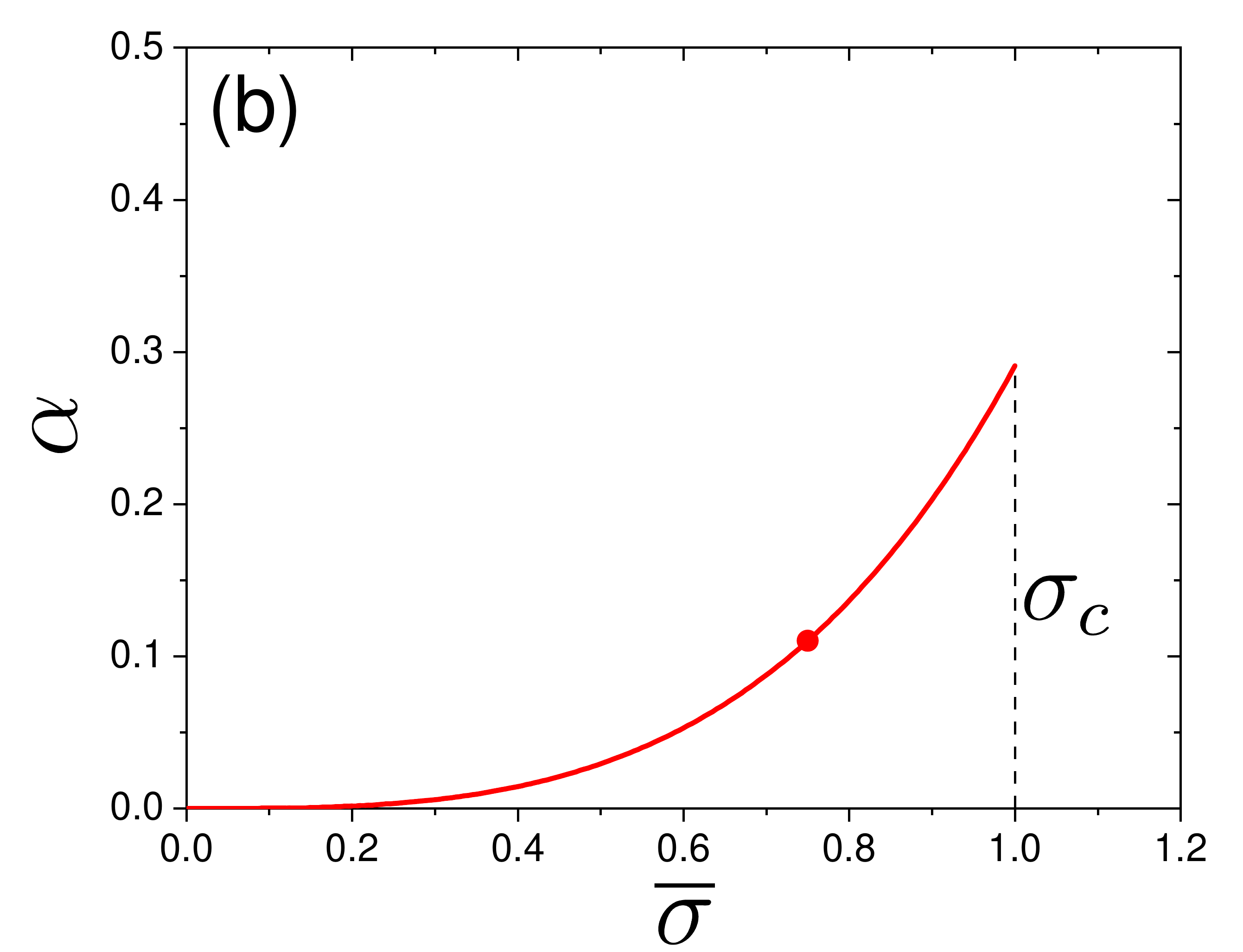}
\caption{Mean-field elastoplastic model results with $\sigma_c=\sigma_0=1$. (a): $P(\sigma)$ obtained by Eq.~(\ref{eq:P_MF}) with $\overline \sigma=0.749$ and 
%$\overline \sigma=0.74948$
$\alpha=0.110$ whose values are
%$\alpha=0.110138$
estimated by KEP. (b): The $\alpha$ versus $\overline \sigma$ plot obtained Eq.~(\ref{eq:alpha_MF}). The circle point corresponds to the prediction by KEP.}
\label{fig:MF}
\end{figure} 

\vspace{0.2cm}
{\it Regime II:} $\overline{\sigma} < \sigma < \sigma_c$. In this case the first term of the RHS of Eq.~(\ref{eq:MF_SS}) can be dropped because it is exponentially smaller than the second one. In consequence, the equation simplifies to 
\begin{equation}
 - \nu(\sigma, \sigma_c)P(\sigma) + \Gamma y(\sigma) = 0.
\label{eq:MF_SS2}
\end{equation}
which implies that in Regime II the steady state distribution reads: 
\begin{equation}
  P(\sigma)= e^{-\frac{(\Delta E(\overline{\sigma})-\Delta E(\sigma))}{T}}y(\sigma) \,\,,
\end{equation}
which is exponentially small at low $T$. 

\vspace{0.2cm}
{\it Regime III:} $\sigma_c \leq \sigma $. In this case the third term of the RHS of Eq.~(\ref{eq:MF_SS}) is zero. In consequence, the equation simplifies to 
\begin{equation}
 \alpha \Gamma \frac{\partial^2 P(\sigma)}{\partial \sigma^2} - \frac{1}{\tau_0} P(\sigma) = 0.
\label{eq:MF_SS3}
\end{equation}
By imposing continuity at $\sigma = \sigma_c$ and normalizability, one therefore finds:
\[
P(\sigma)=e^{-\frac{\Delta E(\overline{\sigma})-\Delta E(\sigma_c)}{T}} y(\sigma_c) \exp \left[ -\frac{\sigma-\sigma_c}{\sqrt{\alpha e^{-\Delta E(\overline \sigma)/T}}}\right]
\,\,,
\]
which implies a fast decay to zero in Regime III from an already very small value. This regime is therefore completely negligible. \\\\
We can finally check self-consistently our initial assumption by computing $\Gamma$ using the solution we found. By decomposing the integral in Eq.~(\ref{eq:yield_rate}) in the three regimes, one finds that the third one gives a negligible contribution. The first one gives  
\begin{equation}
    \int_{-\overline{\sigma}}^{\overline{\sigma}} \mathrm{d} \sigma \ \frac{1}{\tau_0} e^{-\frac{\Delta E(\sigma)}{T}} P(\sigma).
    \label{eq:yield_rate2a}
\end{equation}
This integral is dominated by integration around the edges because the exponentially small term is exponentially larger there. Up to exponential accuracy, it gives a contribution that is of the order $e^{-\frac{\Delta E(\overline{\sigma})}{T}}=e^{-\frac{\overline{E}}{T}}$, which is indeed the one we assumed in the very first place. Physically this contribution corresponds to relaxations due to the sites with the typical smallest barriers. \\
The second regime also gives a contribution $e^{-\frac{\overline{E}}{T}}$ since in the integral,
\[
2\int_{\overline{\sigma}}^{\sigma_c} \mathrm{d} \sigma \ \frac{1}{\tau_0} e^{-\frac{\Delta E(\sigma)}{T}} \,\, e^{-\frac{(\Delta E(\overline{\sigma})-\Delta E(\sigma))}{T}}y(\sigma),
\]
the factor $e^{-\frac{\Delta E(\overline{\sigma})}{T}}=e^{-\frac{\overline{E}}{T}}$ can be plugged out and the all the others terms give a multiplicative contribution of the order of one. This concludes the analysis of the steady state solution. 

\subsection{Relaxation time-scale and Elasticity-induced facilitation}

We discuss the role of elasticity-induced facilitation on the relaxation time-scale within the framework of the mean-field theory presented above.
In particular, we discuss how the magnitude of stress redistribution or elastic interaction characterized by $\alpha$ modifies the energy barrier $\Delta E(\overline \sigma)$ governing the relaxation time. 

First, we consider the case in which there is no elasticity (hence no facilitation) by eliminating the stress redistribution term, i.e., $\mathcal{L}(P,P)=0$ in Eq.~(\ref{eq:KEP}). The corresponding equation on the local stress distribution is therefore:
\begin{eqnarray}
\frac{\partial P_i(\sigma, t)}{\partial t} =  - \nu(\sigma, \sigma_c) P_i(\sigma, t) + \Gamma_i(t)y(\sigma) \,\,.
\label{eq:KEP-ni}
\end{eqnarray}
This is like a trap model equation \cite{monthus1996models}. At low temperature the steady state solution (for each site) is simply
\[
P(\sigma)=\frac{\Gamma}{\nu(\sigma, \sigma_c)} y(\sigma),
\]
which leads to an exponentially peaked distribution around zero (decreasing as fast as $e^{\frac{\Delta E({\sigma})}{T}}$ going away from zero). The value of $\Gamma$ is therefore simply given by $\Gamma \sim e^{-\frac{\Delta E(0)}{T}}$. Recall that $\Delta E(\sigma)$ is a decreasing function of $\sigma$ for $\sigma \geq 0$. In consequence, we find the phenomenon shown in Fig.~\ref{fig:dynamics}(b) and discussed in the main text: elasticity reduces the effective energy barrier and hence the relaxation timescale. Within our analysis this corresponds to the change from $\Delta E(0)$ to $\Delta E(\overline{\sigma})$.
This can also be seen from the general solution presented in Fig.~\ref{fig:MF}. 
With decreasing $\alpha$ (which characterizes the magnitude of elastic interaction), $\overline \sigma$ decreases and hence $P(\sigma)$ becomes more narrow peaked. Concomitantly, the energy barrier $\Delta E(\overline \sigma)$ increases. When $\alpha$ goes to zero, $\overline \sigma$ also becomes zero.

To conclude, our mean-field theoretical analysis naturally explains the facilitation mechanism caused by elasticity.

\end{document}